\documentclass[a4paper]{report}
\usepackage[twoside=false,margin=2.5cm,bindingoffset=2cm]{geometry}
\usepackage[german]{babel}
\usepackage{tikz}
\usetikzlibrary{arrows}
\usepackage{graphicx}
\usepackage{acronym}
\usepackage[frozencache,cachedir=.]{minted}
\usepackage{longtable}
\usepackage{listings}
\usepackage{comment}

\title{
Entwicklung einer Webanwendung\\ zur Generierung von skolemisierten RDF Daten\\ für die Verwaltung von Lieferketten\\~\\
\textbf{Abschlussarbeit}\\~\\
zur Erlangung des akademischen Grades:\\~\\
\textbf{Master of Science (M.Sc.)}\\~\\
an der\\~\\
Hochschule für Technik und Wirtschaft Berlin\\
Fachbereich 4: Informatik, Kommunikation\\ und Wirtschaft\\
Studiengang Angewandte Informatik\\~\\
1. Prüfer: Prof. Dr. Thomas Hoppe\\
2. Prüfer: Dipl.-Inf. Marko Harasic\\~\\~\\
Eingereicht von Roman Laas
\date{2. Oktober 2023}
}

\begin{document}

\maketitle

\pagenumbering{Roman}

\vspace*{\fill}
\section*{\centering Kurzbeschreibung}

Für eine frühzeitige Erkennung von Lieferengpässen müssen Lieferketten in einer geeigneten digitalen Form vorliegen, damit sie verarbeitet werden können. 
Der für die Datenmodellierung benötigte Arbeitsaufwand ist jedoch, gerade IT-fremden Personen, nicht zuzumuten. 

Es wurde deshalb im Rahmen dieser Arbeit eine Webanwendung entwickelt, welche die zugrunde liegende Komplexität für den Benutzer verschleiern soll. 
Konkret handelt es sich dabei um eine grafische Benutzeroberfläche, auf welcher Templates instanziiert und miteinander verknüpft werden können. 
Für die Definition dieser Templates wurden in dieser Arbeit geeignete Konzepte erarbeitet und erweitert. 

Zur Erhebung der Benutzerfreundlichkeit der Webanwendung wurde abschließend eine Nutzerstudie mit mehreren Testpersonen durchgeführt. 
Diese legte eine Vielzahl von nützlichen Verbesserungsvorschlägen offen. 

\section*{\centering Abstract}

For early detection of supply bottlenecks, supply chains must be available in a suitable digital form so that they can be processed. 
However, the amount of work required for data modeling cannot be expected of people who are not familiar with IT topics. 

Therefore, a web application was developed in the context of this thesis, which is supposed to disguise the underlying complexity for the user. 
Specifically, this is a graphical user interface on which templates can be instantiated and linked to each other. 
Suitable concepts for the definition of these templates were developed and extended in this thesis. 

Finally, a user study with several test persons was conducted to determine the usability of the web application. 
This revealed a large number of useful suggestions for improvement. 
\vfill

\tableofcontents

\listoffigures

\renewcommand\listoflistingscaption{Listingverzeichnis}
\listoflistings 

\listoftables

\clearpage
\pagenumbering{arabic}

\chapter{Einleitung} \label{einleitung}

In unserer modernen und vernetzten Welt sind Lieferketten nicht einfach nur das Bindeglied zwischen Produzenten und Konsumenten, sondern gleichzeitig ein entscheidendes Rückgrat der globalen Wirtschaft.
Spätestens die COVID-19-Pandemie hat jedoch gezeigt, wie anfällig viele Lieferketten gegenüber schwerwiegenden Störungen sind. 
Die Resilienz, also die Fähigkeit, solchen Störungen zu widerstehen, sich den veränderten Umständen anzupassen und den normalen Betrieb wiederherzustellen, war und ist in vielen Fällen mangelhaft. \cite[S. V]{Ivanov2023}

Dabei sind es nicht nur globale Pandemien, die sich negativ auf Lieferketten auswirken. 
Auch lokal begrenzte Ereignisse, wie etwa eine blockierte Haupttransportroute, können mitunter globale Lieferengpässe auslösen. 
Als prominentestes Beispiel der letzten Jahre sei hierbei die Havarie des Containerschiffs Ever Given und die damit einhergehende Blockade des Suezkanals für mehrere Tage im März 2021 genannt. 

Doch auch ohne (direktes) menschliches Verschulden ist in nächster Zeit mit Störungen von globalem Ausmaß zu rechnen, wie jüngst Staus am Panamakanal aufgrund von Niedrigwasser belegen. 
Hauptursache hierfür dürfte der Klimawandel sein, der sich auch mit anderen Naturkatastrophen wie Waldbränden, Überschwemmungen oder Stürmen immer stärker negativ bemerkbar macht. 

Längst gibt es Berufsgruppen, die sich im Rahmen des Logistikmanagements auch um Themen wie Resilienz kümmern. 
Diese als \textit{Supply Chain Management} (Lieferkettenmanagement) bezeichnete Disziplin setzt hierbei selbstredend auf die Unterstützung durch Computer, etwa für Datenmodellierung und -analyse sowie Simulationen. \cite[S. 245 ff.]{Hohmann2022}

Insgesamt betrachtet scheint es daher nur logisch, verstärkt Forschungsanstrengungen im Bereich der Lieferketten-Resilienz zu tätigen und dabei neue Technologien und Werkzeuge zu schaffen, die von der Datenmodellierung über Simulationen bis hin zu Visualisierungen reichen. 

\section{Hintergrund}

Im Rahmen des ResKriVer Projekts entsteht eine wissensbasierte Informationsplattform, die bei der Verbesserung der Resilienz von Lieferketten beitragen soll \cite{bfwk}.
Lieferketten werden in diesem Projekt als \textit{Knowledge Graphs} (Wissensgraphen) erfasst und dokumentiert. 
Sie sind die Basis für Vorhersagen und Analysen, um Lieferengpässe und ihre Auswirkungen frühzeitig zu erkennen \cite{reskriver}.
Dabei tritt jedoch die Tatsache, dass viele Informationen von Lieferketten nicht bekannt sind, als Problem auf. 
Es wurde nach einer Möglichkeit gesucht, wie man Lieferketten trotz vieler unbekannter Daten dennoch für Analysen verwenden und nachträglich bekannt gewordene Informationen ergänzen kann. 

Die Idee, Graphen anhand von Templates zu generieren und dabei unbekannte Entitäten mit skolemisierten Identifizierern auszustatten, stammt aus diesem Projekt. 
Die Shape Expressions Sprache wurde dabei als potenziell geeignete Technologie identifiziert, mithilfe derer Graphstrukturen für diesen Zweck beschrieben werden können. 
Im Rahmen von zwei Studienarbeiten (\cite{shex-1} und \cite{shex-2}) wurde sie bereits kennengelernt und erste Möglichkeiten zur Generierung von Graphen ausgelotet. 

\section{Motivation und Problemstellung}

Wenngleich das Konzept des \textit{Semantic Web} und seine Technologien, wie das Resource Description Framework, seit Jahrzehnten bekannt sind, so sind diese dennoch nicht weit verbreitet. 
Selbstredend gilt das genauso oder sogar noch mehr auch für die Shape Expressions Sprache. 

Personen, die sich mit Lieferketten befassen und gewillt sind, diese für eine weiterführende Verarbeitung zu digitalisieren, kann jedoch nicht zugemutet werden, sich in diese Themenbereiche einzuarbeiten. 
Das gilt erst recht, wenn diese Personen IT-fremd sind, was dabei ein durchaus denkbares Szenario ist. 

Es muss also eine Möglichkeit gefunden werden, wie semantische Technologien auf der einen Seite mit Lieferketten-Experten auf der anderen Seite zusammengebracht werden können. 
Es scheint sich anzubieten, dafür eine grafische Benutzeroberfläche zu schaffen, welche die zugrundeliegende Komplexität für den Benutzer verschleiert. 

\section{Ziel der Arbeit}

Im Rahmen dieser Arbeit soll eine Webanwendung entwickelt werden, die es ermöglichen soll, Lieferketten mithilfe vordefinierter Templates auf einer grafischen Benutzeroberfläche modellieren zu können. 
Zunächst müssen dafür jedoch offene Fragen bei der für diesen Zweck vorgesehenen Shape Expressions Sprache zur Modellierung von Templates geklärt werden. 
Erst dann kann die folgende Forschungsfrage gestellt werden: 

\begin{center}
\textit{Lässt sich auf der Basis von Shape Expressions eine Webanwendung entwickeln, die auch IT-fremden Personen die Modellierung von Lieferketten als RDF Graph ermöglicht?}
\end{center}

\section{Aufbau der Arbeit}

Diese Arbeit gliedert sich in insgesamt 8 Kapitel und orientiert sich dabei an den Vorgaben für den inhaltlichen Aufbau nach der \textit{Richtlinie zur Anfertigung einer Abschlussarbeit im Studiengang Angewandte Informatik} aus \cite[S. 7]{richtlinien}. 

Bereits gegeben wurde mit Kapitel \ref{einleitung} eine Einleitung in das Thema sowie die Hintergründe und das Ziel der Arbeit. 

Als nächstes folgt ein umfassendes Kapitel zu konzeptionellen und technischen Hintergründen. 
Dabei werden Grundbegriffe erörtert und die im Rahmen dieser Arbeit benötigten Konzepte und Technologien betrachtet. 
Dazu zählen etwa das Resource Description Framework in Abschnitt \ref{rdf-grundlagen} und die Shape Expressions Sprache in Abschnitt \ref{shex-grundlagen}. 
Doch auch Usability-Interaktionsprinzipien und -Testmöglichkeiten müssen mit Abschnitt \ref{usability-grundlagen} erarbeitet werden, um später die Benutzerfreundlichkeit der entwickelten Webanwendung erfassen zu können. 
Mit einer Übersicht über Bibliotheken und Frameworks, die für die Entwicklung vorgesehen sind, wird das Kapitel \ref{grundlagen} abgeschlossen. 

In Kapitel \ref{templates mit shex} wird erläutert, wie \acs{shex} Shapes bzw. Schemas für die Generierung von \ac{rdf} Graphen verwendet werden können und welche Methoden dafür zusätzlich definiert werden müssen. 

Anschließend werden in Kapitel \ref{anforderungen} die funktionalen und nicht-funktionalen Anforderungen an die zu entwickelnde Webanwendung nach Benutzergruppen erhoben und analysiert. 

Darauf folgt mit Kapitel \ref{konzept} die Konzeption. 
Es wird ein Überblick über die geplante Architektur auf Komponenten-, Datenbank- und Code-Ebene gegeben, gefolgt von den Design-Ideen für die Benutzeroberfläche. 

Diese werden sogleich mit der in Kapitel \ref{implementierung} beschriebenen Implementierung umgesetzt. 
Dabei werden nacheinander das entwickelte Backend, Frontend und schließlich die Konfiguration der Containervirtualisierung beschrieben. 

Anschließend kann die Nutzerstudie durchgeführt werden, welche in Kapitel \ref{durchführung nutzerstudie} dargestellt wird. 
Es wird dabei ein Überblick über den Ablauf und die verpflichteten Testpersonen gegeben. 
Anschließend werden zwei Usability-Test-Methoden durchgeführt und die Ergebnisse vorgestellt. 

Abschließend wird eine Zusammenfassung, bestehend aus Fazit, Limitationen und Ausblick, in Kapitel \ref{zusammenfassung} gegeben. 

\chapter{Grundlagen} \label{grundlagen}

\section{Lieferketten}

Für ein besseres Verständnis über die Gestalt von Lieferketten sollen diese im Folgenden aus verschiedenen Blickwinkeln betrachtet werden. 
Dazu wird auch der Begriff \textit{Lieferkette} allgemein definiert. 

Nach Gudehus ist eine Lieferkette (allgemeiner auch als Logistikkette bezeichnet) eine Aneinanderreihung von Transportverbindungen und Zwischenstationen.
Diese wird, ausgehend von einer Lieferstelle, von angeforderten Waren und Gütern durchlaufen und endet an einer Empfangsstelle. 
Unterschieden werden muss dabei nach internen und externen Lieferketten. 
Mit ersteren sind jene Lieferketten gemeint, die sich innerhalb eines Betriebsgeländes, also beispielsweise vom Wareneingang zur Produktionsstelle oder vom Lager zum Warenausgang, erstrecken. 
Bei externen Lieferketten handelt es sich hingegen um Transportverbindungen zwischen Unternehmen, etwa vom Warenausgang des Lieferanten zum Wareneingang des Kunden. \cite[S. 937]{Gudehus2012}

Eine ähnliche Definition liefert das am 1. Januar 2023 in Kraft getretene Lieferkettensorgfaltspflichtengesetz (LkSG), welches auch als Lieferkettengesetz bezeichnet wird.
Es beschreibt den Umfang der Lieferketten eines Unternehmens in \S 2 Abs. 5 als "[...] alle Schritte im In- und Ausland, die zur Herstellung der Produkte und zur Erbringung von Dienstleistungen erforderlich sind, angefangen von der Gewinnung der Rohstoffe bis zu der Lieferung an den Endkunden [...]".
Dabei sei das Handeln von unmittelbaren und mittelbaren Zulieferern ebenso Teil einer Lieferkette, wie jenes des vom Gesetz betroffenen Unternehmens. 

\begin{figure}[H]
\centering
\includegraphics[height=.15\textheight]{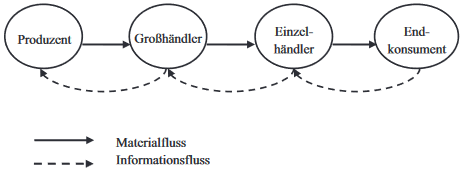}
\caption{Schematische Darstellung einer simplen Lieferkette aus \cite[S. 7]{Hohmann2022}}
\label{lieferkette_schematisch}
\end{figure}

Hohmann liefert mit Abbildung \ref{lieferkette_schematisch} eine schematische Darstellung einer Lieferkette in ihrer einfachsten möglichen Ausprägung. 
Zu sehen ist ein Materialfluss von einem Produzenten über einen Groß- und einen Einzelhändler bis hin zum Endkonsumenten. 
In Kontakt treten die Akteure dabei jeweils nur mit dem in der Lieferkette voranstehenden Teilnehmer. 
In der Praxis gestalten sich Lieferketten meist jedoch ungleich komplexer. 
So besitzen diese nicht selten deutlich mehr als vier Stufen, von denen jede von mehreren Unternehmen gleichzeitig besetzt sein kann. 
Je komplexer eine Lieferkette ausfällt, desto eher sollte diese nach Hohmann dabei als ein Liefernetzwerk bezeichnet werden. 
Dieser Begriff hat sich in der Literatur jedoch nicht durchgesetzt, weshalb man heute sowohl einfache als auch komplexe Lieferstrukturen allgemein als Lieferkette bezeichnet. \cite[S. 6 ff.]{Hohmann2022}

Um die Wettbewerbsfähigkeit zu gewährleisten, müssen Unternehmen ihre Lieferketten stets überwachen und gegebenenfalls optimieren oder sogar komplett neu schaffen \cite[S. 937]{Gudehus2012}. 
Dieser Umstand trifft jedoch nicht nur auf Gesellschaften mit einem primären Gewinninteresse zu, sondern auch auf jene, bei denen die Sicherstellung der eigenen Tätigkeit im Vordergrund steht. 
Dies ist etwa bei öffentlichen Rettungs-, Ordnungs- oder Gesundheitseinrichtungen der Fall. 

\section{Templates}

Ein Template ist in der Informatik eine Art Vorlage oder Muster, das als Ausgangsbasis für die Erstellung verschiedener Elemente dient, seien es Dokumente, Programme oder Datenstrukturen. 
Im Bereich der Datenverarbeitung können Templates etwa als strukturelle Vorlagen für Daten dienen, beispielsweise in Form von XML-Templates, welche die Struktur eines XML-Dokuments vorgeben, das dann mit konkreten Daten gefüllt wird. 
Das vereinfacht die Datenerfassung und -analyse, da die Daten in einer vordefinierten und somit leichter verständlichen Weise organisiert sind. \cite{template1}

In der Webentwicklung stellen Templates in der Regel \acs{html}-, \acs{css}- oder JavaScript-Dateien dar, die das Grundgerüst für eine Webseite bilden. 
Diese Dateien enthalten häufig Platzhalter, die zur Laufzeit durch tatsächliche Daten ersetzt werden, um eine vollständige und funktionsfähige Webseite zu erstellen. 
Dies erleichtert nicht nur die Webentwicklung erheblich, sondern sorgt auch für eine konsistente Benutzererfahrung, da das Design und die Struktur der Webseite durch das Template standardisiert sind. \cite{template1}

Im Kontext der Softwareentwicklung ermöglichen Templates die Erstellung von generischen Typen, was es erlaubt, Code zu schreiben, der für viele verschiedene Datentypen geeignet ist, ohne den Code für jeden einzelnen Datentyp neu schreiben zu müssen. 
Diese Flexibilität ist wichtig, da sie nicht nur die Wiederverwendbarkeit des Codes fördert, sondern auch die Wartbarkeit und Skalierbarkeit von Softwareprojekten verbessert. \cite{template1}

In der Textverarbeitung und bei der Dokumenterstellung bieten Templates vorformatierte Dokumente an, die als Ausgangspunkt für neue Dokumente dienen können. 
Solche Templates enthalten in der Regel bereits spezifische Formatierungen, Abschnitte und manchmal auch Standardtexte, die dann nach Bedarf angepasst werden können. 
Dies spart Zeit und Mühe und sorgt für ein einheitliches Erscheinungsbild der Dokumente. \cite{template2}

Obwohl Templates also in verschiedenen Bereichen der Informatik auftreten und sich deshalb zwangsläufig mehr oder weniger voneinander unterscheiden, so eint sie doch das Konzept, Strukturen vorzugeben, welche wiederverwendet werden können und damit Aufwand sparen. 

Abschließend lässt sich zudem festhalten, dass es keine bekannten Template-Mechanismen oder -Technologien in der \ac{rdf} Domäne gibt, was in Anbetracht der mittlerweile verfügbaren Datenmassen, die zum Teil vermutlich auch händisch erfasst wurden, verwunderlich ist. 

\section{Skolemisation} \label{skolemisation}

In der mathematischen Logik ist die Skolemisation ein Prozess, mit dem Existenzquantoren in einer prädikatenlogischen Formel eliminiert werden, indem sie durch geeignete Funktionssymbole ersetzt werden.
Dieser Vorgang ist nach dem norwegischen Mathematiker Albert Thoralf Skolem benannt.
Eine Formel der Prädikatenlogik erster Stufe, die in Pränexform vorliegt, hat die Eigenschaft, dass alle ihre Quantoren am Anfang der Formel stehen.
Das Ziel der Skolemisation ist es, alle Existenzquantoren zu eliminieren, ohne die Erfüllbarkeit der Formel zu verändern.

Für jede Instanz eines Existenzquantors in der Formel, der durch einen oder mehrere Allquantoren vorausgeht, wird der Existenzquantor durch eine neue Funktion (Skolemfunktion) ersetzt.
Wenn keine Allquantoren vorausgehen, wird der Existenzquantor durch einen neuen Konstanten-Term (Skolemkonstante) ersetzt.

Es sei als Beispiel die folgende Formel gegeben:
\[\forall x \exists y P(x, y)\]
In dieser Formel steht \(\forall\) für den Allquantor, welcher besagt, dass die folgende Aussage für alle Elemente einer gegebenen Menge wahr ist. \(x\) und \(y\) sind Variablen, die Elemente dieser Menge repräsentieren. \(\exists\) ist der Existenzquantor und besagt, dass es mindestens ein Element in der Menge gibt, für das die folgende Aussage wahr ist. Das Prädikat \(P\) könnte irgendeine Beziehung zwischen \(x\) und \(y\) darstellen. \cite{skolem-mpg}

Die Skolemisation dieser Formel führt zu:
\[\forall x P(x, f(x))\]
Hier repräsentiert \(f\) die Skolemfunktion. Sie ersetzt den zuvor durch den Existenzquantor beschriebenen Wert \(y\). Diese Funktion liefert für jeden Wert von \(x\) einen Wert \(y\), sodass das Prädikat \(P(x, y)\) wahr ist. \cite{skolem-mpg}

Es ist wichtig zu beachten, dass die Skolemisation die ursprüngliche Bedeutung der Formel ändert.
Während die ursprüngliche Formel besagt, dass es für jeden Wert von \(x\) einen Wert von \(y\) gibt, so dass \(P(x, y)\) wahr ist, besagt die skolemisierte Formel, dass für jeden Wert von \(x\) die Funktion \(f\) einen Wert liefert, so dass \(P(x, f(x))\) wahr ist. \cite{skolem-mpg}

Skolemisation wird oft in Beweisen und Algorithmen in der mathematischen Logik und in der theoretischen Informatik verwendet, insbesondere bei der Umwandlung von Formeln in die Klauselform, einem Schritt, der oft in automatisierten Deduktionsverfahren erforderlich ist. \cite{skolem-mpg}

\section{Resource Description Framework} \label{rdf-grundlagen}

Das \acf{rdf} dient als Grundlage für die Speicherung und den Austausch von vernetzten Informationen.
Es wurde im Rahmen der Vision für das Semantische Web entwickelt, einem Konzept, das Ende der 1990er Jahre von Tim Berners-Lee, dem Erfinder des World Wide Web, vorgestellt wurde.
Die ursprüngliche \ac{rdf}-Fassung wurde 1999 als \ac{w3c} Recommendation veröffentlicht und später im Jahr 2004 in der Version 1.0 standardisiert.
Im Jahr 2014 wurde mit \ac{rdf} 1.1 die heute aktuellste Fassung veröffentlicht.

\subsection{Konzepte}

\begin{figure}[H]
\centering
\includegraphics[height=.04\textheight]{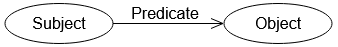}
\caption{Schematische Darstellung eines Triples in \ac{rdf} aus \cite{rdf_concepts}}
\label{rdf-triple-schema}
\end{figure}

In \ac{rdf} besteht die grundlegendste Datenstruktur aus sog. Tripeln, die jeweils aus einem Subjekt, einem Prädikat und einem Objekt bestehen.
Subjekt und Objekt können dabei als Knoten in einem Graph angesehen werden, die über das Prädikat als gerichtete Kante von Subjekt zu Objekt miteinander in Beziehung stehen (siehe Abbildung \ref{rdf-triple-schema}). 
Das Subjekt soll dabei mithilfe von Prädikat und Objekt beschrieben werden, sodass sich eine logische Aussage ergibt. 
Es gibt dabei drei grundlegende Datentypen in \ac{rdf}: \acs{iri}s (\acl{iri}), Literale und Blank Nodes. 
\acs{iri}s können für alle Teile eines Triples verwendet werden und sollen eine Ressource eindeutig identifizieren. 
Sie sorgen also für Eindeutigkeit und ermöglichen damit das Verlinken von Daten über verschiedene Datenquellen und Anwendungen hinweg.
Dagegen wird mit Blank Nodes nur generell das Vorhandensein einer unidentifizierten Ressource angegeben. 
Sie sollen in Abschnitt \ref{blanknodes} näher betrachtet werden. 
Blank Nodes können dabei nur für das Subjekt oder das Objekt verwendet werden. 
Literale sind konkrete Werte, die nur als Objekt eines Triples auftreten können. 

\ac{rdf} kann in einer Vielzahl von Syntaxformaten repräsentiert werden. Zu den bekanntesten zählen RDF/XML, eine XML-basierte Darstellung; JSON-LD, das eine JSON-basierte Repräsentation bietet und vermehrt bei der Webentwicklung eingesetzt wird; und die \ac{turtle}, die für ihre Lesbarkeit und einfache Manipulation bekannt ist und in Abschnitt \ref{turtle} genauer beschrieben wird.

Die Anfragesprache von \ac{rdf} ist die \ac{sparql}. 
Sie ist als eigene Recommendation vom \ac{w3c} standardisiert und ermöglicht es, Daten aus \ac{rdf} Graphen mithilfe von Graph Pattern Matching (Graph Musterabgleich) abzufragen oder zu manipulieren. 
Mit \ac{sparql} 1.1 von 2013 sind auch komplexere Anfragen möglich, die zum Teil Parallelen zu bekannten relationalen Datenbank-Anfragesprachen besitzen. 
Als Ausgabeformat für Ergebnisse unterstützt \ac{sparql} gängige \ac{rdf} Syntaxen, darunter auch die zuvor genannten. \cite{sparql-primer}

Ergänzend zu \ac{rdf} und \ac{sparql} existiert auch \ac{rdfs}, ein Vokabular, das es ermöglicht, Klassen und Eigenschaften von \ac{rdf}-Ressourcen zu beschreiben.
RDFS dient als einfaches Ontologie-Sprachsystem und ist besonders nützlich für die Definition von Beziehungen und Hierarchien.
Für komplexere Ontologien wird hingegen die \ac{owl} \cite{owl} verwendet.
\ac{owl} erlaubt die Definition von komplizierteren Beziehungen zwischen Klassen und Eigenschaften und bietet fortschrittlichere Inferenzmöglichkeiten.
Es kann beispielsweise verwendet werden, um logische Schlussfolgerungen über Klassen von Ressourcen zu treffen oder um Einschränkungen und Regeln zu definieren.

Zusammen bilden \ac{rdf}, \ac{sparql}, \ac{rdfs} und \ac{owl} das Fundament für das Semantische Web, welches im Kern darauf abzielt, Daten im Internet maschinenlesbar und damit intelligenter verarbeitbar zu machen.

\subsection{Terse RDF Triple Language} \label{turtle}

Die \acl{turtle} (\acs{turtle}) \cite{turtle} ist eine Syntax zur Darstellung semantischer Daten in \ac{rdf} und zielt darauf ab, die Lesbarkeit und Kompaktheit von \ac{rdf}-Daten zu erhöhen, ohne dabei die Ausdruckskraft zu beschränken.
\ac{turtle} unterstützt dabei etwa die Verwendung von Präfixen für IRIs, was die Ausgabe erheblich verkürzt und damit die Lesbarkeit erhöht. 
Dies wird in Listing \ref{turtle example} gezeigt, indem diese in den Zeilen 2 bis 4 definiert und dann in den Zeilen 6 bis 8 verwendet werden. 

Zudem bietet \ac{turtle} Mechanismen für die Darstellung von Listen und von mehrfachen Objekten für ein einziges Subjekt-Prädikat-Paar, was die Darstellung von komplexen \ac{rdf}-Graphen weiter vereinfacht.
Zu sehen ist dies in den Zeilen 6 bis 8 des Listings \ref{turtle example}, in denen \texttt{<Roman>} nur einmal definiert wird, jedoch durch die Verwendung von Semikolons für alle drei Triple gilt. 

\begin{listing}[H]
\begin{minted}[xleftmargin=20pt,linenos]{text}
@base <http://fokus.fraunhofer.de/> .
@prefix rdf: <http://www.w3.org/1999/02/22-rdf-syntax-ns#> .
@prefix xsd: <http://www.w3.org/2001/XMLSchema#> .
@prefix foaf: <http://xmlns.com/foaf/0.1/> .

<Roman> rdf:type foaf:Person ;
        foaf:name "Roman Laas"@de ;
        _:age "30"^^xds:integer .
\end{minted}
\caption{\acs{rdf} Graph in \acs{turtle} Syntax} \label{turtle example}
\end{listing}

Datentypen oder Sprach-Tags für Literale können in \ac{turtle} unkompliziert direkt nach einem Literal angegeben werden. 
Im Beispiel zeigt Zeile 7 die Verwendung eines Sprach-Tags und Zeile 8 die eines Datentyps. 

In der Praxis wird Turtle oft in Kontexten verwendet, in denen eine menschliche Interaktion stattfindet, also etwa bei händischer Datenmodellierung oder -manipulation oder auch nur beim Sichten von Datensätzen. 

\subsection{Probleme mit Blank Nodes} \label{blanknodes}

Blank Nodes sollen in \ac{rdf} die Existenz einer Sache angeben, ohne diese zu identifizieren und können sowohl dabei für das Subjekt als auch das Objekt eines Triples verwendet werden. 
Mit dieser Zusammenfassung ist - zumindest die konzeptionelle - Definition von Blank Nodes in den aktuellen \ac{w3c} Recommendations erschöpft. 
In den ursprünglichen Fassungen vor der Version 1.1 aus 2014 hieß es hingegen noch zusätzlich: 
\textit{\glqq Given two blank nodes, it is possible to determine whether or not they are the same.\grqq{}} \cite[6.6]{rdf2004}
Es sei also möglich herauszufinden, ob zwei Blank Nodes gleich sind. 
Diese Passage wurde in der Version 1.1 gestrichen und mit einem Hinweis zu sog. \textit{Blank Node Identifiern} ersetzt. 
Bei diesen handelt es sich um lokale Identifizierer, die vollständig von jener Implementierung zu verwalten sind, die \ac{rdf} verwendet. 
Einzig auf eine korrekte Referenzierung für gleiche Blank Node Instanzen sei dabei zu achten. 

Diese vage Definition von Blank Nodes widerspricht dem eigentlichen Tenor aus \cite{rdf-primer}, nach dem \ac{rdf} ein Framework für den Austausch von Informationen zwischen Applikationen ohne den Verlust der Bedeutung ist. 
Auch dem Grundsatz, dass erfasste Informationen für Applikationen nutzbar sein sollen, für die sie ursprünglich nicht vorgesehen waren, wird damit nicht Rechnung getragen.

Mallea et al. betrachten diesen Umstand in \cite{Mallea2011OnBN} kritisch und beleuchten den Umgang mit Blank Nodes in \ac{rdf} Standards und deren Verbreitung in populären Datensätzen. 
Für alle bekannten \ac{rdf} Syntaxe konstatieren sie dabei eine volle Unterstützung von Blank Nodes und bemerken zudem deren Nutzung z.B. bei \textit{internen} Mechanismen, wie etwa bei Standardwerten in \textit{Collections} der \ac{turtle} Syntax. 
Jedoch fallen auch Probleme auf, etwa bei der Serialisierung. 
So gibt es zum einen keinen Algorithmus, welcher Graphen mit Blank Nodes in Polynomialzeit auf Isomorphie, wie in \cite[3.6]{rdf-concepts} beschrieben, prüfen kann. 
Zum anderen lassen sich Änderungen an einem Graph aufgrund der beliebigen \textit{Blank Node Identifier} meist nicht sinnvoll verfolgen. \cite[1]{hoganSkol} \cite[4.1]{Mallea2011OnBN}

Auch in \ac{rdfs}, \ac{owl} und \ac{sparql} finden Blank Nodes Anwendung. 
Für \ac{rdfs} und \ac{owl} auf Basis der Umsetzung der \textit{Simple Entailment} Regeln aus \cite[5.2]{rdf-semantics} und damit etwa als Ersatz für Literale bei der Erzeugung von ansonsten ungültigen Statements. 
Bei \ac{sparql} können Blank Nodes hingegen als Variablen in einer Anfrage auftauchen. 
Doch auch für diese Standards finden Mallea et al. kleine Widersprüchlichkeiten und Ungenauigkeiten bei der Verwendung von Blank Nodes. 

Aufgrund der Probleme mit Blank Nodes raten Heath und Bizer in \cite{Heath2011} dazu, diese vollständig zu vermeiden. 
Sie gehen dabei auf die angesprochene lokal begrenzte Gültigkeit und die Probleme beim Zusammenführen von Graphen ein und raten, allen Knoten einen referenzierbare \ac{iri} zuzuweisen. 
Dennoch wird in \cite{everything} im \textit{BTC-2012} Korpus bestehend aus 1,23 Milliarden Triplen aus 8,37 Millionen \ac{rdf} Dokumenten erhoben, dass 44,9 Prozent dieser Dokumente Blank Nodes beinhalten. 

Um zu verstehen, weshalb eine so große Zahl an Blank Nodes in den betrachteten Dokumenten vorhanden ist, haben Mallea et al. sowohl Beitragende als auch Mitglieder der entsprechenden Mailverteiler beim \ac{w3c} zu ihrer Ansicht über die Rolle von Blank Nodes befragt. 
In dieser Befragung (88 Teilnehmende, \textit{Multiple Choice}) gaben die Teilnehmer mit 46,9 Prozent an, dass Blank Nodes für eine unbekannte Sache verwendet werden können und mit 23,9 Prozent, dass Blank Nodes auch für eine versteckte Sache (z.B. wegen Datenschutz) verwendet werden können. 
Auch gaben die Teilnehmer mit 2,3 Prozent an, dass Blank Nodes für eine nur eventuell existierende Sache verwendet werden können. 
Dieses Ergebnis zeigt, dass sich die Bedeutung von Blank Nodes selbst unter Experten deutlich unterscheidet und sich die Ansichten zum Teil komplett von der eigentlichen Definition entfernt haben. \cite[6.3]{everything}

In \cite[3.5]{rdf-concepts} wird theoretisch das Vorgehen zum Ersetzen von Blank Nodes mit \ac{iri}s vorgestellt. 
Hierzu soll die in Kapitel \ref{skolemisation} beschriebene Skolemisation angewandt werden. 
Ein Graph mit Blank Nodes soll also zu einem Graphen transformiert werden, in welchem die Blank Nodes durch einzigartige \ac{iri}s ersetzt wurden. 
Diese damit als \textit{Skolem \ac{iri}s} bezeichneten Identifikatoren müssen dabei jedoch global einzigartig sein. 
Denn nur damit ist sichergestellt, dass sich die Bedeutung des \ac{rdf} Graphen global nicht ändert. 

In \cite[3.5]{rdf-concepts} wird weiter ausgeführt, dass die erzeugten Skolem \ac{iri}s, um auch außerhalb des eigenen Systems als eben solche erkannt zu werden, den Beschreibungen einer \textit{well-known \ac{iri}}, wie durch \cite{rfc5785} vorgegeben, folgen und den Zusatz \texttt{genid} verwenden sollten.
Eine solche generierte \ac{iri} als Ersatz für eine Blank Node könnte demnach folgendermaßen aussehen: 
\\
\centerline{\texttt{http://fokus.fraunhofer.de/.well-known/genid/d26a2d0e93413374ad70a677abc1f6}}

\section{Shape Expressions} \label{shex-grundlagen}

\subsection{Überblick}

\ac{shex} \cite{shex-primer} ist eine Sprache für die Beschreibung von Graph-Strukturen in \ac{rdf}. 
Sie ermöglicht das Aufstellen von sog. \ac{shex} Schemas, mithilfe derer Restriktionen an einen Graphen gestellt werden können. 
Ein Graph gilt als \textit{konform} wenn er alle Bedingungen eines \ac{shex} Schemas erfüllt. 
Die Konformitätsprüfung ist dabei ein zentraler Mechanismus in der technischen Beschreibung der \ac{shex} Sprache, nimmt in dieser Arbeit jedoch nur eine untergeordnete Rolle ein, weshalb im Folgenden nur vereinzelt Bezug darauf genommen wird. 

\ac{shex} ist aufgrund der robusten \ac{shexj} für Anwendungen geeignet, in denen eine maschinenlesbare Beschreibung von Datenstrukturen und -typen nützlich ist. 
Das können etwa Anwendungen zur Validierung von Daten, Generierung von Nutzerinterfaces oder Transformierung von \ac{rdf} Graphen in andere Datenformate oder -strukturen sein. 
Jedoch eignet sich \ac{shex} durch die \ac{shexc} auch für einen händischen Umgang. 
Im Rahmen dieser Arbeit werden alle \ac{shex} Angaben deshalb in der \ac{shexc} Syntax angegeben. 

Das zentralste Element in \ac{shex} ist die \textbf{Shape Expression}.
Sie besteht aus einer logischen Kombination von Node Constraints und Shapes, welche in den folgenden Abschnitten jeweils genauer beschrieben werden. 
Gebündelt werden Shape Expressions wiederum zu einem bereits angesprochenen \ac{shex} Schema. 
Dieses umfasst neben Shape Expressions üblicherweise in den ersten Zeilen des Dokuments die Angabe von \texttt{BASE}-, \texttt{PREFIX}- und \texttt{IMPORT}-Direktiven. 
Die ersten beiden sind dabei in ihrer Funktion analog zu den gleichnamigen Direktiven in der \ac{rdf} \ac{turtle} Syntax. 
Die \texttt{IMPORT}-Angabe ermöglicht hingegen das Importieren von anderen \ac{shex} Schemas, was ebenfalls im Nachfolgenden erläutert wird. 
In Listing \ref{Shex Schema Bsp} seien in den ersten drei Zeilen die unterschiedlichen Direktiven beispielhaft aufgeführt. 
Daneben befindet sich in der Zeile 5 ein Kommentar. 

\begin{listing}[H]
\begin{minted}[xleftmargin=20pt,linenos]{text}
BASE <http://fokus.fraunhofer.de/>
PREFIX foaf: <http://xmlns.com/foaf/0.1/>
IMPORT <MySchema>

# Shape Expressions below
\end{minted}
\caption{\acs{shex} Schema mit den drei Direktiven und einem Kommentar} \label{Shex Schema Bsp}
\end{listing}

\subsection{Node Constraints} \label{Node Constraints}

Node Constraints \cite[3.1]{shex-primer} legen fest, welche Werte oder Datentypen für einen bestimmten Knoten zulässig sind. 
Sie können dabei nicht alleine auftreten, sondern sind immer Teil eines Triple Constraints (siehe Abschnitt \ref{triple constraints}). 
Nachfolgend werden alle Node Constraint-Arten beschrieben und mit Beispielen dargestellt, wobei der Node-Constraint-Anteil jeweils hervorgehoben ist. 

Die grundlegendste Variante von Node Constraints sind die \textbf{Node Kind Constraints}, mit welchen die Art eines Knotens eingeschränkt werden kann. 
Mögliche Werte für Node Kind Constraints sind \texttt{\ac{iri}}, \texttt{BNode} (Blank Node), \texttt{Literal} oder \texttt{NonLiteral} (\ac{iri} oder Blank Node). 
Im Beispiel in Listing \ref{Node Kind Constraints Bsp} zeigt Zeile 4 einen Node Kind Constraint, welcher für Objekte, die in einem Triple mit dem Prädikat \texttt{foaf:knows} stehen, eine IRI vorgibt. 

\begin{listing}[H]
\begin{minted}[xleftmargin=20pt,linenos,escapeinside=||]{text}
BASE <http://fokus.fraunhofer.de/>
PREFIX foaf: <http://xmlns.com/foaf/0.1/>

<MyIriConstraint> foaf:knows |\colorbox{gray!20}{IRI}|
\end{minted}
\caption{\acs{shex} Schema mit Node Kind Constraint} \label{Node Kind Constraints Bsp}
\end{listing}

Mit \textbf{Datatype Constraints} können die für Literale zulässigen Datentypen eingeschränkt werden. 
Üblich sind hierbei beispielsweise \ac{xsd} Datentypen wie \texttt{xsd:string}, \texttt{xsd:integer} oder \texttt{xsd:boolean}. 
Doch auch das Vorhandensein eines Sprach-Tags, wie in Listing \ref{Datatype Constraints Bsp} Zeile 8 zu sehen, kann vorgegeben werden. 

\begin{listing}[H]
\begin{minted}[xleftmargin=20pt,linenos,escapeinside=||]{text}
BASE <http://fokus.fraunhofer.de/>
PREFIX rdf: <http://www.w3.org/1999/02/22-rdf-syntax-ns#>
PREFIX xsd: <http://www.w3.org/2001/XMLSchema#>
PREFIX rdfs: <http://www.w3.org/2000/01/rdf-schema#>
PREFIX ex: <http://example.com/>

<MyDateConstraint> ex:submittedOn |\colorbox{gray!20}{xsd:date}|
<MyLangStringConstraint> rdfs:label |\colorbox{gray!20}{rdf:langString}|
\end{minted}
\caption{\acs{shex} Schema mit Datentyp und Sprach-Tag Datatype Constraint} \label{Datatype Constraints Bsp}
\end{listing}

\textbf{XML Schema String Facet Constraints} können an die String-Repräsentation von Literalen, \ac{iri}s oder Blank Nodes (sofern diese eine solche besitzen) gerichtet werden. 
Neben Restriktionen für die Länge (siehe Listing \ref{XML Schema String Facet Constraints Bsp} Zeilen 4 und 5) ist es auch möglich, reguläre Ausdrücke anzugeben, denen der jeweilige String entsprechen muss (Zeile 6).

\begin{listing}[H]
\begin{minted}[xleftmargin=20pt,linenos,escapeinside=||]{text}
BASE <http://fokus.fraunhofer.de/>
PREFIX ex: <http://example.com/>

ex:IssueShape ex:submittedBy |\colorbox{gray!20}{LENGTH 10}|
ex:IssueShape ex:submittedBy |\colorbox{gray!20}{MINLENGTH 10 MAXLENGTH 20}|
ex:IssueShape ex:submittedBy |\colorbox{gray!20}{/genuser[0-9]+/i}|
\end{minted}
\caption{\acs{shex} Schema mit XML Schema String Facet Constraints aus \cite{shex-primer}} \label{XML Schema String Facet Constraints Bsp}
\end{listing}

Analog zu den beschriebenen String Facet Constraints handelt es sich bei \textbf{XML Schema Numeric Facet Constraints} um Beschränkungen, die an numerische Literale gerichtet werden können. 
Neben den in Listing \ref{XML Schema Numeric Facet Constraints Bsp} Zeile 4 und 5 gezeigten \texttt{MININCLUSIVE} und \texttt{MAXINCLUSIVE} Facet-Bezeichnungen, welche für Kleiner-Gleich respektive Größer-Gleich Vergleiche stehen, gibt es auch noch die selbsterklärenden \texttt{MINEXCLUSIVE} und \texttt{MAXEXCLUSIVE} Varianten. 
In Zeile 7 und 8 ist mit \texttt{FRACTIONDIGITS} die maximale Anzahl der Nachkommastellen auf 6 begrenzt. 
Zusätzlich gehört zu den XML Schema Numeric Facet Constraints noch \texttt{TOTALDIGITS}, womit die maximale Anzahl an Stellen eines numerischen Literals restriktiert werden kann. 

\begin{listing}[H]
\begin{minted}[xleftmargin=20pt,linenos,escapeinside=||]{text}
BASE <http://fokus.fraunhofer.de/>
PREFIX geo: <http://www.w3.org/2003/01/geo/wgs84_pos#>

<MyLatRangeConstraint> geo:lat |\colorbox{gray!20}{MININCLUSIVE -90 MAXINCLUSIVE 90}|
<MyLongRangeConstraint> geo:long |\colorbox{gray!20}{MININCLUSIVE -180 MAXINCLUSIVE 180}|
<MyLatFractionConstraint> geo:lat |\colorbox{gray!20}{FRACTIONDIGITS 6}|
<MyLongFractionConstraint> geo:long |\colorbox{gray!20}{FRACTIONDIGITS 6}|
\end{minted}
\caption{\acs{shex} Schema mit XML Schema Numeric Facet Constraints} \label{XML Schema Numeric Facet Constraints Bsp}
\end{listing}

Schließlich kann mit \textbf{Values Constraints} ein sog. \textit{Value Set} vorgegeben werden, dem ein Knoten entsprechen muss. 
In einem Value Set können sich dabei ein oder mehrere explizite Werte befinden. 
Das Beispiel aus Listing \ref{Values Constraints Bsp} Zeile 5 zeigt, wie ein Values Constraint etwa die möglichen Sprach-Tags einschränkt. 
Weiter wird in den Zeilen 6 und 7 gezeigt, wie genau ein einziger Wert vorgegeben werden kann. 

\begin{listing}[H]
\begin{minted}[xleftmargin=20pt,linenos,escapeinside=||]{text}
BASE <http://fokus.fraunhofer.de/>
PREFIX foaf: <http://xmlns.com/foaf/0.1/>
PREFIX rdf: <http://www.w3.org/1999/02/22-rdf-syntax-ns#>

<MyLanguageConstraint> foaf:name |\colorbox{gray!20}{[ @en @de ]}|
<MyMailConstraint> foaf:mbox |\colorbox{gray!20}{[ "roman.laas@student.htw-berlin.de" ]}|
<MyPersonConstraint> rdf:type |\colorbox{gray!20}{[ foaf:Person ]}|
\end{minted}
\caption{\acs{shex} Schema mit Values Constraints} \label{Values Constraints Bsp}
\end{listing}

\subsection{Triple Constraints} \label{triple constraints}

Im vorherigen Abschnitt wurden Triple Constraints \cite[3.2]{shex-primer} bereits genannt und auch in den gegebenen Beispielen gezeigt. 
Sie sollen in diesem Abschnitt nun aber genauer erläutert werden. 
Dafür muss auch das Konzept der sogenannten \textit{Focus Nodes} beschrieben werden. 

Mit \textbf{Triple Constraints} werden Restriktionen an alle Triple gestellt, in denen eine bestimmte Focus Node als das Subjekt eines Triples auftritt. 
Die \textbf{Focus Node} wird dabei im Rahmen der Konformitätsprüfung, auf welche in dieser Arbeit wie angesprochen nicht ausführlich eingegangen wird, bestimmt und meint den zu prüfenden Knoten. 
Ein Triple Constraint besteht aus bis zu drei Bestandteilen, die in Listing \ref{Triple Constraint Bsp} Zeile 5 gezeigt werden.
Diese sind ein Prädikat (gelb), ein Node Constraint (grün) und optional die Angabe der Kardinalität (blau). 
Der mit Rot markierte Teil wird weiter im folgenden Kapitel \ref{shapes} erläutert. 

\begin{listing}[H]
\begin{minted}[xleftmargin=20pt,linenos,escapeinside=||]{text}
BASE <http://fokus.fraunhofer.de/>
PREFIX xsd: <http://www.w3.org/2001/XMLSchema#>
PREFIX foaf: <http://xmlns.com/foaf/0.1/>

|\colorbox{red!20}{<MyNameConstraint>}| |\colorbox{yellow!20}{foaf:name}| |\colorbox{green!20}{xsd:string}| |\colorbox{blue!20}{+}|
\end{minted}
\caption{\acs{shex} Schema mit Triple Constraint} \label{Triple Constraint Bsp}
\end{listing}

Für die Angabe der Kardinalität haben sich die Autoren an bekannten Quantoren orientiert, wie man sie etwa auch von regulären Ausdrücken kennt. 
Im Beispiel aus Listing \ref{Triple Constraint Bsp} gibt also das Pluszeichen an, dass die Fokus Node als Subjekt in einem oder mehreren Triplen mit \texttt{foaf:name} als Prädikat vorkommen und das Objekt dabei jeweils vom Typ \texttt{String} sein muss. 
Alternativ gültige Quantoren wären ein Asterisk (beliebig viele), ein Fragezeichen (genau null oder eins) oder exakte Zahlenangaben in geschweiften Klammern (\{m\} für genau m, \{m,n\} für mindestens m und maximal n).
Wird die Kardinalität mit \texttt{\{0\}} auf null gesetzt, spricht man dabei von einem \textbf{Negative Triple Constraint} \cite[4.2]{shex-primer}. 
Gibt es keine Kardinalitätsangabe, greift automatisch der Standardfall von genau eins (\{1\}). 

Ebenso möglich ist die Angabe von sogenannten \textbf{Inverse Triple Constraints} \cite[4.1]{shex-primer}. 
Diese unterscheiden sich von normalen Triple Constraints mit einem vorangestellten Zirkumflex und setzen die Fokus Node bei der Prüfung damit nicht mehr als Subjekt, sondern als Objekt. 
Im Beispiel aus Listing \ref{Inverse Triple Constraint Bsp} wird damit definiert, dass die Fokus Node beliebig viele eingehende \texttt{foaf:knows} Beziehungen von Knoten mit einer \ac{iri} haben darf. 

\begin{listing}[H]
\begin{minted}[xleftmargin=20pt,linenos,escapeinside=||]{text}
BASE <http://fokus.fraunhofer.de/>
PREFIX foaf: <http://xmlns.com/foaf/0.1/>

^<MyNameConstraint> foaf:knows IRI *
\end{minted}
\caption{\acs{shex} Schema mit Inverse Triple Constraint} \label{Inverse Triple Constraint Bsp}
\end{listing}

\subsection{Triple Expressions und Shapes} \label{shapes}

In den bis hierhin gezeigten \ac{shex} Beispielen wurden jeweils nur einzelne Triple Constraint besprochen. 
Es gibt jedoch auch die Möglichkeit, diese zu gruppieren. 
Man spricht in diesem Fall von sogenannten \textbf{Triple Expressions} \cite[3.3]{shex-primer}. 
Listing \ref{Triple Expression Bsp} zeigt in den Zeilen 6 und 7 eine Triple Expression, die aus zwei Triple Constraints besteht. 
Die Triple Constraints sind dabei mit einem Semikolon getrennt und gelten damit beide für die während der Konformitätsprüfung betrachtete Fokus Node. 
Die Angabe von Triple Constraints ist darüber hinaus nur innerhalb von geschweiften Klammern möglich. 

\begin{listing}[H]
\begin{minted}[xleftmargin=20pt,linenos,escapeinside=||]{text}
BASE <http://fokus.fraunhofer.de/>
PREFIX rdf: <http://www.w3.org/1999/02/22-rdf-syntax-ns#>
PREFIX foaf: <http://xmlns.com/foaf/0.1/>

<MyPersonShape> {
    foaf:name xsd:string ;
    ^foaf:knows IRI *
}
\end{minted}
\caption{\acs{shex} Schema mit Shape und Triple Expression} \label{Triple Expression Bsp}
\end{listing}

Anhand des Beispiels aus Listing \ref{Triple Expression Bsp} soll zudem nun erstmals das Konzept der \textit{Shapes} erläutert werden. 
Als \textbf{Shape} wird in \ac{shex} eine Gruppierung von Triple Expressions bezeichnet. 
Eine Shape besitzt dabei immer eine Bezeichnung, die eine valide \ac{iri} sein muss. 
In Listing \ref{Triple Expression Bsp} Zeile 5 ist diese Bezeichnung als \texttt{<MyPersonShape>}, bzw. mit der \texttt{BASE} aufgelöst zu \texttt{http://fokus.fraunhofer.de/MyPersonShape}, angegeben. 
Die mit den Listings \ref{Node Kind Constraints Bsp} bis \ref{Triple Expression Bsp} gegebenen \ac{shex} Beispiele zeigen demnach also alle valide Shapes. 
Denn da eine Shape nur aus einer Triple Expression bestehen kann und diese Triple Expression ebenfalls nur aus einem Triple Constraint bestehen kann, ist bei diesen die Kurzform ohne geschweifte Klammern gültig. 

Shapes können darüber hinaus jedoch auch anonym, also ohne Bezeichnung angegeben werden. 
Dies ist möglich, wenn sie anstelle eines Node Constraints verwendet werden, um damit eine Verschachtelung zu erreichen. 
So können also Restriktionen an das Objekt, bzw. im inversen Fall an das Subjekt, gestellt werden. 
Im Beispiel in Listing \ref{Nested Shape Bsp} wird dies in den Zeilen 7 bis 9 gezeigt. 
Die Fokus Node darf hier eine Beziehung mit dem Prädikat \texttt{foaf:knows} zu einem Knoten haben, welcher wiederum einen \texttt{foaf:name} vom Typ \texttt{xsd:string} haben muss. \cite[3.5]{shex-primer}

\begin{listing}[H]
\begin{minted}[xleftmargin=20pt,linenos,escapeinside=||]{text}
BASE <http://fokus.fraunhofer.de/>
PREFIX rdf: <http://www.w3.org/1999/02/22-rdf-syntax-ns#>
PREFIX foaf: <http://xmlns.com/foaf/0.1/>

<MyPersonShape> {
    foaf:name xsd:string ;
    foaf:knows {
        foaf:name xsd:string
    }
}
\end{minted}
\caption{\acs{shex} Schema mit verschachtelter anonymer Shape} \label{Nested Shape Bsp}
\end{listing}

\subsection{Referenzierung, Import und Wiederverwendung}

Abgeschlossen werden soll die Beschreibung der Konzepte in \ac{shex} mit Erläuterungen zu Referenzierung, Import und Wiederverwendung von Shapes. 
Innerhalb eines Schemas lassen sich beliebig viele Shapes definieren und dann mithilfe des At-Zeichens referenzieren. 
Dadurch wird der gleiche Effekt erreicht, wie bei der im Vorherigen beschriebenen Verwendung von anonymen verschachtelten Shapes. 
Zu sehen ist die Verwendung einer solchen Referenzierung innerhalb eines Schemas in Listing \ref{Import Ref Reuse Bsp} Zeile 8. 

\begin{listing}[H]
\begin{minted}[xleftmargin=20pt,linenos,escapeinside=||]{text}
BASE <http://fokus.fraunhofer.de/>
PREFIX rdf: <http://www.w3.org/1999/02/22-rdf-syntax-ns#>
PREFIX foaf: <http://xmlns.com/foaf/0.1/>
IMPORT <AnotherSchema>

<MyPersonShape> {
    foaf:name xsd:string ;
    foaf:knows @<MyFriendShape> ;
    \$<MyTripleExpression> (
        foaf:title LITERAL ;
        foaf:mbox LITERAL +
    ) ;
    foaf:homepage @<HomepageShape>
}

<MyFriendShape> {
    foaf:name xsd:string ;
    &<MyTripleExpression>
}
\end{minted}
\caption{\acs{shex} Schema mit verschachtelter anonymer Shape} \label{Import Ref Reuse Bsp}
\end{listing}

Weiter kann auch einer Triple Expression eine Bezeichnung zum Zweck der Referenzierung gegeben werden. 
In Listing \ref{Import Ref Reuse Bsp} wird in den Zeilen 9 bis 12 eine Triple Expression als \texttt{<MyTripleExpression>} bezeichnet und in Zeile 18 dann mit einem Et-Zeichen referenziert. \cite[5.1]{shex-primer}

In Zeile 4 des gleichen Beispiels wird zudem die Import-Direktive verwendet. 
Diese bewirkt eine rekursive Verschmelzung aller importierter Schemas zu einem Dokument, weshalb die Referenzierung der \texttt{ @<HomepageShape>} in Zeile 13 möglich ist. \cite[5.2]{shex-primer}

\subsection{Hintergründe}

Mit der \ac{shex} Sprache soll eine Lücke in der \ac{rdf} Domäne geschlossen werden. 
Denn mit \ac{rdf} und \ac{owl} wurden Technologien geschaffen, die der \textit{Open World Assumption} folgen. 
Fehlende Daten in einem Graph gelten bei ihnen nicht als Fehler, sondern als (noch) unbekannt. 
Dieser Ansatz ist jedoch, gerade in produktiven Umgebungen, nicht immer gewünscht. 
Vielmehr muss hier die Integrität der Daten sichergestellt werden, was nun mit einer Sprache wie \ac{shex} und den Mechanismen zur Konformitätsprüfung gegeben ist. 

Entwickelt wird \ac{shex} von der \textit{Shape Expressions Community Group} am \ac{w3c}. 
Damit wird eine \ac{w3c} nahe Arbeit an neuen Technologien ermöglicht, was schließlich auch zu einer Übernahme als \textit{\ac{w3c} Recommendation} münden kann. 
Neben \ac{shex} steht jedoch die \ac{shacl}, welche ähnliche Konzepte verfolgt und darüber hinaus bereits seit 2015 eine offizielle Recommendation ist. 

Obwohl beide für ähnliche Anwendungsfälle konzipiert sind, nähern sie sich dem Problem der \ac{rdf}-Validierung aus verschiedenen Blickwinkeln: \ac{shex} ist hauptsächlich schemenorientiert und verwendet einen grammatikbasierten Ansatz. \ac{shacl} setzt hingegen auf \ac{sparql} als Kernabfragesprache für die Einschränkungsvalidierung.
Beide Sprachen haben potenzielle Anwendungen in der zukünftigen Forschung und Entwicklung. 
Zum Beispiel könnten sie verwendet werden, um grafische Darstellungen wie UML-Diagramme formal zu untermauern, wenn \ac{rdf}-Wortschätze entworfen werden. 
Darüber hinaus besteht die Notwendigkeit, diese Spezifikationen für die Handhabung großer Datensätze zu optimieren und effiziente Implementierungsstrategien zu entwickeln. \cite{Gayo2018}

Ein weiterer interessanter Forschungsbereich ist die automatische Ableitung von Schemas für vorhandene \ac{rdf}-Daten, was die Verfügbarkeit solcher Schemas beschleunigen würde. 
Sowohl ShEx als auch SHACL könnten auch verwendet werden, um Benutzeroberflächen für die Bearbeitung von \ac{rdf}-Daten zu verbessern, indem Formulare auf der Grundlage bekannter Datenstrukturen generiert werden.

Ob oder welche der beiden Sprachen sich in Zukunft durchsetzten wird, kann in dieser Arbeit nicht geklärt werden. 
Jedoch ist zu vermuten, dass Technologien zur Beschreibung von Graphstrukturen eine entscheidende Rolle in der Zukunft von \ac{rdf} und Semantic-Web-Technologien spielen. 
Sie werden für die Gewährleistung der Datenqualität und der Systeminteroperabilität unerlässlich sein, da die Skala semantischer Daten weiter wächst.

\section{Usability} \label{usability-grundlagen}

Als Usability (auch Benutzerfreundlichkeit oder Gebrauchstauglichkeit) wird allgemein die Qualität der Benutzererfahrung bei der Interaktion mit einem Produkt, System oder Dienst bezeichnet. 
Es handelt sich dabei um ein multidimensionales Konzept, das verschiedene Aspekte wie Effizienz, Effektivität und Zufriedenheit umfasst.
Im Kern geht es darum, wie einfach und intuitiv die Bedienung eines Systems und damit die Erreichung des jeweiligen Ziels für den Benutzer ist.

\subsection{Interaktionsprinzipien} \label{interaktionsprinzipien}

Die prominente DIN EN ISO 9241 Norm-Reihe zur \textit{Ergonomie der Mensch-System-Inter-aktion} liefert mit Teil 110 \cite{DINENISO9241-110} insgesamt sieben Interaktionsprinzipien, die zur Erreichung einer hohen Usability berücksichtigt werden sollten und im Folgenden zusammengefasst werden:

Das erste Prinzip der \textbf{Aufgabenangemessenheit} besagt, dass die Benutzeroberfläche so gestaltet sein sollte, dass sie dem Benutzer ermöglicht, seine Ziele effizient und effektiv zu erreichen.
Unnötige Funktionen, die den Benutzer ablenken könnten, sollten minimiert werden.

Das Prinzip der \textbf{Selbstbeschreibungsfähigkeit} legt fest, dass jeder Schritt in einem Dialog so gestaltet sein sollte, dass er für sich selbst erklärend ist.
Zusätzliche Informationen sollten bereitgestellt werden, um den Benutzer zu unterstützen, falls nötig.

\textbf{Erwartungskonformität} bedeutet, dass die Benutzeroberfläche sich an den Erwartungen und Erfahrungen des Benutzers orientieren sollte.
Das betrifft sowohl den gesamten Ablauf der Interaktion als auch kleinere Details wie die Position von Schaltflächen oder die Verwendung von Farben und Symbolen.

\textbf{Lernförderlichkeit} im Kontext dieser Norm heißt, dass das System dazu beitragen sollte, dass der Benutzer rasch lernt, wie es effizient und effektiv genutzt werden kann.
Das kann durch gute Dokumentation, schrittweise Einführungen und ähnliche Maßnahmen erreicht werden.

\textbf{Steuerbarkeit} meint, dass der Benutzer die Kontrolle über den Dialogprozess haben sollte.
Der Benutzer sollte in der Lage sein, Aktionen zu initiieren, zu ändern oder abzubrechen.

\textbf{Fehlertoleranz} im Kontext dieser Norm heißt, dass das System so gestaltet sein sollte, dass Fehler minimale Auswirkungen haben und leicht korrigiert werden können.
Möglichkeiten zur Korrektur und klare Rückmeldungen über die Art des Fehlers sind hierbei von Vorteil.

Das letzte Prinzip \textbf{Individualisierbarkeit} schließlich bedeutet, dass die Benutzeroberfläche anpassbar sein sollte, um verschiedenen Benutzerbedürfnissen gerecht zu werden.
Das kann durch Optionen für die Anpassung von Layout, Textgröße oder Funktionstasten realisiert werden.

Abschließend lässt sich jedoch sagen, dass diese Prinzipien keine strikten Regeln sind, sondern vielmehr Leitlinien, die je nach Kontext und spezifischen Anforderungen angepasst werden können.
Eine Einhaltung dieser Prinzipien wird jedoch generell die Benutzererfahrung verbessern und dazu beitragen, dass die Benutzer ihre Ziele effektiver und zufriedenstellender erreichen können.

\subsection{Usability-Tests} \label{usability tests}

Zur Überprüfung der Usability eines Produkts werden spezielle Usability-Tests durchgeführt. 
Diese Tests können während der Entwicklung (formative Tests) oder gegen Ende des Entwicklungsprozesses (summative Tests) erfolgen. 
Formative Tests sind dabei weniger formal und zielen darauf ab, grobe Schwachstellen zu erkennen. 
Summative Tests hingegen messen die Usability in einer möglichst realitätsnahen Umgebung und entscheiden, ob das Produkt marktreif ist. 
Diese Tests tragen zu einem kontinuierlichen Verbesserungsprozess bei, in dem Fehler und Schwachstellen identifiziert und in der nächsten Entwicklungsphase behoben werden. \cite[S. 220 f]{Moser2012}

Wenn aus verschiedenen Gründen keine Tests mit echten Benutzern durchgeführt werden können, gibt es die Möglichkeit von Expertentests. In diesen Tests überprüfen Usability-Spezialisten das Produkt anhand bestimmter Kriterien. Diese Methode ist schneller und kostengünstiger, allerdings weniger präzise, da sie auf allgemeinen Erfahrungen und Statistiken basiert. Expertentests sollten daher als ergänzende Methode betrachtet werden, nicht als vollständiger Ersatz für Nutzertests. \cite[S. 221]{Moser2012}

Um einen Usability-Test erfolgreich durchzuführen und nützliche Erkenntnisse zu gewinnen, sind eine gründliche Vorbereitung und strukturierte Umsetzung essentiell. Der Prozess lässt sich in sechs Hauptphasen unterteilen:

\begin{enumerate}
  \item Ziele definieren: Im ersten Schritt werden die Untersuchungsziele festgelegt. Es wird entschieden, welche Aspekte der Benutzbarkeit, wie beispielsweise eine neu entwickelte Funktion, überprüft werden sollen.
  \item Untersuchungsplan entwickeln: Anschließend wird ein Untersuchungsdesign erarbeitet, das dem Projektstatus, den verfügbaren Ressourcen und dem Zeitrahmen entspricht. Dabei werden Methoden, Zeitplan und Teilnehmerzahl bestimmt.
  \item Teilnehmerauswahl: Nun werden die Testteilnehmer rekrutiert. Die Auswahl sollte eine repräsentative Bandbreite an Nutzertypen abdecken. Zeitfenster für die Tests und Pausen zur Auswertung sollten eingeplant werden, und es ist ratsam, mehr Teilnehmer einzuplanen, als eigentlich benötigt, um Ausfälle zu kompensieren.
  \item Vorbereitung der Evaluation: In dieser Phase werden die benötigten Materialien wie Prototypen, Aufgabensets und Szenarien vorbereitet. Für Expertentests werden spezifische Kriterien festgelegt.
  \item Durchführung der Evaluation: Der eigentliche Test wird durchgeführt, wobei Teilnehmer je nach Szenario Aufgaben mit dem Prototyp erfüllen müssen. Abhängig von der Art des Tests (formativ oder summativ) können während der Durchführung auch Verbesserungsvorschläge diskutiert werden.
  \item Auswertung der Ergebnisse: Zum Abschluss werden die gesammelten Daten analysiert. Diese Analyse sollte zeitnah erfolgen, um keine Details zu vergessen. Die erkannten Usability-Probleme werden nach ihrer Bedeutung sortiert und dem Entwicklerteam für weitere Anpassungen übergeben. \cite[S. 222]{Moser2012}
\end{enumerate}

\begin{figure}[H]
    \centering
    \includegraphics[width=.85\linewidth]{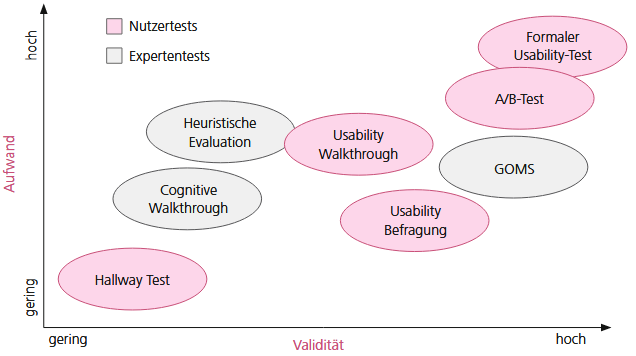}
    \caption{Übersicht und Einordnung bekannter Usability-Test-Methoden aus \cite[S. 225]{Moser2012}}
    \label{fig:enter-label}
\end{figure}

Es existieren dabei eine Vielzahl von Usability-Test-Methoden, welche unterschiedliche Ansätze verfolgen und eigene Vor- und Nachteile besitzen. 
Es stellt deshalb eine Herausforderung dar, gut einzuschätzen, welche dieser Methoden für den aktuellen Projektstand am geeignetsten ist. 
In der Regel liefern einfache und schnelle Ansätze in frühen Stadien ausreichend gute Ergebnisse, wohingegen gegen Ende des Projekts umfangreichere Tests durchgeführt werden sollten, um überhaupt noch Schwachstellen erkennen zu können. 
In Abbildung \ref{fig:enter-label} wird ein Überblick über bekannte Methoden gegeben, von denen der formale Usability-Test und die Usability-Befragung im Folgenden beschrieben werden sollen. 

\subsubsection{Formaler Usability-Test}

Beim formalen Usability-Test werden Testpersonen realistische Aufgaben gestellt und dabei beobachtet, ob und wie sie diese mit dem zu testenden Produkt lösen können. 
Diese Methodik zielt darauf ab, einen Großteil der Usability-Probleme zu erkennen und wird meist in einer späten Phase der Entwicklung eingesetzt, da sowohl Vorbereitung als auch Durchführung vergleichsweise umfangreich ausfallen. \cite[S. 230]{Moser2012}

Bei einem strikten formalen Usability-Test erhält die Testperson die Aufgabenstellung schriftlich und soll diese ohne Hilfe bearbeiten. 
Es soll keine Kommunikation mit den Beobachtern stattfinden, weshalb die Testperson unter Umständen sogar räumlich getrennt arbeitet und dabei von einer Videokamera aufgezeichnet wird. 
Sollten die Aufgaben nicht gelöst werden können, wird der Test abgebrochen. \cite[S. 230]{Moser2012}

Für den Rahmen dieser Arbeit ist eine abgeschwächte Form denkbar, in welcher etwa die Aufgabenstellung mündlich und die Beobachtung der Durchführung persönlich vorgenommen wird. 
Auch von einem Abbruch des Tests soll abgesehen und stattdessen Hilfestellungen gegeben werden. 
Damit verschiebt sich der Charakter des Tests in Richtung des \textit{Usability Walkthrough}, unterscheidet sich von diesem aber weiterhin z.B. durch die Verpflichtung von Testpersonen, die keine Usability-Experten sind. 

Zur Vorbereitung auf einen formalen Usability-Test müssen zunächst Aufgaben formuliert werden, welche die zu überprüfenden Funktionen des Produkts abdecken. 
Dabei sollten die Aufgaben in der Sprache der Testperson formuliert werden und keine Hinweise auf die Lösung eben dieser enthalten. 
Außerdem sollte die Person für den Test nicht länger als eine Stunde in Anspruch genommen werden, da ansonsten mit einem Abfall der Konzentration zu rechnen ist. \cite[S. 230]{Moser2012} 

Vor der Durchführung des formalen Usability-Tests ist es darüber hinaus wichtig, ein Briefing durchzuführen. 
Dabei soll die Testperson in das Thema eingeführt und ihr relevante Hintergrundinformationen gegeben werden. 
Auch wird ihr mitgeteilt, dass es sich um einen Test des Produkts und nicht der Person handelt, womit evtl. vorhandene Anspannung genommen werden soll. 
Zudem wird die Testperson aufgefordert, während der Durchführung Gedanken laut auszusprechen, was als \textit{Concurrent Think Aloud} bekannt ist.

Bei der Durchführung achten die Beobachter auf jede Interaktion der Testperson mit dem Produkt und notieren alle relevanten Details und Aussagen. 
Besonders relevant sind dabei selbstredend auftretende Usability-Probleme. 
Doch auch positives Feedback sollte festgehalten werden, um die daraus erhaltenen Erkenntnisse auch auf andere Teile des Produkts anzuwenden. 

Abschließend gibt es eine Nachbesprechung, in welcher die Testperson ihre Eindrücke zur Durchführung schildern kann. 
Ebenfalls macht es Sinn, der Testperson Fragen zu den notierten Beobachtungen zu stellen, um diese auszuführen. 

\subsubsection{Usability-Befragung} \label{usability befragung}

Bei der Usability-Befragung handelt es sich um eine quantitative Methode, mit welcher die Usability eines Produkts durch einen Fragebogen ermittelt wird. 
Dazu muss die befragte Testperson das Produkt zunächst selbstredend verwendet haben, z.B. im Rahmen eines Usability-Tests. 
Der Fragebogen umfasst in der Regel Fragen oder Aussagen zu Themenbereichen wie Aufgabenangemessenheit oder Fehlertoleranz, welche mit einer numerischen Skala beantwortet werden. 
Dabei muss entschieden werden, ob ein standardisierter oder ein eigener Fragebogen verwendet wird. 
Der Vorteil von standardisierten Fragebögen ist - neben der Zeitersparnis -, dass sie von Usability-Experten aufgestellt wurden. 
Ein eigener Fragebogen kann hingegen viel genauer auf das zu testende Produkt zugeschnitten werden, erfordert jedoch auch eine gewisse Expertise. \cite[S. 236]{Moser2012} 

Willumeit et al. definieren in \cite{isometrics} mit \textit{IsoMetrics} einen der populärsten deutschsprachigen Fragebogen. 
Es gibt diesen in einer kurzen Version, in welcher Aussagen nur anhand der numerischen Skala beantwortet werden, und einer langen Version, in welcher zusätzlich noch die Wichtigkeit der Frage und weitere Anmerkungen gegeben werden können. 
Die Grundlage des IsoMetrics Fragebogens sind dabei die sieben in \ref{interaktionsprinzipien} beschriebenen Interaktionsprinzipien der DIN EN ISO 9241 Norm. 
So ergeben sich die Themenbereiche Aufgabenangemessenheit, Selbstbeschreibungsfähigkeit, Steuerbarkeit, Erwartungskonformität und Fehlerrobustheit. 

\begin{figure}[H]
    \centering
    \includegraphics[width=1\linewidth]{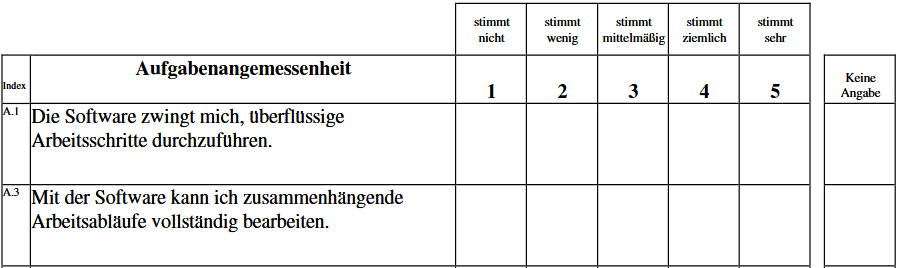}
    \caption{Beispielfragen und Layout aus dem IsoMetrics Fragebogen nach \cite{isometrics}}
    \label{isometrics-bild}
\end{figure}

Abbildung \ref{isometrics-bild} zeigt beispielhaft den Aufbau des Fragebogens. 
Zu sehen sind zwei von der Testperson zu bewertende Aussagen zur Aufgabenangemessenheit, gefolgt von der numerischen Skala und einem Feld für die Option \textit{Keine Angabe}. 

Die Auswertung der Ergebnisse erfolgt schließlich in der Regel durch die Errechnung von Mittelwert, Maximum und Minimum für jede Frage und (sofern vorhanden) Themenbereich. 
Anschließend kann eine Diskussion einzelner Fragen oder ein Abgleich mit Normwerten erfolgen. 

\section{Backend-Technologien} \label{backend-technologien}

\subsection{SQLAlchemy} \label{sqlalchemy}

SQLAlchemy \cite{sqlalchemy-basics} ist ein \acs{sql}-Toolkit und Object-Relational Mapping (ORM) Framework für die Programmiersprache Python.
Es bietet eine Schnittstelle für den Zugriff und die Manipulation relationaler Datenbanken.
Dabei minimiert es den Bedarf an manuellem \acs{sql}-Code und steigert zudem die Effizienz der Datenbankinteraktion. \cite{sqlalchemy-features}

Eines der Kern-Merkmale von SQLAlchemy ist die Abstraktionsebene, die es in Form des \textit{Data Mapper Patterns} bietet.
Im Gegensatz zum \textit{Active Record Pattern}, das in vielen anderen ORM-Tools verwendet wird, ermöglicht das Data Mapper Pattern eine Trennung der Objektlogik vom Datenbankcode, was die Flexibilität erhöht.

Die Core-Komponente von SQLAlchemy bietet Möglichkeiten für die schematische Definition von Datenbanktabellen und Beziehungen zwischen ihnen.
Dies geschieht durch das \texttt{MetaData}-Objekt, welches als Container für die Schemainformationen dient.
Hier können Tabellen und ihre Beziehungen zueinander auf eine sehr strukturierte Weise definiert werden.
Darüber hinaus bietet SQLAlchemy Core eine \acs{sql}-abstrakte Sprache, bekannt als \textit{SQL Expression Language}.
Diese Sprache ermöglicht die Erstellung von \acs{sql}-Abfragen in einer Syntax, die eng an Python angelehnt ist.
Der Vorteil hierbei ist, dass die resultierenden \acs{sql}-Abfragen damit nicht nur ausführbar, sondern auch leicht verständlich, überprüfbar und modifizierbar sind. \cite{sqlalchemy-core}

Die SQLAlchemy ORM-Komponente bietet hingegen eine noch höhere Abstraktionsebene.
Sie ermöglicht es, Datenbanktabellen als Python-Klassen zu definieren.
Diese Klassen, oft als Modelle bezeichnet, können instanziiert und manipuliert werden, als wären sie reguläre Python-Objekte.
Das ORM-Modul erleichtert zudem \acs{crud}-Operationen erheblich, indem es die zugrunde liegenden \acs{sql}-Operationen automatisch abwickelt.
Die ORM-Komponente bietet auch die Fähigkeit, komplexe Beziehungen zwischen Tabellen wie Ein-zu-Viele, Viele-zu-Viele und Ein-zu-Ein über Python-Code zu definieren.
Darüber hinaus gibt es Unterstützung für Vererbung und Polymorphismus, was die Modellierung komplexer Datensätze erleichtert. \cite{sqlalchemy-orm}

Darüber hinaus bietet SQLAlchemy ORM das Konzept der \textit{Unit of Work}.
Innerhalb dieser werden alle Änderungen, die an den Datenobjekten vorgenommen wurden, in einer \textit{Session} konsolidiert.
Diese Session fungiert als Arbeitsbereich, in dem alle Änderungen entweder in einem Rutsch auf die Datenbank angewendet oder bei einem Fehler komplett verworfen werden.
Diese Methode der Transaktionsverwaltung ist äußerst effizient und minimiert das Risiko von Dateninkonsistenzen. \cite{sqlalchemy-features}

Die Bibliothek enthält zudem Funktionen für die Abfragekonstruktion, darunter die Möglichkeit, Abfragen \textit{lazy} oder \textit{eager} zu laden.
Lazy Loading bezieht Daten erst dann, wenn sie tatsächlich benötigt werden, was die Anfangsperformance verbessern kann.
Eager Loading lädt hingegen alle Daten im Voraus, was in bestimmten Szenarien ebenso von Vorteil sein kann. \cite{sqlalchemy-loading}

\subsection{FastAPI}

FastAPI \cite{fastapiOfficial} ist ein Framework für die Erstellung von REST-APIs mit Python.
Es basiert auf Starlette \cite{starletteOfficial} für die Web-Routing-Funktionalität und Pydantic \cite{pydanticOfficial} für die Datenvorverarbeitung.
Beide Bibliotheken sind für ihre Schnelligkeit und Effizienz bekannt und bilden die Grundlage für die vergleichsweise hohe Performance von FastAPI.

Eine der besonderen Eigenschaften von FastAPI ist die Unterstützung der Python-Typ-Hinweise.
Diese Typ-Hinweise ermöglichen eine bessere Codequalität und können für die Validierung der Eingabedaten verwendet werden.
Durch die Verwendung von Typ-Hinweisen kann eine automatische Vervollständigung und Code-Analyse in modernen \acs{ide}s genutzt werden.
Darüber hinaus erlauben Typ-Hinweise die automatische Generierung von Dokumentationen, Tests und sogar Client-Bibliotheken in verschiedenen Programmiersprachen, wodurch der gesamte Entwicklungsprozess unterstützt wird.

Wie bereits kurz genannt, nutzt FastAPI Pydantic-Modelle für die Validierung von Anfragedaten.
Pydantic erlaubt dabei eine breite Palette von automatischen Validierungen und Transformationen.
Es ist jedoch auch möglich, eigene Validierungsdekorateure zu schreiben, um komplexere Logiken oder benutzerdefinierte Fehlermeldungen einzuführen.
Das ermöglicht eine Python-untypische starke Typüberprüfung und sorgt für aussagekräftige Fehlermeldungen, wenn Daten nicht den Anforderungen entsprechen. 

Im Vergleich zu REST-API-Lösungen in anderen Programmiersprachen wie Java (z.B. Spring Boot) oder Go bietet FastAPI Vorteile vor allem im Entwicklungsaufwand.
Gerade Java-Frameworks können in Bezug auf die Performance überlegen sein, jedoch erfordern ihre statische Typisierung oft mehr Boilerplate-Code, was den Entwicklungsaufwand erhöht.
FastAPI bietet durch die Nutzung moderner Python-Features und Code für Datenvalidierung, Serialisierung und Generierung von Dokumentationen, eine insgesamt schnellere Entwicklungszeit.

In Leistungstests, insbesondere für \textit{IO-bound} und \textit{High-Concurrency} Anwendungen, hat FastAPI gezeigt, dass es - je nach Konfiguration - gute Ergebnisse leisten kann, was es zu einer attraktiven Alternative für die API-Entwicklung macht.
Gleichwohl gibt es als Beleg dafür keine umfassenden wissenschaftlichen Studien, die diese spezifischen Behauptungen direkt unterstützen, aber anekdotische Evidenzen und Benutzerberichte, wie aus \cite{luongcomparison}, deuten auf die Richtigkeit dieser Aussagen hin. 

\subsection{RDFLib}

Die Python-Bibliothek RDFLib \cite{rdflib} stellt verschiedene Werkzeuge für die Arbeit mit \ac{rdf}-Daten bereit.
Dazu zählt etwa die Unterstützung für diverse \ac{rdf}-Syntaxe, darunter auch die in dieser Arbeit erwähnten.
Außerdem zeichnet sich RDFLib durch eine intuitive API aus, die eine Erstellung und Manipulation von \ac{rdf}-Graphen vereinfacht.
Durch eine native SPARQL-Unterstützung können Abfragen direkt auf dem \ac{rdf}-Graphen ausgeführt werden.

Die Flexibilität der Bibliothek ist ein entscheidender Faktor.
Sie bietet Erweiterungspunkte für benutzerdefinierte Speicher-Backends und Parser, was RDFLib zu einer vielseitigen Option, auch für spezialisierte Anwendungsfälle, macht.
Dies ist besonders relevant für Projekte, die proprietäre Datenhaltungssysteme oder spezielle Parsing-Logiken verwenden.

Angesichts der Rolle von \ac{rdf} und SPARQL in der Modellierung und Abfrage semantischer Daten hat RDFLib eine zunehmend wichtige Position in der Entwicklergemeinschaft erlangt, was sich bereits positiv auf eine kontinuierliche Weiterentwicklung der Bibliothek ausgewirkt hat.

\subsection{PyShEx} \label{pyshex-grundlagen}

PyShEx \cite{pyshex} ist eine Python-Bibliothek, welche die Verarbeitung von \ac{shex} in Python-Projekten ermöglicht.
Die Bibliothek ist darauf ausgerichtet, \ac{rdf}-Graphen gegen ShEx-Schemas zu validieren, und erfüllt somit eine kritische Funktion bei der Gewährleistung der Datenintegrität in \ac{rdf}-basierten Anwendungen.

Die Flexibilität von PyShEx zeigt sich in der Unterstützung für verschiedene \ac{rdf}-Datenformate und -Quellen.
Das macht die Bibliothek zu einem vielseitigen Instrument, insbesondere wenn sie in einem Ökosystem von heterogenen \ac{rdf}-Datenquellen eingesetzt wird.

Ein wichtiges Feature von PyShEx ist der SchemaLoader.
Er ermöglicht es, ShEx-Schemas von verschiedenen Quellen wie URLs, Dateisystemen oder sogar direkt aus Strings zu laden und überführt diese dabei in das \ac{shexj} Format. 
Dies erleichtert nicht nur den Import von Schemas, sondern bietet erst die Möglichkeit, diese weiterzuverarbeiten.

\section{Frontend-Technologien} \label{frontend technologien}

\subsection{React} \label{react}

\textit{React} ist eine JavaScript-Bibliothek, die sich auf die Erstellung von Benutzeroberflächen konzentriert.
Sie wurde ursprünglich bei Facebook entwickelt und erstmals 2013 veröffentlicht.
Seitdem hat sie sich zu einer der verbreitetsten Technologien im Frontend-Entwicklungs-bereich entwickelt, mit einer aktiven Community und einer großen Zahl von Zusatzbibliotheken. \cite{reactjs}

Eine der innovativsten Funktionen von React ist das virtuelle DOM (Document Object Model).
Ein DOM ist - vereinfacht beschrieben - eine programmatische Repräsentation einer Webseite im Speicher.
Das virtuelle DOM ist hingegen eine in-Memory-Repräsentation des realen DOM-Elements.
Alle Änderungen am virtuellen DOM werden erst dann auf das reale DOM angewendet, wenn die sog. Reconciliation (Abstimmung) erfolgt.
Diese Abstimmung wird durch einen effizienten Algorithmus realisiert, der die Differenz zwischen dem aktuellen und dem neuen virtuellen DOM berechnet.
So werden nur die tatsächlichen Änderungen im realen DOM vorgenommen.
Diese Technik minimiert die Notwendigkeit teurer DOM-Operationen und ermöglicht so reibungslose Interaktionen, selbst bei rechenintensiven Anwendungen. \cite{virtualDOM}

Die Architektur von React ist zudem komponentenbasiert.
Das bedeutet, dass die Benutzeroberfläche in unabhängige, wiederverwendbare Teile unterteilt ist, die als Komponenten bezeichnet werden.
Jede Komponente hat ihren eigenen Zustand und ihre eigene Logik, die sie damit von anderen Komponenten unabhängig macht.
Diese Modularität erleichtert die Entwicklung und Wartung von großen Webanwendungen erheblich, indem sie die Einhaltung des DRY-Prinzips (\textit{Don't Repeat Yourself}) fördert.
Jedoch ist es, gerade bei komplexeren Benutzeroberflächen, häufig notwendig, Zustände, Funktionen und andere Daten zwischen Komponenten zu synchronisieren.
Da die Komponenten hierarchisch im sogenannten \textit{Component Tree} miteinander verknüpft sind, löst man dies, indem die zu verteilenden Objekte beim nächsten gemeinsamen Eltern-Komponent liegen und an die Kind-Komponenten lediglich als Properties weitergegeben werden.
In komplexeren Anwendungsfällen kann man auch auf State-Management-Lösungen zurückgreifen, welche in Abschnitt \ref{redux-grundlagen} genauer erläutert werden, um einen globalen Zustand zwischen verschiedenen Komponenten zu teilen. \cite{statemanagement}

Weiter verwendet React \ac{jsx}, eine Syntaxerweiterung für JavaScript.
\ac{jsx} sieht aus wie \acs{html} und kann als vereinfachte Möglichkeit betrachtet werden, das virtuelle DOM mit React-Komponenten zu beschreiben.
Es wird jedoch während des Build-Prozesses in reinen JavaScript-Code umgewandelt.
Das ermöglicht eine intuitivere und sauberere Syntax, insbesondere für Entwickler, die bereits mit \acs{html} und JavaScript vertraut sind.
Die \ac{jsx}-Dateien werden während des Build-Prozesses zu JavaScript transpiliert. \cite{jsx}

Eine der eher neuerlichen Ergänzungen zu React sind die sog. \textit{Hooks}.
Mit Hooks können Zustände und Lebenszyklusmethoden in funktionalen Komponenten verwendet werden, was zuvor nur in Klassenkomponenten möglich war.
Das macht den Code in vielen Fällen schlanker und einfacher zu verstehen.
Die Verwendung von Hooks fördert auch die Wiederverwendbarkeit von Code, da sie es ermöglichen, Zustandslogik zwischen verschiedenen Komponenten zu teilen, ohne den Component Tree umzustrukturieren zu müssen.
Außerdem haben Hooks die Entwicklung von wiederverwendbaren Custom Hooks ermöglicht, die Zustandslogik und Nebeneffekte kapseln können. \cite{reacthooks}

\subsection{Redux} \label{redux-grundlagen}

React Redux ist eine State-Management-Bibliothek.
Redux selbst ist eigentlich unabhängig von React und kann mit jeder Bibliothek oder jedem Framework genutzt werden, aber sie wird am häufigsten in Verbindung mit React verwendet.
Die Bibliothek ermöglicht es, Zustände einer Anwendung in einem zentralen Speicher, dem sogenannten \textit{Store}, zu verwalten. 
Neben diesem sind die sog. \textit{Actions} und \textit{Reducer} weitere Hauptbestandteile.
Aktionen sind dabei Objekte, die beschreiben, was passiert ist. 
Reducer hingegen sind reine Funktionen, die den vorherigen Zustand und eine Aktion entgegennehmen und einen neuen Zustand zurückgeben. \cite{redux}

In einer React-Redux-Anwendung folgen die Daten einem unidirektionalen Fluss: 
Aktionen werden durch Benutzerinteraktionen oder Systemereignisse ausgelöst und an den Reducer gesendet.
Der Reducer aktualisiert dann den Store basierend auf der Aktion.
Die Komponenten \textit{hören} auf Änderungen im Store und aktualisieren sich entsprechend. \cite{reactredux}

React Redux bietet den \texttt{Provider}-Wrapper und die \texttt{useSelector} und \texttt{useDispatch} Hooks, um React-Komponenten mit dem Redux Store zu verbinden.
Der Provider wird meistens in der Wurzelkomponente der Anwendung platziert und ermöglicht es allen Kindkomponenten, auf den Store zuzugreifen.
Mit \texttt{useSelector} können diese Daten aus dem \texttt{Store} abrufen, und \texttt{useDispatch} gibt ihnen Zugriff auf die Dispatch-Funktion des Stores, um Aktionen auszulösen. \cite{reactredux}

Ein weiteres wichtiges Tool in der \textit{Redux-Welt} ist \textit{Redux Persist}.
Dies ist eine Bibliothek, welche die Möglichkeit bietet, den Redux-Store in verschiedenen Speicheroptionen wie dem lokalen Speicher des Browsers zu persistieren.
Das ist besonders nützlich für Fälle, in denen der Benutzer die Webseite aktualisiert, die Zurück- oder Vorwärts-Funktion des Browsers verwendet oder die Webseite gar gänzlich verlässt und zu einem späteren Zeitpunkt zurückkehrt.
Mit Redux Persist können Anwendungsdaten wiederhergestellt werden, ohne dass der Benutzer seinen Arbeitsfortschritt verliert. \cite{reduxpersist}

\subsection{React Router DOM} \label{router-dom}

React Router DOM \cite{reactrouter} ist eine Routing-Bibliothek für React-Anwendungen. 
Die Bibliothek ermöglicht die Navigation und das Rendern von Komponenten in Abhängigkeit von der URL und stellt somit einen essenziellen Baustein für sog. \textit{Single Page} Anwendungen dar. 
React Router DOM ermöglicht es, clientseitiges Routing durchzuführen, was für eine schnellere Seitennavigation sorgt, da nicht jedes Mal ein kompletter Serveraufruf erforderlich ist.

Die Grundlage der Bibliothek bilden einige Hauptkomponenten wie \texttt{BrowserRouter}, \texttt{Route} und \texttt{Link}. 
Der \texttt{BrowserRouter} ist dabei eine Wrapper-Komponente, welche die Routing-Funktionalität für alle darin eingebetteten Komponenten bereitstellt und deshalb in der Regel in der Wurzelkomponente der Anwendung eingebunden wird. 
\texttt{Route} ermöglicht es, Komponenten basierend auf dem aktuellen Pfad zu rendern, während \texttt{Link} für die Navigation zwischen verschiedenen Routen verwendet wird.

Ein weiteres Feature von React Router DOM sind die dynamischen Routen. 
Hierbei können Platzhalter in den Routen-Definitionen verwendet werden, sodass URL-Parameter einfach abgegriffen und innerhalb der Komponenten verwendet werden können. 
Das ist besonders hilfreich, um detailreiche Inhalte, wie z.B. Produktseiten in einem Webshop, zu rendern.

Die Bibliothek unterstützt auch die Verschachtelung von Routen, wodurch die Komposition von Komponenten einfacher wird. 
Dies ist besonders nützlich, um komplexe UI-Strukturen zu realisieren, bei denen beispielsweise Seitenlayouts, Menüs und Subrouten kombiniert werden müssen.

Mit dem Erscheinen von Hooks in React bietet React Router DOM auch einige spezielle Hooks wie \texttt{useParams}, \texttt{useHistory} und \texttt{useLocation}, die das Abrufen von Routenparametern, die Manipulation des Browserverlaufs und den Zugriff auf den aktuellen \textit{Standort} einfacher machen.

\subsection{React Flow} \label{react flow}

React Flow \cite{reactflow} ist eine React-Bibliothek für die Erstellung von knotenbasierten Editoren, Diagrammen und Netzwerkgrafiken.
Entwickelt von Webkid, wird sie oft für Anwendungsfälle wie Flussdiagramme, Workflow-Diagramme oder allgemeine Graphenvisualisierungen in einer React-Anwendung eingesetzt.
Die Bibliothek ist darauf ausgelegt, einfach zu verwenden, aber gleichzeitig hochgradig anpassbar zu sein.

Zu den Hauptfunktionen gehört die Möglichkeit, Knoten und Kanten auf einer Arbeitsoberfläche zu rendern.
React Flow unterstützt die Interaktion mit diesen Elementen durch einfache Drag-and-Drop-Aktionen.
Sie ermöglicht auch weitere Bedienmöglichkeiten wie Zoomen und Panning, wodurch eine komplexe Navigation innerhalb des Diagramms möglich ist.

Die Bibliothek ist flexibel genug, um eine breite Palette von vordefinierten Knoten- und Kantentypen zu unterstützen, von einfachen Formen bis zu komplexeren, benutzerdefinierten Visualisierungen.
React Flow bietet eine Reihe von Helferfunktionen und Hooks, um die Erstellung benutzerdefinierter Knoten und Kanten zu vereinfachen.

In Sachen Zustandsmanagement und Ereignishandhabung bietet React Flow eine umfangreiche \ac{api}.
Sie enthält zahlreiche Events, die während der Interaktion mit dem Diagramm ausgelöst werden, wie beispielsweise beim Hinzufügen oder Entfernen eines Knotens.
Auf diese Events kann \textit{gelauscht} werden, um dann darauf zu reagieren, etwa um spezifische Aktionen auszulösen, wie das Aktualisieren des Anwendungszustands oder das Ausführen von Seiteneffekten.

\subsection{Material-UI} \label{material-ui}

Material-UI \cite{mui} ist eine UI-Bibliothek für React, die sich auf die Material Design-Richtlinien von Google \cite{google-mui} stützt.
Die Bibliothek bietet eine umfangreiche Auswahl an vorkonfigurierten, komponentenbasierten Design-Elementen, die es ermöglichen, benutzerfreundliche und optisch ansprechende Benutzeroberflächen zu schaffen.
Diese Vielfalt an Komponenten ist dabei nicht nur bei der Entwicklung praktisch, sondern stellt auch eine intuitive Erfahrung für die Endbenutzer sicher.

Ein wichtiger Aspekt von Material-UI ist dabei die hohe Flexibilität in der Personalisierung der verschiedenen UI-Komponenten.
Durch diese Anpassungsfähigkeit können Erscheinungsbild und Verhalten der UI-Elemente anhand der spezifischen Anforderungen gestaltet werden.
Dies macht Material-UI zu einem vielseitigen Werkzeug, das sowohl für einfache als auch für komplexe Webanwendungen geeignet ist.

Material-UI ist zudem für responsive Design-Ansätze optimiert.
Dies bedeutet, dass die UI-Elemente automatisch ihre Größe und Anordnung anpassen, abhängig von der Bildschirmgröße und -auflösung des Endgeräts.
Dies ist besonders wichtig in einer Zeit, in der mobiles Browsing immer mehr an Bedeutung gewinnt.

Die Bibliothek unterstützt außerdem ein einfaches Theming.
Das heißt, dass Farbschemata, Schriftarten und andere Design-Elemente konsistent im gesamten Anwendungsdesign angewendet werden können.
Dies erleichtert nicht nur die Entwicklung, sondern sorgt auch für ein einheitliches und angenehmes Benutzererlebnis.
Bekannte UI-Elemente reduzieren die Lernkurve für die Endanwender und fördern so die Benutzerakzeptanz und -zufriedenheit.
Die Material Design-Richtlinien, auf denen Material-UI basiert, sind das Ergebnis umfangreicher Usability-Studien.
Daher kann man davon ausgehen, dass die in Material-UI implementierten Design-Prinzipien eine hohe Benutzerfreundlichkeit und Zugänglichkeit gewährleisten.

\section{Containervirtualisierung} \label{container}

Unter dem Begriff \textit{Containervirtualisierung} versteht man Technologien, die es ermöglichen, Anwendungen samt ihren Abhängigkeiten in einer isolierten Umgebung, dem sogenannten Container, auszuführen \cite{goldberg1974}.
Bei diesen Anwendungen handelt es sich in der Regel um einzelne Komponenten, etwa einer Webanwendung, bestehend aus z.B. Frontend, Backend, Datenbanken und Proxy.
Ein Vorteil gegenüber der Nutzung ohne Containervirtualisierung ist dabei die erhöhte Sicherheit. 
Ein Angreifer, der Zugriff auf einen isolierten Container erlangt, hat nicht notwendigerweise Zugriff auf andere Container oder gar das Host-System \cite{bernstein2014}. 
Gegenüber traditionellen Virtualisierungslösungen wie VMs ist zudem die Nutzung einer gemeinsamen Instanz auf dem Betriebssystem, während VMs jeweils eine eigene Betriebssysteminstanz benötigen.
Dies führt zu einer insgesamt leichtgewichtigeren Virtualisierung mit weniger Overhead. \cite{soltesz2007}

\subsection{Docker} \label{docker}

Docker \cite{docker} ist eine Open-Source-Plattform für Containervirtualisierung, die sich seit der ersten Version im Jahr 2013 rasch zu einer der prominentesten Technologien in diesem Bereich etabliert hat.
Ihre Hauptkomponenten sind der Docker-Client, der Docker-Host, der Docker-Daemon und die Docker-Images.
Während der Docker-Client die Befehle vom Benutzer annimmt, ist es der Docker-Daemon, der auf dem Hostsystem läuft und die Erstellung, den Betrieb und die Überwachung von Containern steuert.
Docker-Images sind im Wesentlichen Vorlagen, von denen Container instanziiert werden.
Diese Images sind in Schichten organisiert, wodurch die Wiederverwendung von Code und die schnelle Bereitstellung von Container-Instanzen ermöglicht wird.

Ein weiterer Bestandteil von Docker sind die sog. Dockerfiles.
Ein Dockerfile ist ein Skript, das Anweisungen enthält, anhand derer ein Docker-Image erstellt wird.
Durch das Dockerfile kann der Entwickler also genau definieren, welche Software, Bibliotheken und Konfigurationen in einem Image enthalten sein sollen.
Die Images sind dabei in Schichten organisiert und jede Änderung an ihrer Definition führt zu einer oder mehreren neuen Schichten. 
Diese schichtbasierte Architektur ermöglicht eine effiziente Speicherung und Übertragung von Images, da bei Änderungen nur die modifizierten Schichten übertragen oder gespeichert werden müssen.
Zudem fördert dies die Wiederverwendung von Image-Schichten über verschiedene Projekte hinweg, wodurch redundante Daten vermieden und der Speicherverbrauch minimiert wird. \cite{docker-schichten}

Ein Werkzeug im Docker-Ökosystem ist \textit{Docker Compose}, womit mehrere Container als eine Multi-Container-Anwendung definiert und orchestriert werden können. 
Dazu wird eine in YAML geschriebene Datei verwendet, in welcher Services, Netzwerke und Volumes definiert werden können.
Mit wenigen Befehlen können dann alle oder einzelne Services gestartet werden, was besonders in Entwicklungs- und Testumgebungen nützlich ist. 
Docker Compose erleichtert also die Verwaltung von Anwendungen, die aus mehreren miteinander interagierenden Containern bestehen, indem es ermöglicht, die Konfiguration in einem einzigen File zu zentralisieren und gleichzeitig eine Portabilität zu gewährleisten.
So können komplexe Applikationen, die auf mehreren Containern basieren, einfach reproduziert und verteilt werden. \cite{docker-compose}

Abschließend sei die breite Community und die Vielzahl von verfügbaren Docker-Images auf dem Docker Hub, einer zentralen Plattform, auf der Entwickler und Unternehmen ihre Docker-Images teilen können, als Vorteil der Docker-Domäne zu nennen.
So können Anwendungen und Services mit wenigen Befehlen heruntergeladen und in Betrieb genommen werden.

\chapter{Definition von Templates mit ShEx} \label{templates mit shex}

Nachdem im Grundlagen-Kapitel alle relevanten Begriffe und Technologien, wie \ac{rdf} und die \ac{turtle} Syntax, \ac{shex} und Templates im Allgemeinen, genauer betrachtet wurden, soll in diesem Kapitel nun beschrieben werden, wie anhand eines \ac{shex} Schemas, welches als Template fungiert, konkrete \ac{rdf} Graphen generiert werden können. 

\section{Übertragen der Konzepte} \label{template1}

Es wird angenommen, dass in einer \ac{shex} Shape genügend Informationen vorhanden sind, um einen gültigen \ac{rdf} Graph ableiten zu können. 
Im Fokus steht dabei die Bezeichnung einer Shape in Verbindung mit ihren Triple Constraints. 
Da die Shape-Bezeichnung zu Referenzierungszwecken eine valide \ac{iri} sein muss, sei diese als das Subjekt und die Bestandteile der Triple Constraints jeweils als Prädikate und Objekte zu verstehen. 

Zur Verdeutlichung sollen die Listings \ref{asd1} und \ref{asd2} dienen. 
In Listing \ref{asd1} ist ein \ac{shex} Schema zu sehen, welches neben \texttt{BASE}- und \texttt{PREFIX}-Direktiven zudem in den Zeilen 6 bis 8 eine Shape \texttt{<Roman>} mit zwei Triple Constraints zeigt. 
Listing \ref{asd2} zeigt einen \ac{rdf} Graphen, von welchem angenommen wird, dass dieser anhand des Schemas aus Listing \ref{asd1} generiert wurde. 
Dieser zeigt die gleichen \texttt{@base}- und \texttt{@prefix}-Direktiven und besitzt zudem zwei Triples mit dem Subjekt \texttt{<Roman>}. 
Aus der Shape-Bezeichnung wurde also das Subjekt, aus den Prädikaten der Triple Constraints die Prädikate und aus den Node Constraints die Objekte der Triples. 
Die farbliche Markierung soll die beschriebenen Übereinstimmungen ergänzend hervorheben. 

\begin{listing}[H]
\begin{minted}[xleftmargin=20pt,linenos,escapeinside=||]{text}
|\colorbox{blue!20}{BASE <http://fokus.fraunhofer.de/>}|
|\colorbox{violet!20}{PREFIX rdf: <http://www.w3.org/1999/02/22-rdf-syntax-ns#>}|
|\colorbox{violet!20}{PREFIX foaf: <http://xmlns.com/foaf/0.1/>}|

|\colorbox{red!20}{<Roman>}| {
    |\colorbox{yellow!20}{rdf:type}| [|\colorbox{green!20}{foaf:Person}|] ;
    |\colorbox{yellow!20}{foaf:name}| [|\colorbox{green!20}{"Roman Laas"}|]
}
\end{minted}
\caption{Simples \acs{shex} Schema als Basis zur Generierung eines \acs{rdf} Graphen} \label{asd1}
\end{listing}

\begin{listing}[H]
\begin{minted}[xleftmargin=20pt,linenos,escapeinside=||]{text}
|\colorbox{blue!20}{@base <http://fokus.fraunhofer.de/> .}|
|\colorbox{violet!20}{@prefix rdf: <http://www.w3.org/1999/02/22-rdf-syntax-ns#> .}|
|\colorbox{violet!20}{@prefix foaf: <http://xmlns.com/foaf/0.1/> .}|

|\colorbox{red!20}{<Roman>}| |\colorbox{yellow!20}{rdf:type}| |\colorbox{green!20}{foaf:Person}| ;
         |\colorbox{yellow!20}{foaf:name}| |\colorbox{green!20}{"Roman Laas"}| .
\end{minted}
\caption{\acs{rdf} Graph abgeleitet aus dem \acs{shex} Schema aus Listing \ref{asd1}} \label{asd2}
\end{listing}

Dieses Beispiel ist sehr simpel gehalten und beinhaltet nur konkrete Werte, etwa \textit{Roman Laas} für \texttt{foaf:name}, was die Ableitung eines \ac{rdf} Graphen ebenso simpel gestaltet. 
Tatsächlich eignen sich, zumindest was die verschiedenen Node Constraints Arten angeht, lediglich Value Constraints für die hier beschriebene Nutzung von \ac{shex} Shapes als Templates. 
Alle übrigen Node Constraints Arten sind zu abstrakt, können aber dennoch einen Nutzen haben, was in Abschnitt \ref{ausblick} besprochen wird. 

\section{Verwendung von skolemisierten Identifizierern} \label{gen ident}

Da die \ac{shex} Sprache aufgrund ihres eigentlichen Zwecks zur Beschreibung von Graph Strukturen für die Konformitätsprüfung entgegen dem Beispiel aus \ref{template1} auch in der Lage ist, Shapes ohne Bezeichnung beschreiben zu können, muss für diesen Fall ebenso eine Lösung gefunden werden. 
Aus diesem Grund soll ein weiteres Beispiel gegeben werden, erneut bestehend aus einem \ac{shex} Schema (Listing \ref{asd3}) und dem davon abgeleiteten \ac{rdf} Graphen (Listing \ref{asd4}). 
Zu sehen ist hierbei in den Zeilen 9 bis 12 von Listing \ref{asd3} die Verwendung einer anonymen verschachtelten Shape. 
Ebenso wird innerhalb der Zeilen 7 bis 13 eine referenzierbare Triple Expression definiert, was sich abzüglich der anonymen Shape jedoch technisch nicht vom Beispiel aus \ref{template1} unterscheidet und nur der Demonstration und Übersicht halber verwendet wird. 

\begin{listing}[H]
\begin{minted}[xleftmargin=20pt,linenos,escapeinside=||]{text}
BASE <http://fokus.fraunhofer.de/>
PREFIX rdf: <http://www.w3.org/1999/02/22-rdf-syntax-ns#>
PREFIX foaf: <http://xmlns.com/foaf/0.1/>

<Roman> {
    foaf:name ["Roman Laas"] ;
    $<PersonTripleExpression> (
        rdf:type [foaf:Person] ;
        rdf:knows {
            rdf:type [foaf:Person] ;
            foaf:name ["Alice Muster"]
        }
    )
}

<Bob> {
    foaf:name ["Bob Muster"] ;
    &<PersonTripleExpression>
}
\end{minted}
\caption{\acs{shex} Schema als Template mit anonymer Shape und referenzierbarer Triple Expression als Basis für die Generierung eines \acs{rdf} Graphen} \label{asd3}
\end{listing}

Da es für die verschachtelte anonyme Shape keine Bezeichnung gibt, kann also auch keine \ac{iri} für die Verwendung als Subjekt abgeleitet werden. 
Hier kommen die in Kapitel \ref{blanknodes} beschriebenen Skolem \ac{iri}s zum Tragen. 
Statt für das unbekannte Subjekt eine Blank Node zu verwenden, wird stattdessen eine Skolem \ac{iri} generiert und verwendet. 
Das Ergebnis ist in Listing \ref{asd4} in den Zeilen 7, 11 und 13 zu sehen, wo der Knoten entsprechend des Templates aus Listing \ref{asd3} als Subjekt oder Objekt auftaucht.  

\begin{listing}[H]
\begin{minted}[xleftmargin=20pt,linenos,escapeinside=||]{text}
@base <http://fokus.fraunhofer.de/> .
@prefix rdf: <http://www.w3.org/1999/02/22-rdf-syntax-ns#> .
@prefix foaf: <http://xmlns.com/foaf/0.1/> .

<Roman> rdf:type foaf:Person ;
        foaf:name "Roman Laas" ;
        foaf:knows <.well-known/genid/d26a2d0e98334696f4ad70a677abc1f6> .

<Bob> rdf:type foaf:Person ;
      foaf:name "Bob Muster" ;
      foaf:knows <.well-known/genid/d26a2d0e98334696f4ad70a677abc1f6> .

<.well-known/genid/d26a2d0e98334696f4ad70a677abc1f6> rdf:type foaf:Person ;
        foaf:name "Alice Muster" .
\end{minted}
\caption{\acs{rdf} Graph abgeleitet aus dem \acs{shex} Schema aus Listing \ref{asd3} zur Demonstrierung einer Skolem \acs{iri} bei anonymen Shapes} \label{asd4}
\end{listing}

\section{Existenzquantifizierende Variablen}

Die Verwendung von Skolem \ac{iri}s kann jedoch auch für andere Zwecke nützlich sein, etwa wenn in einem \ac{shex} Template das Vorhandensein einer unbekannten Sache definiert werden soll. 
Dabei stellt sich jedoch die Frage, wie dies zu signalisieren ist, ohne die Validität des entsprechenden \ac{shex} Schemas zu verletzen und so eine maschinelle Verarbeitung mit bekannten Algorithmen (siehe z.B. \ref{pyshex-grundlagen}) ohne Anpassungen zu ermöglichen. 
Im Folgenden soll ein Ansatz hierfür erläutert werden, welcher das Konzept der \textit{existenzquantifizierenden Variablen (\acs{exvar})} einführt. 

Es sei ein neuer Namensraum \texttt{http://exVar/} definiert, welcher als \texttt{PREFIX}-Direktive in einem \ac{shex} Schema bekannt gemacht werden soll. 
Mit der Angabe einer Bezeichnung im Pfad-Anteil dieses Namensraums sollen dabei unterschiedliche existenzquantifizierende Variablen unterscheidbar sein. 
Zur Verdeutlichung des Konzepts sollen die Listings \ref{asd5} und \ref{asd6} dienen. 
Zu sehen ist in Zeile 4 des Listings \ref{asd5} die Bekanntmachung des \texttt{exVar} Namensraums und die Verwendung von diesem in der Zeile 9 als Value Constraint und in der Zeile 12 als Shape Bezeichnung. 

\begin{listing}[H]
\begin{minted}[xleftmargin=20pt,linenos,escapeinside=||]{text}
BASE <http://fokus.fraunhofer.de/>
PREFIX rdf: <http://www.w3.org/1999/02/22-rdf-syntax-ns#>
PREFIX foaf: <http://xmlns.com/foaf/0.1/>
PREFIX exVar: <http://exVar/>

<Roman> {
    rdf:type [foaf:Person] ;
    foaf:name ["Roman Laas"] ;
    rdf:knows exVar:Bob
}

exVar:Bob {
    rdf:type [foaf:Person] ;
    foaf:name ["Bob Muster"]
}
\end{minted}
\caption{\acs{shex} Schema mit Templates zur Demonstration von existenzquantifizierenden Variablen als Basis für die Generierung eines \acs{rdf} Graphen} \label{asd5}
\end{listing}

Der von dem \ac{shex} Schema aus Listing \ref{asd5} zu erwartende \ac{rdf} Graph ist in Listing \ref{asd6} gezeigt. 
Zu sehen ist eine Skolem \ac{iri}, die sowohl als Subjekt in der Zeile 9 als auch als Objekt in der Zeile 7 verwendet wird. 

\begin{listing}[H]
\begin{minted}[xleftmargin=20pt,linenos,escapeinside=||]{text}
@base <http://fokus.fraunhofer.de/> .
@prefix rdf: <http://www.w3.org/1999/02/22-rdf-syntax-ns#> .
@prefix foaf: <http://xmlns.com/foaf/0.1/> .

<Roman> rdf:type foaf:Person ;
        foaf:name "Roman Laas" ;
        foaf:knows <.well-known/genid/d26a2d0e98334696f4ad70a677fjf73v> .

<.well-known/genid/d26a2d0e98334696f4ad70a677fjf73v> rdf:type foaf:Person ;
      foaf:name "Bob Muster" .
\end{minted}
\caption{\acs{rdf} Graph abgeleitet aus dem \acs{shex} Schema aus Listing \ref{asd5} zur Demonstrierung einer existenzquantifizierenden Variable} \label{asd6}
\end{listing}

An dieser Stelle sei erwähnt, dass durchaus auch die Verwendung von Blank Nodes anstelle des neuen Namensraums \texttt{http://exVar/} für die Signalisierung von unbekannten Entitäten denkbar ist. 
Unterschiedliche Bezeichnungen könnte man dabei statt über den Pfad-Anteil mithilfe des Blank Node Labels angeben (\texttt{\_:Bob} statt \texttt{http://exVar/Bob}). 
Jedoch wurde in Kapitel \ref{blanknodes} bereits ausführlich auf die Probleme beim Umgang mit Blank Nodes hingewiesen, weshalb dem Rat von Heath und Bizer gefolgt wird und mit dem soeben Beschriebenen eine alternative Lösung vorgestellt wurde. 

\section{Eingangs- und Ausgangsvariablen} \label{iovars}

Erweitert werden soll der in diesem Kapitel vorgestellte Mechanismus zur Nutzung von \ac{shex} für Templates um \textit{Eingangs- und Ausgangsvariablen}. 
Dabei handelt es sich um eine Möglichkeit, die innerhalb eines Templates genutzten \ac{exvar} von oder nach außen zu referenzieren. 
Damit soll es ermöglicht werden, Templates miteinander in Beziehung zu setzen und gleiche \ac{exvar}s dabei korrekt zu verknüpfen. 
Dies ist besonders dann hilfreich, wenn Templates unabhängig voneinander, also etwa von unterschiedlichen Personen oder zu unterschiedlichen Zeiten, angelegt wurden und \ac{exvar}s mit unterschiedlichen Bezeichnern gleichgesetzt werden müssen. 
Diese Funktionalität ist also primär für eine Visualisierung gedacht, wie sie im Rahmen des Kapitels \ref{implementierung} beschrieben wird. 

Zu sehen ist dieser Ansatz in den Listings \ref{asd7} und \ref{asd8}, jeweils in den Zeilen 7 und 8. 
Mit \texttt{\#in} ist dabei die Eingangs- und mit \texttt{\#out} die Ausgangsvariablen-Liste dekoriert. 
Da es sich bei diesen um in \ac{shex} gültige Kommentare handelt, bleibt das jeweilige \ac{shex} Schema dabei valide. 

\begin{listing}[H]
\begin{minted}[xleftmargin=20pt,linenos,escapeinside=||]{text}
BASE <http://fokus.fraunhofer.de/>
PREFIX rdf: <http://www.w3.org/1999/02/22-rdf-syntax-ns#>
PREFIX ex: <http://example.com/>
PREFIX exVar: <http://exVar/>

<Production> {
    #in: exVar:location
    #out: exVar:product, exVar:location
    
    rdf:type [ex:Production] ;
    ex:produces [exVar:product] ;
    ex:locatedAt [exVar:location]
}
\end{minted}
\caption{\acs{shex} Schema mit Template einer Produktion zur Demonstration von Eingangs- und Ausgangsvariablen} \label{asd7}
\end{listing}

\begin{listing}[H]
\begin{minted}[xleftmargin=20pt,linenos,escapeinside=||]{text}
BASE <http://fokus.fraunhofer.de/>
PREFIX rdf: <http://www.w3.org/1999/02/22-rdf-syntax-ns#>
PREFIX ex: <http://example.com/>
PREFIX exVar: <http://exVar/>

<TruckTransport> {
    #in: exVar:good, exVar:from
    #out: exVar:good, exVar:to
    
    rdf:type [ex:Transport] ;
    ex:transportedGood [exVar:good] ;
    ex:from [exVar:from] ;
    ex:to [exVar:to]
}
\end{minted}
\caption{\acs{shex} Schema mit Template eines LKW Transports zur Demonstration von Eingangs- und Ausgangsvariablen} \label{asd8}
\end{listing}

Das Resultat der beiden Templates ist der in Listing \ref{asd9} gezeigte \ac{rdf} Graph. 
Es sei dabei angenommen, dass die Ausgangsvariablen \texttt{exVar:product} und \texttt{exVar:location} des \texttt{<Production>} Templates mit den Eingangsvariablen \texttt{exVar:good} respektive \texttt{exVar:from} des \texttt{<TruckTransport>} Templates verbunden wurden. 
Zu sehen sind Skolem \ac{iri}s für alle unbekannten Entitäten, was bereits aus den vorherigen Abschnitten bekannt ist. 
Neu ist hingegen die Verknüpfung der gleichen \ac{exvar}s über das \texttt{sameAs} Prädikat aus dem \ac{owl} Namensraum. 
Wenn auch (noch) unbekannt, so ist damit insgesamt ersichtlich, dass das von der Produktion hergestellte Produkt in diesem Beispiel das gleiche ist, wie das transportierte Gut des LKW-Transports. 
Das gleiche gilt für den Standort der Produktion, welcher durch die Verknüpfung zum Startpunkt des Transports wurde. 

\begin{listing}[H]
\begin{minted}[xleftmargin=20pt,linenos,escapeinside=||]{text}
@base <http://fokus.fraunhofer.de/> .
@prefix rdf: <http://www.w3.org/1999/02/22-rdf-syntax-ns#> .
@prefix ex: <http://example.com/> .
@prefix owl: <http://www.w3.org/2002/07/owl#> .

<Production> rdf:type ex:Production ;
             ex:produces <.well-known/genid/fjf73j> ;
             ex:locatedAt <.well-known/genid/dskl33> .

<TruckTransport> rdf:type ex:Transport ;
                 ex:transportedGood <.well-known/genid/482mdf> ;
                 ex:from <.well-known/genid/fkfjww> ;
                 ex:to <.well-known/genid/loi653> .

<.well-known/genid/fjf73j> owl:sameAs <.well-known/genid/482mdf> .
<.well-known/genid/dskl33> owl:sameAs <.well-known/genid/fkfjww> .
\end{minted}
\caption{\acs{rdf} Graph abgeleitet aus den Templates aus den Listings \ref{asd7} und \ref{asd8}} \label{asd9}
\end{listing}

Es sei an dieser Stelle erwähnt, dass ebenso eine \textit{Gleichsetzung} der Skolem \ac{iri}s erfolgen könnte, um die Triple mit \texttt{owl:sameAs} Prädikat nicht angeben zu müssen. 
Dies könnte direkt während der Generierung des \ac{rdf} Graphen oder im Anschluss durch Manipulierung von eben diesem erfolgen.

\chapter{Anforderungserhebung und -analyse} \label{anforderungen}

Im Rahmen dieser Arbeit soll eine Webanwendung entwickelt werden, die es ihren Benutzern ermöglicht, Lieferketten anhand von vordefinierten Templates zu erstellen und als \ac{rdf} Graph zu exportieren. 
Dabei wird die folgende Forschungsfrage gestellt:

\begin{center}
\textit{Lässt sich auf der Basis von Shape Expressions eine Webanwendung entwickeln, die auch IT-fremden Personen die Modellierung von Lieferketten als RDF Graph ermöglicht?}
\end{center}

In diesem Kapitel sollen dafür alle funktionalen und nicht-funktionalen Anforderungen an die zu entwickelnde Webanwendung beschrieben und analysiert werden. 

\section{Funktionale Anforderungen} \label{funktionale anforderungen}

Die zu entwickelnde Webanwendung richtet sich an zwei Nutzergruppen, namentlich den \textit{normalen} Benutzern, die mit vordefinierten Templates Lieferketten zusammenstellen, und den Experten, die sich mit \ac{shex} auskennen und Templates modellieren. 
Aus diesem Grund sei dieser Abschnitt zu funktionalen Anforderungen in zwei Teile gegliedert, für jede Nutzergruppe ein eigener. 

\subsection{Benutzer-Anforderungen}

Ein Benutzer soll in der Lage sein, neue Lieferketten zu erstellen oder bereits bestehende Lieferketten zu bearbeiten. 
Dafür muss es also eine Art Haupt- oder Auswahlmenü geben, in dem entweder zwischen den Lieferketten gewechselt oder eine neue Lieferkette erstellt werden kann. 
Bei der Erstellung einer neuen Lieferkette soll der Benutzer dabei Meta-Daten, wie Name oder Beschreibung, angeben können. 
Weiter muss der Benutzer auch die Möglichkeit haben, zu einem späteren Zeitpunkt die Meta-Daten von bestehenden Lieferketten ändern zu können. 

Hat der Benutzer eine Lieferkette zur Bearbeitung ausgewählt, so muss sich die Oberfläche verändern. 
Das Haupt- bzw. Auswahlmenü soll nicht mehr zu sehen sein und stattdessen soll dem Benutzer eine Liste der vorhandenen Templates und die grafische Oberfläche, auf welcher diese instanziiert und verknüpft werden können, gezeigt werden. 
Sofern zu einem früheren Zeitpunkt bereits Templates instanziiert und verknüpft wurden, muss dieser Zustand beim Aufrufen der Lieferkette wiederhergestellt werden. 
Der Benutzer soll in der Lage sein, die aktuell aufgerufene Lieferkette zu teilen, indem er die aufgerufene \ac{url} kopiert und weiterleitet bzw. diese selbst zu einem späteren Zeitpunkt aufruft, ohne die Lieferkette über das Haupt- bzw. Auswahlmenü öffnen zu müssen. 

Für die Erstellung einer Lieferkette aus den vordefinierten Templates soll der Benutzer in der Lage sein, Templates zu instanziieren. 
Diese erscheinen daraufhin auf der grafischen Oberfläche und können vom Benutzer verschoben oder wieder entfernt werden. 

Template-Instanzen sollen die in der Template-Definition angegebenen Eingangs- und Ausgangsvariablen anzeigen, sodass diese miteinander verknüpft werden können. 
Der Benutzer muss in der Lage sein, mit der Maus eine Verbindung von einem Eingangs- zu einem Ausgangspunkt (oder andersherum) zu legen. 
Bestehende Verbindungen muss der Benutzer genau so erkennen können, wie deren Richtung. 
Zudem muss es dem Benutzer ermöglicht werden, bestehende Verbindungen wieder zu entfernen. 

Da eine Lieferkette aus beliebig vielen Template-Instanzen bestehen kann, muss dem Benutzer, neben der beschriebenen Möglichkeit zur händischen Verschiebung eben dieser, die Möglichkeit für ein automatisches Anordnen der Template-Instanzen auf der grafischen Oberfläche gegeben werden. 
Diese Funktion muss die Template-Instanzen dabei logisch anhand der bestehenden Verknüpfungen anordnen. 

Zu jeder Zeit muss ein Benutzer in der Lage sein, den aktuellen Zustand der modellierten Lieferkette herunterladen zu können. 
Das Format soll dabei \ac{rdf} \ac{turtle} sein. 

\subsection{Experten-Anforderungen}

Ein Experte soll zunächst einmal alle Möglichkeiten besitzen, die auch ein Benutzer hat. 
Demnach hat er also zumindest die gleichen Anforderungen, wie jene des soeben beschriebenen Benutzers. 

Zusätzlich soll er in der Lage sein, neue Templates zu erzeugen. 
Dafür muss es eine entsprechende Funktion geben, die eine Eingabemaske öffnet, in welcher der Experte Meta-Daten und die Template-Definition eingeben kann. 
Neu erzeugte Templates sollen dabei automatisch der Liste der bestehenden Templates hinzugefügt werden. 

Außerdem muss der Experte in der Lage sein, Änderungen an den Meta-Daten oder der Template-Definition vornehmen zu können. 
Diese sollen sich dabei auf das entsprechende Template, nicht aber auf bereits davon instanziierte Templates auswirken, da die Änderungen theoretisch nicht mit diesen kompatibel sind. 
Das Gleiche gilt für die Möglichkeit, bestehende Templates zu löschen. 
Die auf Basis dieser Templates erzeugten Template-Instanzen sollen auf der grafischen Oberfläche verbleiben, sodass die modellierte Lieferkette nicht \textit{zerfällt}. 

\section{Nicht-funktionale Anforderungen} \label{nichtfunktionale anforderungen}

Im Zentrum der nicht-funktionalen Anforderungen steht die Benutzerfreundlichkeit. 
Der Benutzer (hier inklusive Experte) muss in der Lage sein, alle oder zumindest einen Großteil der zur Verfügung stehenden Funktionen ohne externe Hilfe selbst zu erkennen und damit zu erlernen. 
Dafür sollen standardisierte und damit bekannte Elemente für das Design der Benutzeroberfläche verwendet werden. 
Mögliche Aktionen auf der grafischen Benutzeroberfläche sollen dabei durch die Veränderung des Cursor-Icons signalisiert werden. 

Nutzeraktionen sollen zudem zu einer möglichst schnellen und für den Benutzer ersichtlichen Reaktion auf der Benutzeroberfläche führen, also wenn z.B. etwas hinzugefügt, geändert oder gelöscht wird. 
Diese nicht-funktionale Anforderung ist eher an die Performance des Backends gerichtet, da vermieden werden soll, dass das Frontend dem Benutzer einen Zustand anzeigt, welcher nicht mit der Datenbank konsistent ist. 

Eine weitere Anforderung ist eine modulare Gestaltung der für die Webanwendung benötigten Dienste. 
Diese sollten möglichst unabhängig voneinander funktionieren, was eine Wartbarkeit bzw. allgemeine Änderbarkeit ermöglicht. 

Hinzu kommen Anforderungen an die Portabilität und damit auch an die Konfigurationsmöglichkeiten. 
Die Webanwendung muss ohne großen Aufwand auf einem neuen System aufsetzbar sein. 
Dafür eignen sich Techniken der Containervirtualisierung, die es ermöglichen sollen, alle oder auch nur einzelne Komponenten der Webanwendung betreiben zu können. 
Da unter Umständen bereits ähnliche Dienste auf einem System ausgeführt werden, muss es möglich sein, Standardports mit wenigen Anpassungen zu ändern. 

\chapter{Konzeption und Entwurf} \label{konzept}

\section{Architekturentwurf}

Dass die zu entwickelnde Webanwendung aus mehreren Komponenten bestehen soll, wurde bereits in Abschnitt \ref{nichtfunktionale anforderungen} definiert. 
In diesem Abschnitt soll ein Entwurf für die Architektur dieser Komponenten vorgestellt werden. 
Dabei seien noch keine expliziten Technologien, Services oder ähnliches erwähnt, da dies für den nachfolgenden Abschnitt \ref{systementwurf} vorgesehen ist. 

Eine Webanwendung besteht üblicherweise aus einem Frontend- und einem Backend-Anteil. 
Diese benötigen dabei jeweils zumindest einen Server-Dienst, welcher ihre Inhalte ausliefert. 
Im Falle des Frontends ist damit die Auslieferung der für die Visualisierung im Browser des Benutzers benötigten Dokumente (HTML, CSS, JavaScript) gemeint. 
Bei dem Backend-Anteil handelt es sich hingegen meist um einen Service, welcher die \textit{Geschäftslogik} umfasst und über ein \ac{rest} \ac{api}  vom Browser des Benutzers aus erreichbar ist. 
Außerdem wird vom Backend die Interaktion mit der Datenbank übernommen, sodass diese nicht von außen erreichbar sein muss, was selbstredend die Sicherheit erhöht. 

Diesem bewährten Modell zur Modularisierung und zum Datenfluss einer modernen Webanwendung soll auch in dieser Arbeit gefolgt werden. 
Es soll also ein Backend-Service aufgesetzt werden, welcher die benötigte Logik zur Verarbeitung von \ac{shex} Templates und zur Generierung von \ac{rdf} Graphen beherbergt. 
Außerdem soll das Backend die Datenhaltung und damit jegliche Kommunikation mit der Datenbank übernehmen. 
Es stellt für den Zugriff auf vom Frontend benötigte \ac{crud} Operationen ein geeignetes REST-API zur Verfügung. 

Das Frontend besteht aus den in Abschnitt \ref{funktionale anforderungen} beschriebenen Oberflächen und der Logik zur Interaktion mit dem Benutzer. 
Dazu gehört aber auch die Logik zur Kommunikation mit dem Backend, zur Verarbeitung und zum Umgang mit dem Routing und zur Speicherung von Zuständen. 

Als Vermittler zwischen Server- und Client-Seite soll zudem eine Reverse-Proxy-Kompo-nente verwendet werden. 
Diese sorgt für eine erhöhte Transparenz zwischen Browser und Diensten, was unter anderem wichtig für die Einhaltung der CORS (Cross-Origin Resource Sharing) Richtlinien ist. 
Zudem erhöht sich dadurch die Sicherheit, da nur der Reverse-Proxy und der von ihm verwendete Port nach außen hin erreichbar sein müssen. 

\begin{figure}[H]
    \centering
    \includegraphics[width=0.6\linewidth]{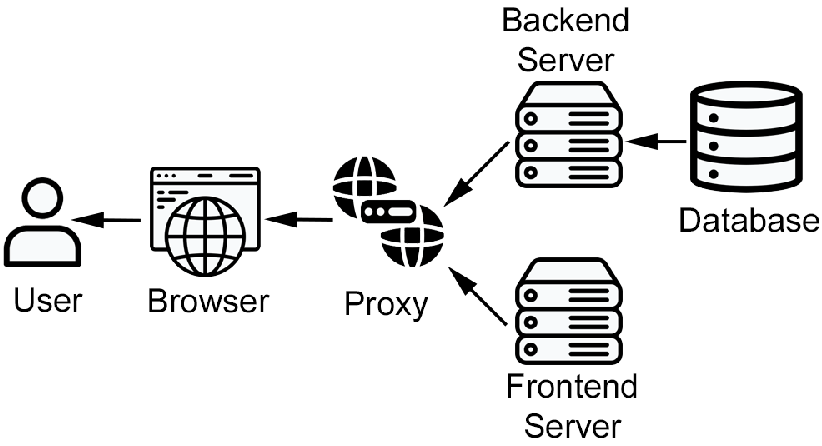}
    \caption{Architekturdiagramm mit Komponenten der geplanten Webanwendung}
    \label{architekturdiagramm}
\end{figure}

Die beschriebenen Komponenten und ihre Beziehung untereinander sollen mit dem in Abbildung \ref{architekturdiagramm} gegebenen Architekturdiagramm visualisiert werden. 
Die Richtung der Pfeile zeigt dabei den Datenfluss zum Nutzer. 

\section{Systementwurf} \label{systementwurf}

In diesem Abschnitt sollen, wie angekündigt, die spezifischen Komponenten genauer beschrieben werden. 
Dabei wird auch auf jene Technologien eingegangen, die zur Entwicklung der Webanwendung vorgesehen sind. 

\subsubsection{Backend}

Zunächst soll die Backend-Seite betrachtet werden. 
Aufgrund der in Abschnitt \ref{backend-technologien} beschriebenen Frameworks und Bibliotheken ist bereits abzusehen, dass dieses Python-basiert sein soll, was verschiedene Gründe hat. 
Zum einen gibt es sowohl für \ac{rdf} als auch für \ac{shex} Bibliotheken, die aufgrund ihres \textit{Open-Source-Charakters} und der aktiven Entwicklercommunity mächtige Werkzeuge im Umgang mit der jeweiligen Technologie sind. 
Zum anderen steht mit der Kombination aus SQLAlchemy und FastAPI eine für Python-Verhältnisse ausgesprochen performante REST-API-Grundlage mit Datenbankanbindung zur Verfügung. 
Hinzu kommt der vergleichsweise geringe Programmieraufwand durch die Nutzung von Python im Allgemeinen. 

Tatsächlich würde es alle benötigten Unterkomponenten für die Entwicklung des Backends auch mit der Programmiersprache Java geben, jedoch würde dabei, durch die Nachteile gegenüber Python, wie z.B. erhöhte Verbosity, zu starke Typisierung und mehr Boilerplate-Code, der Entwicklungsaufwand deutlich umfangreicher ausfallen. 
Der Vorteil mit Java wäre vermutlich zwar eine bessere Performance, was jedoch in Anbetracht der zu erwartenden Last an das Backend stark zu vernachlässigen ist. 

Das Backend stellt also mit FastAPI ein REST-API zur Verfügung, mithilfe dessen Lieferketten, Templates, Template-Instanzen und weitere vom Frontend benötigten Daten (siehe hierfür Abschnitt \ref{Datenbankentwurf}) über gängige \ac{crud}-Operationen durch SQLAlchemy in einer Datenbank persistiert werden. 
Zusätzlich nutzt es für die Verarbeitung der \ac{shex} Schemas die PyShEx-Bibliothek und erstellt anhand des geparsten \ac{shex} Schemas in Verbindung mit der RDFLib-Bibliothek einen \ac{rdf} Graphen. 

\subsubsection{Frontend}

Auch beim Frontend lassen sich die zur Umsetzung geplanten Frameworks und Bibliotheken bereits durch die in Abschnitt \ref{frontend technologien} beschriebenen Technologien ableiten. 
Konkret soll React mit Material-UI als Basis für die Erstellung der Benutzeroberflächen verwendet werden. 
Hinzu kommen soll React Flow für die graphische Oberfläche, auf welcher die Lieferketten modelliert werden können. 
Für Anfragen an das Backend soll der Axios \ac{http} Client \cite{axios} verwendet werden. 
Dieser wurde aufgrund der eher trivialen Features und Verwendung in diesem Projekt in Abschnitt \ref{frontend technologien} nicht weiter beschrieben. 
Erläutert wurden jedoch bereits Redux für das Zustands-Management und React Router DOM für das clientseitige Routing, welche ebenfalls für diesen Prototypen verwendet werden sollen. 

\section{Datenbankentwurf} \label{Datenbankentwurf}

Die geplante Webanwendung soll eine relationale Datenbank zur Haltung von Daten über Lieferketten, Templates und Template-Instanzen haben. 
In diesem Abschnitt sollen deshalb die dafür geplanten Tabellen vorgestellt werden. 
Sie werden später mithilfe der in Abschnitt \ref{sqlalchemy} beschriebenen \ac{orm}-Komponente von SQLAlchemy als Python-Klassen im Code definiert und damit automatisch in der angebundenen Datenbank erstellt. 

Zunächst muss es eine Tabelle für Lieferketten, konkreter für die Meta-Daten eben dieser geben. 
Dafür genügt eine Tabelle mit Feldern für das Label und die Beschreibung. 
Am wichtigsten ist jedoch die eindeutige ID, die als Primärschlüssel dient und später im Frontend beim Routing und bei REST-Anfragen zur Referenzierung benutzt werden soll. 

Weiter soll es eine Tabelle für Templates geben. 
Diese soll ebenso eine eindeutige ID zu Referenzierungszwecken, ein Label und eine Beschreibung aufnehmen können. 
Hinzu kommt noch ein Feld für das unverarbeitete \ac{shex} Schema. 

Außerdem soll es eine Tabelle für Template-Instanzen geben. 
Diese soll die gleichen Felder besitzen, wie die Tabelle für die Templates, abzüglich der Beschreibung und zuzüglich eines Fremdschlüssels, der die ID der Lieferkette, in welcher das Template instanziiert wurde, enthalten soll. 
Damit ergibt sich zwischen den Tabellen eine 1:n-Beziehung. 
Jede Lieferkette darf mehrere Template-Instanzen haben, aber jede Template-Instanz nur eine Lieferkette.

Zudem soll es eine Tabelle für die Eingangs- und Ausgangsvariablen geben. 
Die Einträge dieser Tabelle sollen beim Instanziieren eines Templates generiert werden. 
Aus diesem Grund soll diese Tabelle über einen Fremdschlüssel eine 1:n-Beziehung zur Template-Instanz-Tabelle besitzen. 
Als Felder soll es den Typ (Eingangs- oder Ausgangsvariable), die \ac{iri} sowie die bereits skolemisierte \ac{iri} geben. 

Abschließend soll es noch eine Tabelle für auf der graphischen Oberfläche erzeugte Kanten geben. 
Damit soll es vereinfacht werden, den letzten Stand der darin modellierten Lieferkette beim Aktualisieren oder späteren Weiterarbeiten wiederherzustellen. 

\begin{figure}[H]
    \centering
    \includegraphics[width=1\linewidth]{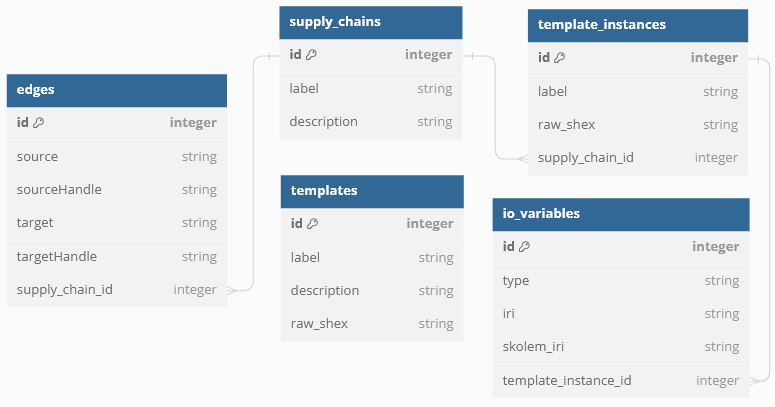}
    \caption{Datenbank Diagramm der geplanten Webanwendung}
    \label{datenbank diagramm}
\end{figure}

Zur Veranschaulichung soll das Datenbank-Diagramm aus Abbildung \ref{datenbank diagramm} dienen. 
Dieses zeigt die beschriebenen Tabellen und ihre Beziehungen zueinander, inklusive der jeweiligen Datentypen. 

\section{Design der Benutzeroberfläche} \label{design}

Die Benutzeroberfläche soll mittels der in Abschnitt \ref{material-ui} beschriebenen Material-UI-Bibliothek entwickelt werden.
Damit können direkt Komponenten verwendet werden, die den erprobten Google Material Design-Richtlinien folgen. 

Eine dieser Komponenten ist die sog. \textit{Top App Bar} \cite{appbar}. 
Bei dieser handelt es sich um eine schmale Leiste, die am oberen Bildrand von einer Seite zur anderen durchgängig verläuft und in der Regel den Titel der Web-Oberfläche und Schaltflächen für wichtige Aktionen enthält. 
Üblicherweise ist diese dabei entweder dauerhaft angezeigt oder verschwindet nach oben beim Navigieren auf der Webseite nach unten. 
Für den zu entwickelnden Prototypen im Rahmen dieser Arbeit wird die Leiste dauerhaft zu sehen sein, da es gar nicht möglich sein soll, nach unten zu scrollen. 

Denn auf der Fläche unter der Appbar soll sich die graphische Oberfläche für die Modellierung der Lieferkette befinden. 
Bei dieser soll es sich um eine mithilfe der in Abschnitt \ref{react flow} beschriebenen React Flow Bibliothek erstellte interaktive Oberfläche für Graph-Editoren handeln. 
Eine Verwendung des Mausrads soll also nicht zu einem \textit{scrolling} der Inhalte führen, sondern die Verwendung der Zoom-Funktion der graphischen Oberfläche ermöglichen. 
Sollte sich der Benutzer noch im Haupt- bzw. Auswahlmenü befinden, soll nicht einfach eine leere graphische Benutzeroberfläche angezeigt werden, sondern ein Hinweistext, der auf das Auswählen einer Lieferkette hinweisen soll. 

Bezüglich des Haupt- bzw. Auswahlmenüs soll ein sog. \textit{Navigation Drawer} \cite{nav-drawer} verwendet werden. 
Dieser soll an der linken Seite des Bildschirmrands zu finden sein und in der Menü-Ansicht dabei die verfügbaren Lieferketten und in der Lieferketten-Ansicht die verfügbaren Templates als Liste anzeigen. 
Bei letzterem Fall soll es zudem die Möglichkeit geben, den Drawer \textit{einzufahren}, also mit einer Animation nach links verschwinden zu lassen, um so den Arbeitsbereich der grafischen Oberfläche zu vergrößern. 

\begin{figure}[H]
    \centering
    \fbox{\includegraphics[width=1\linewidth]{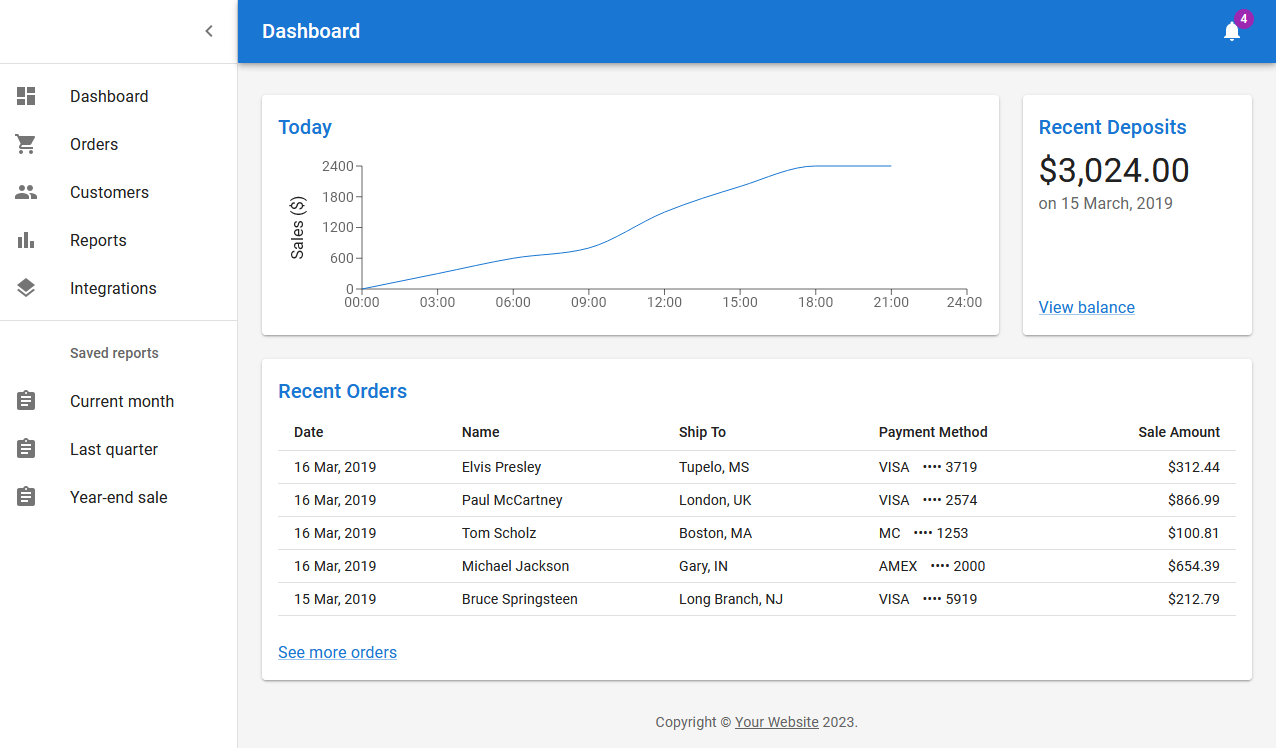}}
    \caption{Beispiel einer Weboberfläche mit Material-UI Elementen aus \cite{mui-example-template}}
    \label{material example template}
\end{figure}

Insgesamt soll sich das Design der Weboberfläche an dem in Abbildung \ref{material example template} gezeigten Beispiel von Material-UI aus \cite{mui-example-template} orientieren. 
Zu sehen ist die App Bar in Blau am oberen Bildrand, sowie der ausgeklappte Drawer auf der linken Seite. 
Anstelle der mit grauem Hintergrund unterlegten weißen Boxen soll sich dabei jedoch die Arbeitsfläche der graphischen Oberfläche befinden. 

\chapter{Implementierung} \label{implementierung}

In diesem Kapitel soll der im Rahmen dieser Arbeit entwickelte Programmcode des Frontends und Backends sowie die Konfigurationsdateien für die Containervirtualisierung beschrieben werden. 

\section{Backend}

Der Programmcode für das Backend, welcher in Python geschrieben wurde, gliedert sich in drei Module. 
Im Modul \texttt{db} befindet sich dabei die Konfiguration für die Verbindung zur Datenbank, die Definition der Datenbankmodelle und der Code für \ac{crud}-Operationen. 
Das Modul \texttt{routers} beinhaltet die Definition der \ac{rest}-\ac{api} Schnittstellen sowie die Schemas, welche die Struktur der eingehenden und ausgehenden Datenobjekte vorgeben. 
Das dritte Modul \texttt{processing} beinhaltet den Programmcode für die Verarbeitung der \ac{shex} Templates sowie jenen zur Generierung des \ac{rdf} Graphen anhand derer. 
In den nachfolgenden Unterabschnitten werden die Module jeweils erläutert. 

\subsection{Datenhaltung}

In der Datei \texttt{config.py} des Moduls \texttt{db} befinden sich alle für die Verbindung und Kommunikation mit der Datenbank benötigten Funktionsaufrufe und Objekte, die zum Teil an anderen Stellen des Programmcodes wiederverwendet werden. 
In Listing \ref{db-config} sei der gesamte Inhalt dieser Datei gezeigt. 
Hinter der Konstante \texttt{SQLALCHEMY\_DATABASE\_URL} in Zeile 5 verbirgt sich dabei die \ac{url} der Datenbank, inklusive dem zu verwendenden Protokoll. 
Für den Prototypen dieser Arbeit sei eine \textit{SQLite} Datenbank verwendet, weshalb die Konstante nicht wie üblich Login-Daten, Ports oder ähnliches enthalten muss. 

In den Zeilen 7 bis 9 wird die \texttt{engine} von SQLAlchemy initialisiert. 
Dabei handelt es sich um den \textit{Mittler} zwischen Python-Code und Datenbank. 
Die \texttt{engine} verwaltet etwa die aktiven Verbindungen zur Datenbank, wählt den zu verwendenden \acs{sql}-Dialekt aus und kümmert sich um das Ausführen der Transaktionen. 

Als nächstes wird in Zeile 11 die Sitzung mit der Datenbank aufgesetzt. 
Diese wird mit der Variable \texttt{SessionLocal} referenziert und später bei der Definition der \ac{rest}-\ac{api} Schnittstellen als Abhängigkeit eingesetzt werden (siehe dazu den nachfolgenden Unterabschnitt). 
Als letztes wird in Zeile 13 eine neue Basis-Klasse initialisiert, von welcher alle \textit{Model}-Klassen, welche die Tabellen und Beziehungen in der Datenbank repräsentieren, abgeleitet werden. 
Das von \texttt{declarative\_base()} zurückgegebene Objekt beinhaltet dabei auch ein \texttt{MetaData}-Objekt, welches alle von der \texttt{Base} abgeleiteten Tabellen-Definitionen enthält. 
Es kann deshalb für eine automatische Erstellung der Datenbankstruktur verwendet werden. 

\begin{listing}[H]
\begin{minted}[xleftmargin=20pt,linenos,escapeinside=||]{python}
from sqlalchemy import create_engine
from sqlalchemy.ext.declarative import declarative_base
from sqlalchemy.orm import sessionmaker

SQLALCHEMY_DATABASE_URL = "sqlite:///./sql_app.db"

engine = create_engine(
    SQLALCHEMY_DATABASE_URL, connect_args={"check_same_thread": False}
)

SessionLocal = sessionmaker(autocommit=False, autoflush=False, bind=engine)

Base = declarative_base()
\end{minted}
\caption{Inhalt der \texttt{config.py} Datei aus dem \texttt{db} Modul zur Aufsetzung der Verbindung und Interaktion mit der Datenbank} \label{db-config}
\end{listing}

Wie in Abschnitt \ref{Datenbankentwurf} beschrieben, werden die Tabellen der relationalen Datenbank mithilfe der \ac{orm}-Komponente von SQLAlchemy als Python Klassen definiert. 
Diese Klassen werden bei SQLAlchemy als \textit{Models} bezeichnet. 
Listing \ref{backend-model-example} zeigt beispielhaft diese Klassendefinition anhand der für die Template-Instanzen benötigten Tabelle. 
Neben der Benennung der Tabelle in Zeile 2, zeigen die Zeilen 4 bis 7 dabei die in Abschnitt \ref{Datenbankentwurf} beschriebenen Felder. 
Mithilfe der in den Zeilen 9 und 10 durch die \texttt{relationship()} Funktion definierten Beziehung zur Tabelle der Lieferketten wird dabei eine bidirektionale Beziehung erstellt, die es den Instanzen beider Klassen erlaubt, aufeinander zuzugreifen. 
Dadurch wird der für die \ac{crud}-Operationen benötigte Code deutlich schlanker. 

\begin{listing}[H]
\begin{minted}[xleftmargin=20pt,linenos,escapeinside=||]{python}
class TemplateInstance(Base):
    __tablename__ = "template-instances"

    id = Column(Integer, primary_key=True, index=True)
    label = Column(String)
    raw_shex = Column(String)
    supply_chain_id = Column(Integer, ForeignKey("supply-chains.id"))

    supply_chain = relationship("SupplyChain", 
                                back_populates="template_instances")
\end{minted}
\caption{Beispielhafter Auszug aus den mit SQLAlchemy definierten Datenbank-Models} \label{backend-model-example}
\end{listing}

In dem Untermodul \texttt{crud} befindet sich der für die \ac{crud}-Operationen benötigte Programmcode, aufgeteilt in mehrere Dateien nach Typ. 
Listing \ref{crud-example} soll beispielhaft zwei Funktionen der Lieferketten zeigen. 
Die Funktion \texttt{get\_supply\_chain()} aus den ersten vier Zeilen soll eine bestimmte Lieferkette aus der Datenbank zurückliefern und zeigt dabei die intuitiven Möglichkeiten von SQLAlchemy, Anfragen zu formulieren. 
Denn als Übergabeparameter erwartet die Funktion neben der ID der Lieferkette auch die aktive Sitzung mit der Datenbank. 
Die Hintergründe dazu werden dabei im folgenden Abschnitt erläutert. 
Auf dem Datenbankobjekt können beispielsweise Anfrage-, Filter- und Manipulations-Methoden aufgerufen werden. 
Im gezeigten Beispiel werden alle Lieferketten abgefragt, anschließend nach der ID gefiltert und das erste Element als Ergebnis zurückgegeben. 

Die in den Zeilen 6 bis 14 gezeigte Funktion \texttt{create\_supply\_chain()} soll einen neuen Eintrag in der Datenbank anlegen und diesen anschließend zurückgeben. 
Hier ist als Übergabeparameter bereits ein Schema des \ac{rest}-\ac{api}s zu sehen, was im folgenden Abschnitt genauer erläutert wird. 
Zu sehen ist in den Zeilen 7 und 8 die Instanziierung eines neuen \texttt{SupplyChain}-Objekts mit den Daten aus dem Schema-Objekt. 
In den Zeilen 10 bis 12 wird es zur Datenbank hinzugefügt, \textit{committed} (persistiert) und schließlich aktualisiert. 
Das Aktualisieren ist wichtig, da die eindeutige ID von der Datenbank vergeben wird und diese ansonsten im Objekt des Programmcodes nicht bekannt ist. 

\begin{listing}[H]
\begin{minted}[xleftmargin=20pt,linenos,escapeinside=||]{python}
def get_supply_chain(db: Session, supply_chain_id: int):
    return db.query(models.SupplyChain)
             .filter(models.SupplyChain.id == supply_chain_id)
             .first()

def create_supply_chain(db: Session, 
                        supply_chain: schemas.SupplyChainBase):
    db_supply_chain = models.SupplyChain(label=supply_chain.label,
                               description=supply_chain.description)
                      
    db.add(db_supply_chain)
    db.commit()
    db.refresh(db_supply_chain)
    return db_supply_chain
\end{minted}
\caption{Beispielhafter Auszug aus den \acs{crud} Funktionen im Datenbank Modul} \label{crud-example}
\end{listing}

\subsection{REST-API}

Wie bereits angesprochen, werden für das \ac{rest}-\ac{api} sogenannte \textit{Schemas} als Pydantic-Klassen definiert. 
Mit ihnen gibt man die Struktur für Eingabe- und Ausgabeobjekte der Schnittstellen vor. 
Dabei macht es Sinn, eine Basis-Klasse zu definieren, von welcher mehrere Unterklassen erben. 
Dadurch kann noch genauer gesteuert werden, welche Daten von den Schnittstellen erwartet und welche ausgegeben werden. 

Zur Verdeutlichung zeigt Listing \ref{fastapi-schemas-example} die für die Lieferketten definierten Schemas. 
Mit \texttt{SupplyChainBase} wird in den Zeilen 1 bis 3 die Basis-Klasse definiert, womit jede Ausprägung einer Lieferkette mindestens die Felder \texttt{label} und \texttt{description} besitzen muss. 
Verwendet wird diese Klasse als erwarteter Typ bei der Erstellung von neuen Lieferketten. 
Die Basis-Klasse selbst erbt dabei von der \texttt{BaseModel}-Klasse von Pydantic. 
Damit werden automatisch Methoden zur Validierung, Serialisierung und Konfiguration übernommen. 

Mit der Unterklasse \texttt{SupplyChainBase} aus den Zeilen 5 bis 9 kommen nun noch weitere Felder hinzu. 
Dabei spielt vor allem Zeile 9 eine wichtige Rolle, denn mit der Angabe von \texttt{from\_attributes=True} wird Pydantic signalisiert, dass Instanzen dieser Klasse durch arbiträre Instanzen einer anderen Klasse erstellt werden können, sofern die Felder gleich benannt sind. 
Damit wird also ermöglicht, dass Objekte aus der Datenbank automatisch zu Objekten für das Frontend \textit{übersetzt} werden können. 
Dies ist besonders dann hilfreich, wenn nicht alle Felder von Instanzen einer Datenbank-Klasse vom REST-API ausgegeben werden sollen (z.B. Passwörter) oder erst durch das Datenbank-Objekt bestimmte Felder hinzukommen, wie in diesem Fall die \texttt{id} in Zeile 6. 

\begin{listing}[H]
\begin{minted}[xleftmargin=20pt,linenos,escapeinside=||]{python}
class SupplyChainBase(BaseModel):
    label: str
    description: str

class SupplyChain(SupplyChainBase):
    id: int
    edges: List[Edge]
    template_instances: List[TemplateInstance]
    model_config = ConfigDict(from_attributes=True)
\end{minted}
\caption{Auszug aus den für die Schnittstellen erstellten Schemas} \label{fastapi-schemas-example}
\end{listing}

Analog zur Aufteilung des Programmcodes der \ac{crud}-Operationen bei der Datenhaltung gliedern sich auch die Schnittstellen-Definitionen in mehrere Dateien, je nach Typ. 
In Listing \ref{fastapi-router-example} wird exemplarisch ein Auszug aus \texttt{/routers/supply\_chains.py} gezeigt. 
Jede Datei benötigt einen eigenen \texttt{APIRouter}, welcher meist an einer zentralen Stelle im Code mit allen übrigen Routern an die FastAPI-Instanz gebunden bzw. registriert wird. 

Funktionen werden über einen Dekorator in Verbindung mit dem Router an bestimmte \ac{http}-Methoden gebunden. 
In Listing \ref{fastapi-router-example} ist dies in Zeile 10 zu sehen. 
Es handelt sich bei der Funktion \texttt{get\_supply\_chains()} also um Programmcode, der infolge einer GET-Methode ausgeführt wird. 
Innerhalb des Dekorators muss zudem der entsprechende Pfad-Anteil angegeben und optional z.B. der Rückgabetyp, welcher zur Validierung verwendet wird. 

Die Funktion besitzt als Übergabeparameter ein Sitzungs-Objekt der Datenbank. 
Hier wird mit \texttt{Depends(get\_db)} das \textit{Dependency Injection} Entwurfsmuster verwendet. 
Die Funktion \texttt{get\_db()} liefert eine aktive Sitzung mit der Datenbank und ist dabei als eine Abhängigkeit der \texttt{get\_supply\_chains()} Funktion definiert. 
Ohne die Datenbank-Sitzung kann also die Schnittstellen-Funktion nicht ausgeführt werden, was eine entsprechende Fehlermeldung zur Folge hat. 

Für jeden Datentyp und jede benötigte HTTP-Methode wird eine eigene Funktion definiert, wobei diese häufig nur aus einer Zeile, nämlich dem Aufruf der entsprechenden \ac{crud}-Funktion der Datenbank, bestehen, wie auch in diesem Fall in Zeile 12 zu sehen. 

\begin{listing}[H]
\begin{minted}[xleftmargin=20pt,linenos,escapeinside=||]{python}
router = APIRouter()

def get_db():
    db = SessionLocal()
    try:
        yield db
    finally:
        db.close()

@router.get("/supply-chain", response_model=list[schemas.SupplyChain])
def get_supply_chains(db: Session = Depends(get_db)):
    return crud.get_supply_chains(db)
\end{minted}
\caption{Auszug aus einer Schnittstellen-Definition inklusive Abhängigkeit} \label{fastapi-router-example}
\end{listing}

\begin{figure}[H]
    \centering
    \includegraphics[width=.7\linewidth]{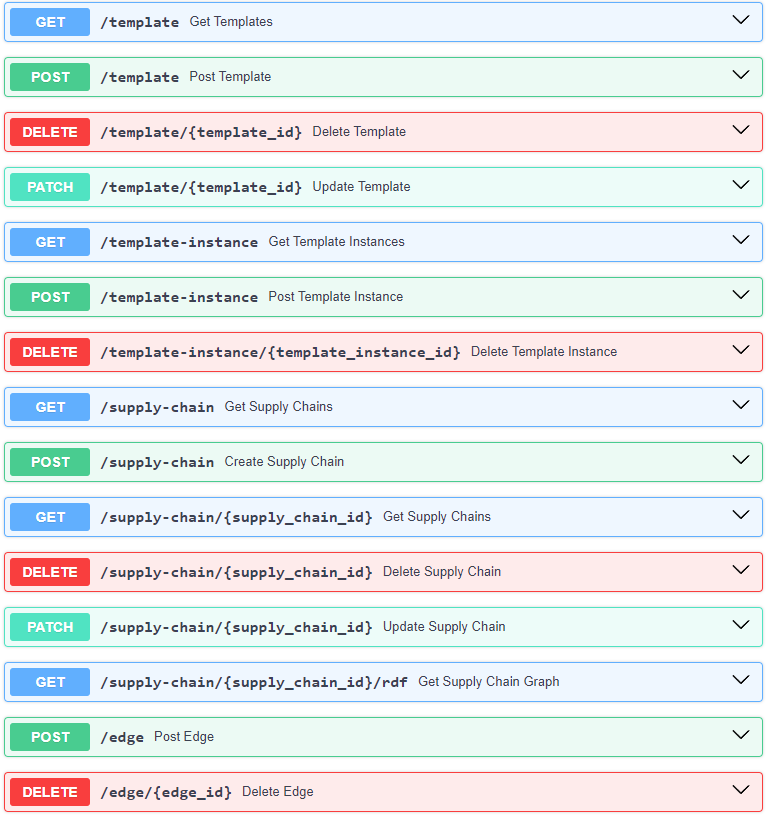}
    \caption{Screenshot aus der \textit{SwaggerUI} Oberfläche des entwickelten Backends}
    \label{api-list}
\end{figure}

Insgesamt besitzt das REST-API für die im Rahmen dieser Arbeit entwickelte Webanwendung 15 Schnittstellen, welche in Abbildung \ref{api-list} aufgelistet sind. 
Die Abbildung ist ein Screenshot aus dem von FastAPI automatisch generiertem und bereitgestelltem \textit{SwaggerUI}, einer interaktiven Oberfläche, in der alle Schnittstellen und Schemas detailliert aufgelistet sind. 
Zudem können die Schnittstellen über diese Oberfläche direkt mit echten Werten ausprobiert und die Rückgabewerte ausgewertet werden, was vor allem bei der Entwicklung hilfreich ist. 

\subsection{Graph Generierung}

Für die Generierung des \ac{rdf} Graphen aus einem \ac{shex} Template sind mehrere Schritte notwendig. 
Im Zentrum steht dabei die Verarbeitung des in die \ac{shexj} Syntax überführten \ac{shex} Schemas. 
Diese sei in gekürzter Fassung in Listing \ref{gen-ausschnitt} gezeigt. 
Zu sehen ist die Funktion \texttt{process\_template()}, welche als Übergabeparameter ein \ac{shex} Schema als String erwartet. 
In Zeile 2 ist die Verwendung des in Abschnitt \ref{pyshex-grundlagen} beschriebenen \texttt{SchemaLoader} zu sehen. 
Dieser wird für das Parsen des \ac{shex} Strings in die \ac{shexj} Syntax verwendet. 

Anschließend folgt in der Zeile 4 das Aufsetzen des \ac{rdf} Graphen mithilfe der RDFLib Bibliothek, gefolgt von einer Übertragung der Prefixes von \ac{shex} zu \ac{rdf} in den Zeilen 6 bis 8. 
Hierfür wurde eine simple Helferfunktion \texttt{get\_shex\_prefixes()} definiert, die diese aus dem \ac{shex} Schema extrahiert. 

Da ein \ac{shex} Schema grundsätzlich aus mehreren Shapes bestehen kann, es im Rahmen dieser Arbeit jedoch erwartet wird, dass pro Schema nur eine Shape definiert ist, wird in der Zeile 10 das erste Element aus der Liste der Shapes des geparsten Schemas zur weiteren Verarbeitung in die Variable \texttt{shape} gespeichert. 
Anschließend werden die vorhandenen Triple Expressions der Shape ab der Zeile 12 iterativ durchlaufen. 
Mithilfe einiger \textit{Guard Clauses} werden nicht berücksichtigte Elemente dabei übersprungen. 
Schließlich werden für alle Einträge im Value Set des Value Constraints in den Zeilen 22 und 23 Triples im Graph erzeugt. 
Die Bezeichnung der Shape ist dabei das Subjekt, Prädikat und Objekt leiten sich hingegen vom jeweiligen Triple Constraint ab. 

\begin{listing}[H]
\begin{minted}[xleftmargin=20pt,linenos,escapeinside=||]{python}
def process_template(shex):
    schema = SchemaLoader().loads(shex)

    graph = Graph(base="http://fokus.fraunhofer.de/")

    prefixes = get_shex_prefixes(shex)
    for prefix_name, prefix_uri in prefixes.items():
        graph.bind(prefix_name, prefix_uri)

    shape = schema.shapes[0]

    for expression in shape.expression.expressions:
        if expression.type != 'TripleConstraint':
            continue

        if type(expression.valueExpr) != NodeConstraint:
            continue

        if not expression.valueExpr.values:
            continue

        for value in expression.valueExpr.values:
            graph.add((URIRef(shape.id), 
                       URIRef(expression.predicate), 
                       URIRef(value)))

    return graph
\end{minted}
\caption{Gekürzte Fassung der Funktion zur Generierung des \acs{rdf} Graphen} \label{gen-ausschnitt}
\end{listing}

Die Erzeugung der Skolem \ac{iri}s geschieht bereits bei der Instanziierung der Template-Instanzen. 
Diese werden konkret mithilfe der \texttt{uuid} Bibliothek \cite{uuid-lib} nach \cite{rfc4122} erzeugt und in der in Abschnitt \ref{blanknodes} beschriebenen Form eingesetzt. 
Zu sehen ist die dafür definierte Funktion in Listing \ref{gen-iri-ausschnitt}. 

\begin{listing}[H]
\begin{minted}[xleftmargin=20pt,linenos,escapeinside=||]{python}
def generate_skolem_iri():
    return f'http://fokus.fraunhofer.de/.well-known/genid/{uuid.uuid4()}'
\end{minted}
\caption{Funktion zur Generierung der Skolem \acs{iri}s} \label{gen-iri-ausschnitt}
\end{listing}

Nachdem der Graph mit dem in Listing \ref{gen-ausschnitt} gezeigten Programmcode generiert wurde, werden die vorhandenen \ac{exvar}s mit den Skolem \ac{iri}s ersetzt. 
Abschließend können diese noch über die durch die \texttt{edges} Tabelle bekannten Verknüpfungen zwischen Template-Instanzen mit \texttt{owl:sameAs} gleichgesetzt werden. 
Da dieser Teil des Codes trivial ist, sei er an dieser Stelle nicht mit einem Listing dargestellt. 

Stattdessen soll mit Listing \ref{ausgabe} eine beispielhafte Ausgabe eines generierten Graphen in \ac{turtle} Syntax gegeben werden. 
Zu sehen ist das Ergebnis aus der Verknüpfung eines Templates \texttt{<TruckTransport>} (Zeilen 5 bis 10) mit einem Template \texttt{<Storage>} (Zeilen 12 bis 15) über \texttt{ex:endPoint} bzw. \texttt{ex:location}, was zu einer Gleichsetzung der entsprechenden Skolem \ac{iri}s in den Zeilen 17 und 18 führt. 

\begin{listing}[H]
\begin{minted}[xleftmargin=20pt,linenos,escapeinside=||]{text}
@base <http://fokus.fraunhofer.de/> .
@prefix ex: <http://example.com/> .
@prefix owl: <http://www.w3.org/2002/07/owl#> .

<TruckTransport> a ex:Activity ;
    ex:affectedBy <.well-known/genid/1742a400-9838-48c1-9bfe-fcbd0ac57e0b> ;
    ex:endPoint <.well-known/genid/9551aa44-4e58-4f7b-9683-fdc05e4d0a70> ;
    ex:startingPoint <.well-known/genid/8bda2641-910d-495c-83ca-f01cf172bcc6> ;
    ex:transports <.well-known/genid/58cf7996-393a-4527-ad38-c52707102f02> ;
    ex:uses <.well-known/genid/ad29a699-322f-4320-ba1e-167003a3b890> .

<Storage> a ex:Storage ;
    ex:affectedBy <.well-known/genid/a481484f-feef-4042-af84-ec1e14cc33a5> ;
    ex:stores <.well-known/genid/34247695-1ab9-48d6-ab3c-fd4ba50cba94> ;
    ex:location <.well-known/genid/c85367c6-4ea5-4a1a-a763-e50f095ddeac> .

<.well-known/genid/9551aa44-4e58-4f7b-9683-fdc05e4d0a70> owl:sameAs
                <.well-known/genid/c85367c6-4ea5-4a1a-a763-e50f095ddeac> .
\end{minted}
\caption{Auf Basis von \acs{shex} Templates generierter \acs{rdf} Graph} \label{ausgabe}
\end{listing}

\section{Frontend}

Das Frontend wurde mithilfe der React-Plattform erstellt. 
Dabei wurde auf \textit{\ac{cra}} zurückgegriffen, um einen initialen Basiszustand herzustellen, von welchem aus die Entwicklung gestartet werden konnte. 
\ac{cra} ist ein von Facebook entwickeltes Werkzeug für eben diesen Zweck und kümmert sich etwa um das Aufsetzen der grundlegenden Ordner- und Dateienstruktur, einen Entwicklungsserver mit Reloading-Funktion und eine ESLint-Integration für eine statische Quellcode-Analyse. 

\subsection{Root-Komponente} \label{react root}

Der Einstiegspunkt des Frontends liegt in der Datei \texttt{src/index.js}. 
Hier wird zunächst der in Abschnitt \ref{router-dom} beschriebene \texttt{BrowserRouter} mithilfe der Funktion \texttt{createBrowserRouter()} initialisiert. 
Diese erhält als Übergabeparameter eine Liste der Routen. 
Listing \ref{frontend-routes} zeigt dabei die in dieser Arbeit verwendete Definition der Routen für das Lieferketten-Menü und die Lieferketten-Ansicht. 
Zu sehen ist in den Zeilen 2 bis 6 dabei die \textit{Root-Route}, welche zur \texttt{<MainMenu />} Komponente führt. 
Außerdem ist eine Komponente angegeben, welche im Fall eines Routing-Fehlers aufgerufen werden soll. 
Sie gilt dabei für alle Routen, sofern nicht an anderen Stellen eigene Error-Komponenten definiert werden. 
In den Zeilen 7 bis 10 ist die Route für die Lieferketten-Ansicht definiert. 
Sie beinhaltet dabei, wie in Abschnitt \ref{router-dom} bezüglich dynamischen Routen beschrieben, einen Platzhalter für die Lieferketten-ID, welche in anderen Komponenten verwendet werden kann. 

\begin{listing}[H]
\begin{minted}[xleftmargin=20pt,linenos,escapeinside=||]{js}
const router = createBrowserRouter([
  {
    path: "/",
    element: <MainMenu />,
    errorElement: <ErrorPage />
  },
  {
    path: "/supply-chain/:supplyChainId",
    element: <SupplyChainView />
  }
])
\end{minted}
\caption{Initialisierung der Routen für das clientseitige Routing} \label{frontend-routes}
\end{listing}

Als nächstes soll die Root-Komponente der React-Anwendung beschrieben werden, die dafür in Listing \ref{index} gezeigt wird. 
Zu sehen ist ihre Initialisierung in Zeile 1 mit einer Referenz auf das \ac{html}-Element \texttt{root} in der \texttt{index.html} Datei.
Diese \ac{html}-Datei ist quasi der \ac{html}-seitige Einstiegspunkt der Webanwendung. 
Ab Zeile 4 befindet sich der Programmcode in der \ac{jsx} Syntax, wie in \ref{react} beschrieben. 
Die Elemente sind dabei wie in \ac{html} verschachtelt und Eltern-Elemente fungieren wie Wrapper, die Daten an ihre Kind-Elemente weitergeben können.

Das \texttt{<React.StrictMode>} Element aktiviert mehrere Funktionen, die helfen sollen, Fehler im Programmcode bereits während der Entwicklung zu finden. 
Etwa werden Kind-Komponenten und ihre Hooks häufiger neu gerendert, um dabei möglicherweise auftretende Fehler zu provozieren. 
Der \textit{StrictMode} ist dabei automatisch nur im Entwicklungsmodus aktiv. \cite{strict-mode}

Bei dem \texttt{<Provider>} Element handelt es sich um den in Abschnitt \ref{redux-grundlagen} beschriebenen Wrapper für die State-Management-Funktionen von Redux. 
Als Property wird diesem die Redux \texttt{storage} Instanz übergeben, unter welcher die vorhandenen Reducer registriert sind. 
Das Frontend der im Rahmen dieser Arbeit entwickelten Webanwendung besitzt nur einen von Redux verwalteten Zustand, welcher angibt, ob der Experten-Modus aktiviert ist. 
Die einzige mögliche Aktion auf diesem Zustand ist es, diesen umzuschalten. 
Aus diesem Grund sei an dieser Stelle nicht weiter auf den \texttt{storage} oder den Reducer eingegangen. 

Der \texttt{<PersistGate>} Wrapper wurde ebenfalls in Abschnitt  \ref{redux-grundlagen} erläutert. 
Bei diesem handelt es sich um ein Element der Redux Persist Bibliothek, welches ermöglicht, Zustände im \textit{Local Storage} des Browsers zu speichern.

Schließlich kommt in Zeile 7 mit dem \texttt{<RouterProvider>} Wrapper das am tiefsten geschachtelte Element. 
Dieser fügt der Root-Komponente von React die mit Listing \ref{frontend-routes} beschriebenen Routing-Funktionen hinzu. 

\begin{listing}[H]
\begin{minted}[xleftmargin=20pt,linenos,escapeinside=||]{text}
const root = ReactDOM.createRoot(document.getElementById('root'))

root.render(
  <React.StrictMode>
    <Provider store={store}>
      <PersistGate persistor={persistor}>
        <RouterProvider router={router} />
      </PersistGate>
    </Provider>
  </React.StrictMode>
)
\end{minted}
\caption{Initialisierung und Einfügen von Wrappern in die React Root-Komponente} \label{index}
\end{listing}

\subsection{Benutzeroberflächen}

Der übrige Programmcode des Frontends, welcher größtenteils die Gestalt und Funktionen der Benutzeroberflächen vorgibt, umfasst rund 1400 Zeilen. 
Dieser soll im Folgenden nicht mehr explizit mit Listings dargestellt werden, sondern es sollen stattdessen Screenshots der Benutzeroberfläche mit Erklärungen zum Gezeigten gegeben werden. 

\begin{figure}[H]
    \centering
    \fbox{\includegraphics[width=.98\linewidth]{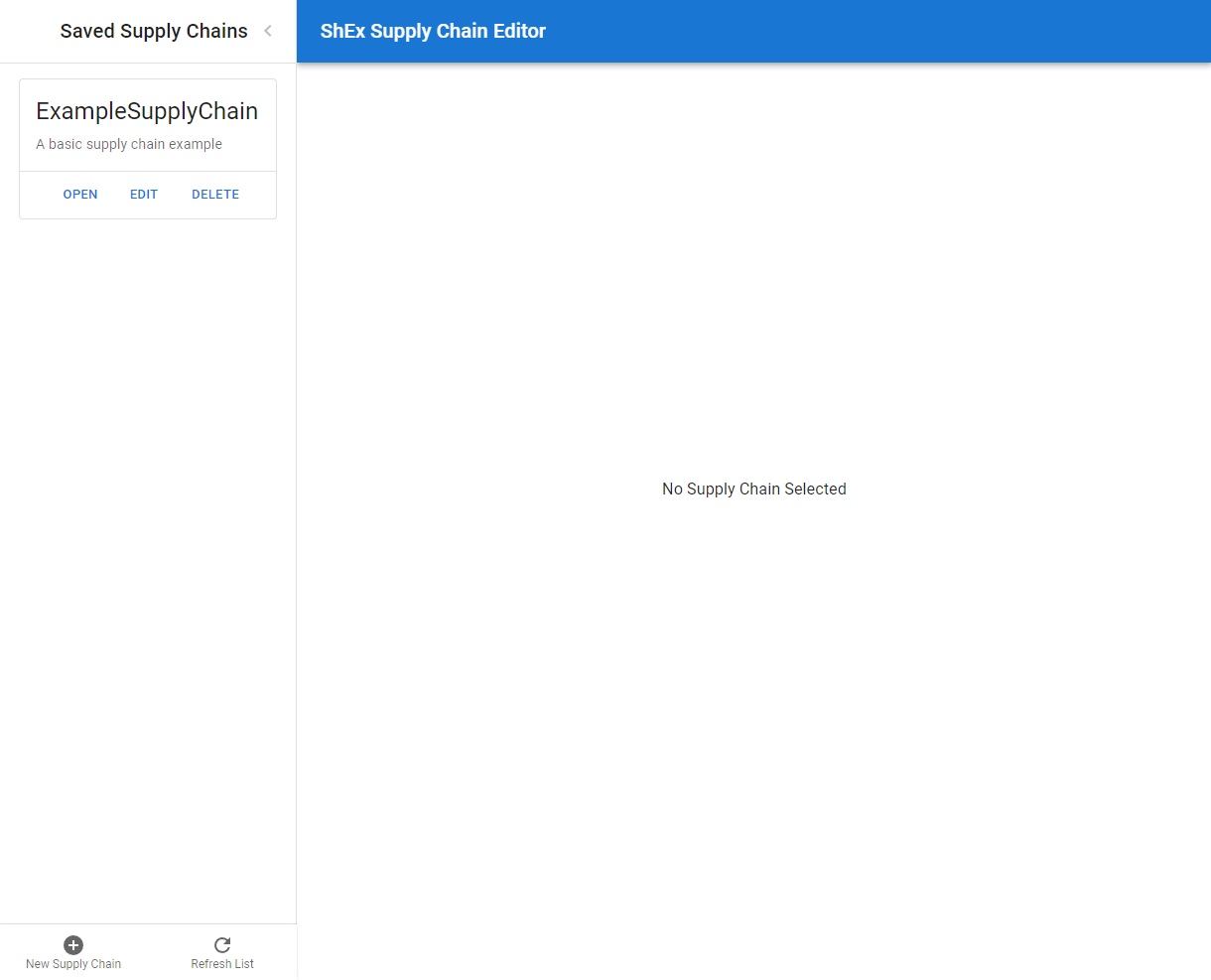}}
    \caption{Screenshot der Lieferketten-Ansicht}
    \label{menu screenshot}
\end{figure}

Begonnen werden soll mit dem Lieferketten-Menü, zu sehen in Abbildung \ref{menu screenshot}. 
Dies ist die Seite, welche erscheint, wenn man die Webanwendung ohne weitere Pfad-Angaben aufruft. 
Zu sehen ist am oberen Bildrand die in Abschnitt \ref{design} beschriebene Material Design App Bar, sowie auf der linken Seite der Material Design Navigation Drawer. 
Der Programmcode hierfür wurde dabei abgeleitet aus einem Beispiel von Material-UI \cite{mui-drawer} und entsprechend der Anforderungen für den Prototypen dieser Arbeit angepasst. 
Eine dieser Anforderungen ist es, eine Liste aller bestehenden Lieferketten anzuzeigen. 
Im gezeigten Beispiel existiert eine Lieferkette, weitere würden dabei unterhalb erscheinen. 

\begin{figure}[H]
    \centering
    \fbox{\includegraphics[width=.98\linewidth]{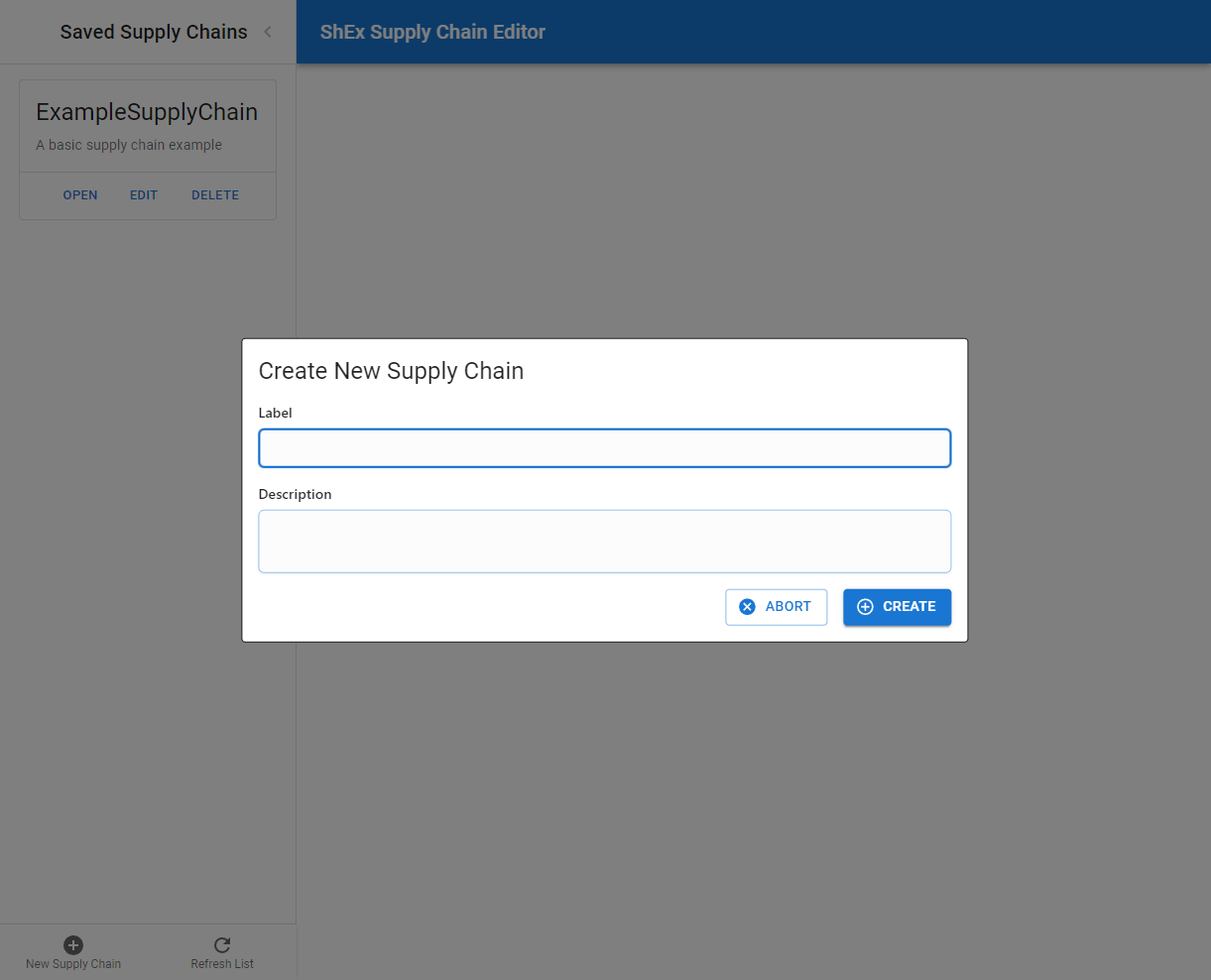}}
    \caption{Screenshot der Lieferketten-Ansicht mit geöffnetem Modal}
    \label{menu modal screenshot}
\end{figure}

Ebenfalls in dem Drawer auf der linken Seite befinden sich am unteren Bildschirmrand zwei Schaltflächen. 
Die rechte Schaltfläche löst eine Aktualisierung der Lieferketten-Liste aus, die linke öffnet hingegen ein sog. \textit{Modal}.
Dabei handelt es sich um ein Fenster, welches sich im Vordergrund der Benutzeroberfläche öffnet und den Hintergrund leicht abdunkelt, sodass der Fokus des Benutzers auf dieses Fenster gerichtet wird. 
Zu sehen ist dies in Abbildung \ref{menu modal screenshot}. 
Der Benutzer hat hier die Möglichkeit, eine neue Lieferkette anzulegen und soll dafür ein Label und eine Beschreibung eingeben. 
Ein Klick auf den \texttt{Abort} Knopf oder auf den abgedunkelten Hintergrund schließen dabei das Modal, ohne dass eine Lieferkette erstellt wird. 
Mit einem Klick auf \texttt{Create} wird in der Datenbank mit den eingegebenen Informationen eine neue Lieferkette angelegt und die Liste der Lieferketten auf der Benutzeroberfläche automatisch aktualisiert. 

Ein solches Modal öffnet sich auch, wenn in einer Lieferketten-Box der \texttt{Edit} Knopf geklickt wird. 
Jedoch sind dann bereits die aktuellen Daten der Lieferkette voreingetragen. 
Sollten die Daten geändert werden, aktualisiert sich auch hier die Liste der Lieferketten automatisch, ebenso bei einem Klick auf den \texttt{Delete} Knopf. 

\begin{figure}[H]
    \centering
    \fbox{\includegraphics[width=.98\linewidth]{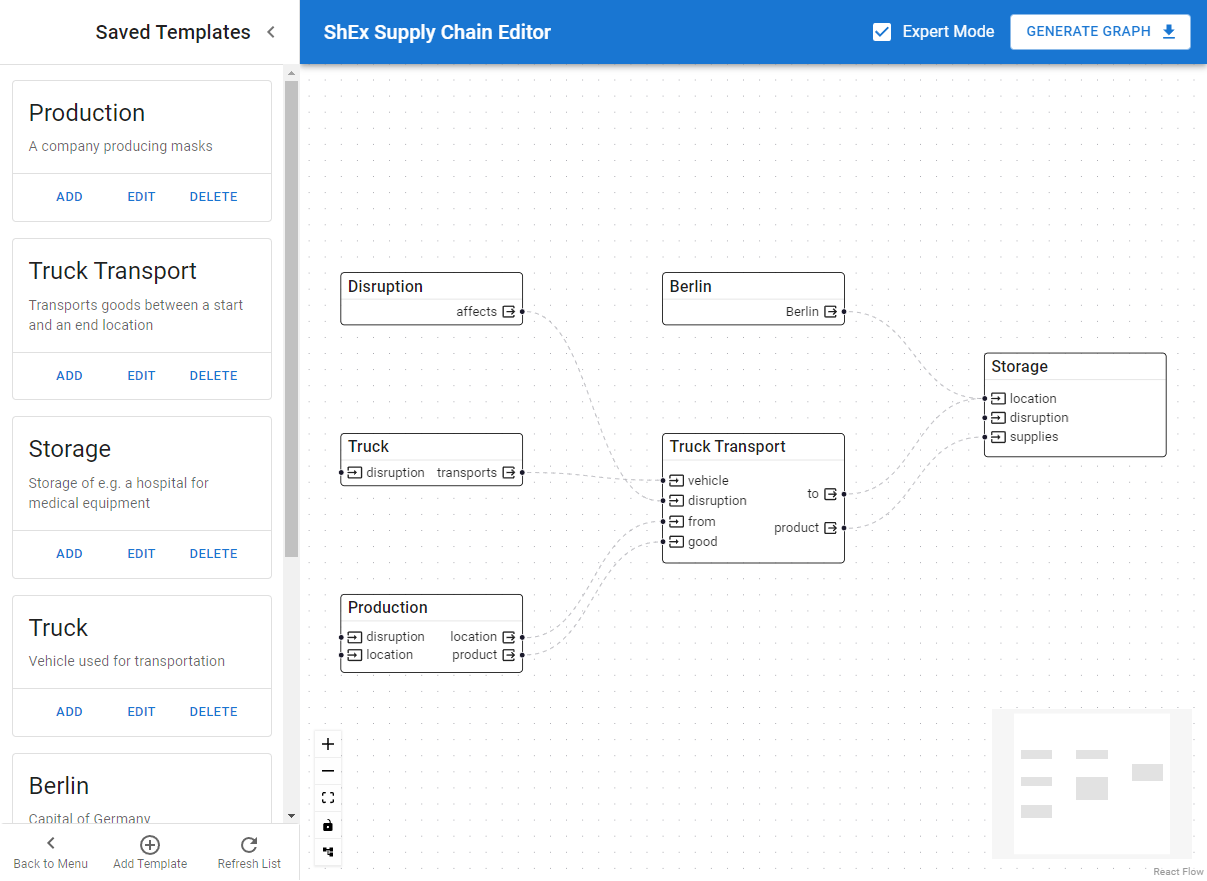}}
    \caption{Screenshot der Template-Ansicht}
    \label{sc screenshot}
\end{figure}

Das Klicken des \texttt{open} Knopfes einer Lieferketten-Box führt hingegen dazu, dass sich die entsprechende Lieferkette \textit{öffnet}. 
Es ändert sich dabei die \ac{url}, wie in \ref{react root} beschrieben, und selbstredend die Oberfläche. 
Zu sehen ist dann die Lieferketten-Ansicht, die beispielsweise wie in Abbildung \ref{sc screenshot} aussieht. 
Im Drawer auf der linken Seite ist nun die Liste der vorhandenen Templates. 
Auch die Schaltflächen am unteren Ende des Drawers haben sich verändert.
Es gibt wieder eine Schaltfläche zur Aktualisierung der Liste und eine Schaltfläche, um ein neues Element zur Liste hinzuzufügen, in diesem Fall jedoch ein Template. 
Als dritte Schaltfläche gibt es einen Knopf, mit dem man zum Lieferketten-Menü gelangt. 

Außerdem hat sich die App Bar geändert. 
Dort ist nun in der oberen rechten Ecke ein Knopf für die Generierung des Graphen zu finden und eine Checkbox, um den Experten-Modus zu aktivieren. 
Ersterer löst dabei einen automatischen Download der in der graphischen Oberfläche gezeigten Lieferkette im \ac{rdf} \ac{turtle} Format aus (siehe dafür Listing \ref{ausgabe}). 

Der Experten-Modus wurde bereits im vorherigen Abschnitt \ref{react root} genannt, soll nun aber noch einmal genauer erklärt werden: 
Dabei handelt es sich um die Umsetzung der in Abschnitt \ref{funktionale anforderungen} erhobenen Anforderungen von Experten und Benutzern. 
Nur wenn diese Checkbox aktiviert ist, lassen sich neue Templates über die Schaltfläche am unteren Rand des Drawers erstellen. 
Auch das Verändern oder Löschen mit den entsprechenden Knöpfen auf den Template-Boxen ist nur dann möglich.
Jederzeit möglich ist hingegen das Hinzufügen einer Template-Instanz auf die graphische Benutzeroberfläche per \texttt{Add} Knopf. 
Ebenfalls unabhängig vom Modus können Template-Instanzen entfernt werden oder Verknüpfungen zwischen Template-Instanzen erzeugt oder entfernt werden. 

Die graphische Oberfläche wurde mit der in Abschnitt \ref{react flow} beschriebenen React Flow Bibliothek realisiert. 
Es handelt sich konkret um eine interaktive Oberfläche, auf welcher die Template-Instanzen und ihre Verknüpfungen zu sehen sind. 
Der Benutzer kann die Elemente verschieben und damit neu anordnen, es gibt jedoch auch eine automatische Layout-Funktion (welche nach \cite{dagre} entwickelt wurde), die über die Funktions-Leiste in der unteren linken Ecke zu sehen ist. 
Dort befinden sich auch Knöpfe für die Zoom-Funktion, die Zentrierung der Ansicht sowie zur Sperrung der Oberfläche. 
In der unteren rechten Ecke befindet sich eine Karte, die bei umfangreichen Lieferketten bei der Orientierung helfen soll. 

Eine Template-Instanz wird als Box visualisiert, die (sofern vorhanden) Eingangs- und Ausgangspunkte besitzt. 
Diese sind mit einem entsprechenden Icon und einer Bezeichnung versehen. 
Die Bezeichnung wird dabei von den im Template als Eingangs- und Ausgangsvariablen (siehe Abschnitt \ref{iovars}) angegebenen Variablen abgeleitet. 
Verknüpfungen zwischen diesen Punkten können erstellt werden, indem mit der Maus eine Linie von einem Eingangs- zu einem Ausgangspunkt (oder andersherum) gezogen wird. 
Template-Instanzen oder Verknüpfungen können von der grafischen Oberfläche (und damit auch in der Datenbank) entfernt werden, indem diese zuerst per Mausklick ausgewählt werden und anschließend das erscheinende Entfernen-Icon geklickt wird. 

\begin{figure}[H]
    \centering
    \fbox{\includegraphics[width=.98\linewidth]{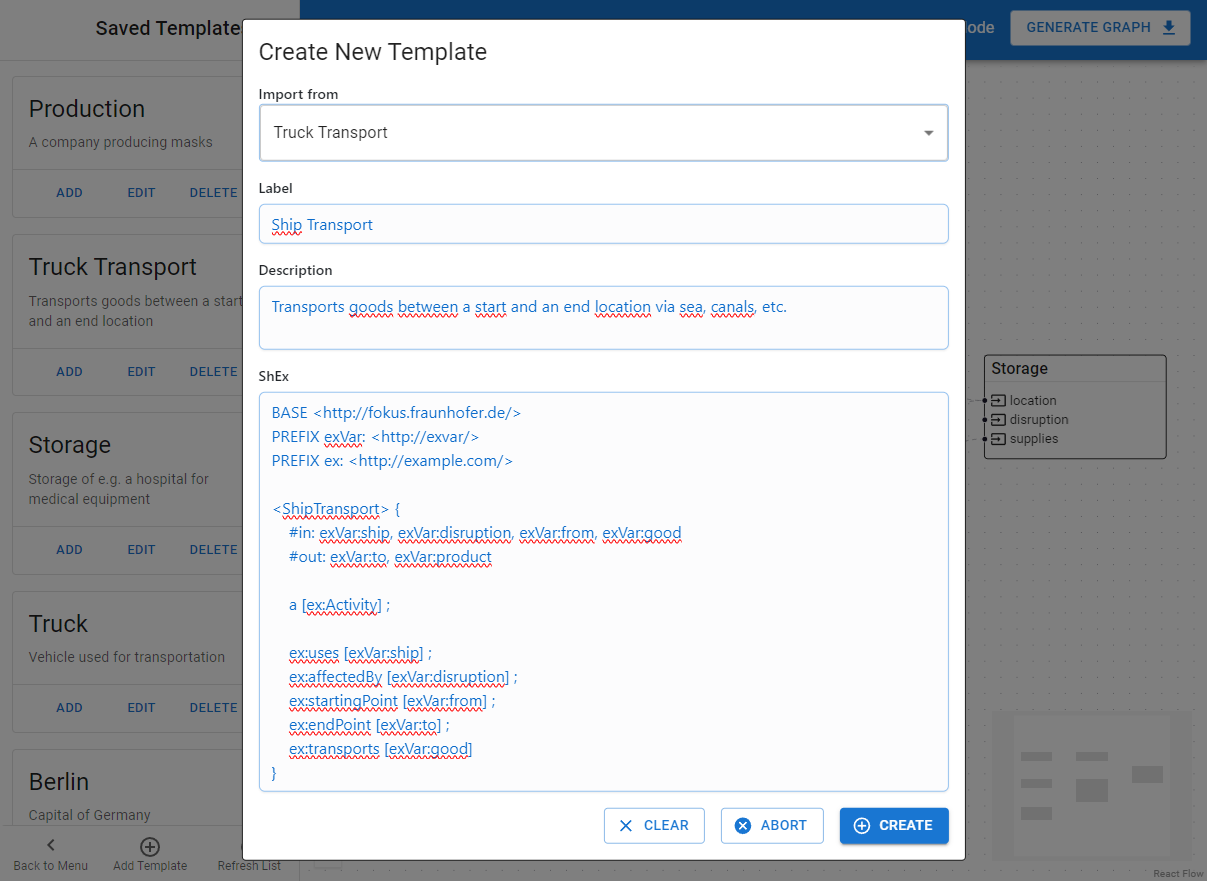}}
    \caption{Screenshot der Template-Ansicht mit geöffnetem Modal}
    \label{sc edit example}
\end{figure}

Wie das Modal zur Erstellung bzw Änderung eines Templates aussieht, soll in Abbildung \ref{sc edit example} gezeigt werden. 
Analog zum bereits erläutertem Lieferketten-Modal gibt es auch hier Felder zur Eingabe des Labels und der Beschreibung. 
Zusätzlich gibt es nun aber auch ein Feld zur Eingabe des \ac{shex} Schemas. 
Außerdem wird im Fall der Erstellung eines neuen Templates angeboten, die Felder mit den Daten eines bestehenden Templates auszufüllen, damit diese anschließend nur noch umgeschrieben werden müssen. 

\section{Containervirtualisierung}

Mit Kapitel \ref{container} wurden die Grundlagen zur Containervirtualisierung, insbesondere mit Blick auf Docker, bereits erläutert. 
In diesem Abschnitt sollen nun die Konfigurationsdateien für die soeben beschriebene Webanwendung vorgestellt werden. 

Begonnen werden soll mit der Datei \texttt{docker-compose.yml}, welche für das Docker Compose Tool verwendet wird.
Docker Compose ist dabei, wie bereits beschrieben, für die Verwaltung von Multi-Container-Anwendungen vorgesehen. 
Für diese Arbeit gibt es drei Container: Backend, Frontend und Proxy. 
In Listing \ref{docker-compose-projekt} sind diese jeweils unter \texttt{services} zu finden. 
Dabei wird mit \texttt{build} der Pfad zum entsprechenden \texttt{Dockerfile} angegeben und mit \texttt{ports} das Mapping der Ports zwischen Container und Host. 
Im Fall des Frontends wird in Zeile 7 der Container-Port 80 auf den Host-Port 8080 gemappt. 

\begin{listing}[H]
\begin{minted}[xleftmargin=20pt,linenos,escapeinside=||]{text}
version: "3"

services:
  frontend:
    build: ./frontend/
    ports:
      - 8080:80

  backend:
    build: ./backend/
    ports:
      - 8187:8187

  proxy:
    build: ./proxy/
    ports:
      - 80:80
\end{minted}
\caption{Docker-Compose Datei für die benötigten Container} \label{docker-compose-projekt}
\end{listing}

Vorausgesetzt die entsprechenden \texttt{Dockerfile}s sind vorhanden (siehe nachfolgende Abschnitte), können mit dem Befehl \texttt{docker-compose up -d} (ausgeführt im Ordner der Datei \texttt{docker-compose.yml}) alle definierten Services gleichzeitig gestartet werden. 
Alternativ kann mit \texttt{docker-compose up <service\_name>} auch nur ein bestimmter Service gestartet werden. 

\subsection{Backend}

Das Backend ist mit der Python Version 3.10 entwickelt und die verwendeten \textit{Requirements} sind dabei in Verbindung mit dem Packaging Tool \textit{Pipenv} definiert worden. 
Für das \texttt{Dockerfile} des Backends (siehe Listing \ref{docker-backend}) wird deshalb das Python-Image in der Version 3.10 als Basis verwendet. 
Anschließend wird Pipenv installiert (Zeile 2) und der Inhalt des Backend-Ordners in das Image kopiert (Zeile 3), bevor auch die Requirements mit Pipenv installiert werden können (Zeile 4). 
Danach wird in Zeile 5 der Befehl angegeben, welcher beim Starten des Containers ausgeführt werden soll. 
In diesem Fall ist es der Aufruf des \textit{Uvicorn} Servers inklusive Einstiegspunkt, Interface und Port. 

\begin{listing}[H]
\begin{minted}[xleftmargin=20pt,linenos,escapeinside=||]{text}
FROM python:3.10
RUN python -m pip install pipenv
COPY . .
RUN pipenv install --dev --system --deploy
CMD uvicorn main:app --host 0.0.0.0 --port 8187
\end{minted}
\caption{Dockerfile des Backends} \label{docker-backend}
\end{listing}

\subsection{Frontend}

Für das \texttt{Dockerfile} des Frontends (siehe Listing \ref{docker-frontend}) wird ein \textit{Node.js} Image als Basis verwendet. 
Konkreter wird dabei eine sog. \textit{Alpine Linux} Variante genutzt, wobei es sich im Kern um eine deutlich schlankere Linux-Distribution handelt. 
Nachdem in den Zeilen 3 bis 5 alle Dateien in das Image kopiert und die benötigten Abhängigkeiten installiert wurden, wird der Befehl \texttt{npm run build} ausgeführt. 
Dieser bewirkt, dass eine für den produktiven Einsatz optimierte Variante des Frontends erstellt wird. 

Anschließend wird eine neue Image-Stufe definiert (Zeile 8), die auf einem \textit{NGINX} Image aufbaut. 
Es werden anschließend in Zeile 9 nur die von der vorherigen Stufe benötigten Dateien kopiert, in diesem Fall das erstellte Frontend. 
Da es sich bei diesem nur um statische \ac{html}- und JavaScript-Dateien handelt, können diese ohne Probleme von einem Webserver wie NGINX bereitgestellt werden. 

\begin{listing}[H]
\begin{minted}[xleftmargin=20pt,linenos,escapeinside=||]{text}
FROM node:14-alpine AS build
WORKDIR /app
COPY package*.json /app/
RUN npm install
COPY . /app/
RUN npm run build

FROM nginx:1.25.2
COPY --from=build /app/build/ /usr/share/nginx/html
CMD ["nginx", "-g", "daemon off;"]
\end{minted}
\caption{Dockerfile des Frontends} \label{docker-frontend}
\end{listing}

\subsection{Proxy}

Auch der Proxy verwendet NGINX, jedoch als Reverse-Proxy. 
Dafür wird eine eigene, in Listing \ref{proxy-config} gezeigte, Konfigurationsdatei in das Image an die Stelle kopiert, an welcher zuvor die Standard-Datei entfernt wurde (Listing \ref{docker-proxy} Zeile 2 und 3). 

\begin{listing}[H]
\begin{minted}[xleftmargin=20pt,linenos,escapeinside=||]{text}
FROM nginx:1.25.2
RUN rm /etc/nginx/conf.d/default.conf
COPY nginx.conf /etc/nginx/conf.d/default.conf
CMD ["nginx", "-g", "daemon off;"]
\end{minted}
\caption{Dockerfile des Proxys} \label{docker-proxy}
\end{listing}

Mithilfe der Konfigurationsdatei können Routen definiert werden. 
In diesem Fall wird die Root-Route zum Frontend weitergeleitet und die Route unter dem Pfad \texttt{/backend} zum Backend. 
Beim Datenverkehr über den Proxy werden dabei gewisse \ac{http}-Header umgeschrieben, um den Effekt eines Reverse-Proxys zu erreichen. 

\begin{listing}[H]
\begin{minted}[xleftmargin=20pt,linenos,escapeinside=||]{text}
client_max_body_size 1000M;
server {
  listen 80;
  server_name host.docker.internal;
  proxy_http_version 1.1;

  location /backend/ {
    proxy_pass         http://host.docker.internal:8187/;
    proxy_set_header   Host $host;
    proxy_set_header   X-Real-IP $remote_addr;
    proxy_set_header   X-Forwarded-For $proxy_add_x_forwarded_for;
    proxy_set_header   X-Forwarded-Host $server_name;
    client_max_body_size 1000M;
  }

  location / {
    proxy_pass         http://host.docker.internal:8080/;
    proxy_set_header   Host $host;
    proxy_set_header   X-Real-IP $remote_addr;
    proxy_set_header   X-Forwarded-For $proxy_add_x_forwarded_for;
    proxy_set_header   X-Forwarded-Host $server_name;
    client_max_body_size 1000M;
  }
}
\end{minted}
\caption{Konfigurationsdatei des Proxys} \label{proxy-config}
\end{listing}

\section{Zugriff auf den Programmcode}

Der vollständige Programmcode der in dieser Arbeit entwickelten Webanwendung, inklusive der für die Containervirtualisierung aufgesetzten Konfigurationsdateien, kann unter Verwendung des Passworts \texttt{mfXqot3K4w} unter folgendem Link abgerufen werden: \\

\noindent \texttt{https://cloud.htw-berlin.de/s/kBy7SKH3ggCzgm4} \\

\noindent Alternativ können die Prüfer auch das firmeninterne GitLab verwenden: \\

\noindent \texttt{https://gitlab.fokus.fraunhofer.de/reskriver/shex-sce}

\chapter{Durchführung der Nutzerstudie} \label{durchführung nutzerstudie}

In diesem Kapitel soll die durchgeführte Nutzerstudie zur Überprüfung der Usability und der Ermittlung von Usability-Problemen dargestellt werden. 
Die verwendeten Usability-Tests wurden dabei bereits mit Abschnitt \ref{usability tests} erläutert. 
Es handelt sich konkret um eine abgewandelte Version des formalen Usability-Tests, gefolgt von einer Usability-Befragung mit dem standardisierten IsoMetrics-Fragebogen. 
Letzterer wurde dafür mit \textit{Google Forms} digitalisiert, um sowohl die Bearbeitung als auch die Auswertung zu vereinfachen. 

\section{Ablauf}

Es wurde folgender Ablauf für die Vorbereitung, Durchführung und Nachbereitung der beiden Usability-Tests festgelegt:

\begin{enumerate}
  \item \textbf{Webanwendung vorbereiten}: Für alle Testdurchläufe muss der Start-Zustand der Webanwendung identisch sein. 
  Es soll eine Lieferkette existieren und bereits einige Lieferketten-Templates und Verknüpfungen vorhanden sein. 
  \item \textbf{Briefing der Testperson}: Die Testperson soll Einblicke in die Hintergründe der zu testenden Webanwendung erhalten. 
  Dabei wird ihr erläutert, weshalb die Anwendung entwickelt wurde und welches Ziel sie hat. 
  Auch sollen die zugrunde liegenden Technologien \ac{rdf} und \ac{shex} kurz angesprochen werden. 
  \item \textbf{Vorbefragung}: Die Testperson wird zu ihren Erfahrungen mit Webanwendungen im allgemeinen, Lieferketten und Datenmodellierung befragt. 
  Außerdem wird um eine Enschätzung der eigenen Englischkenntnisse gebeten (siehe \ref{testpersonen}).
  \item \textbf{Erläuterung des Ablaufs}: Die Testperson erhält eine grobe Übersicht über den Ablauf der beiden Usability-Tests. 
  Dabei wird auch auf den voraussichtlichen zeitlichen Rahmen von 30 bis 60 Minuten und die Tatsache, dass die Anwendung und nicht die Person getestet werden soll, eingegangen. 
  Außerdem wird die Testperson angewiesen, Gedanken während des formalen Usability-Tests laut auszusprechen. 
  \item \textbf{Formaler Usability-Test}: Die Testperson führt im Beisein des Beobachters den formalen Usability-Test durch. 
  Dabei werden die Aufgaben mündlich vom Beobachter gegeben. 
  Sollte die Testperson eine Aufgabe nicht lösen können, wird eine Hilfestellung gegeben. 
  Mögliche Verbesserungsvorschläge können während der Durchführung des Tests besprochen werden. 
  \item \textbf{Nachbesprechung}: Der Testperson werden Frage gestellt, die auf den notierten Beobachtungen basieren. 
  Außerdem wird die Testperson nach ergänzenden Eindrücken zur Webanwendung gefragt. 
  Auch hier können Verbesserungsvorschläge diskutiert werden, was ebenfalls vom Beobachter verschriftlicht wird. 
  \item \textbf{Usability-Befragung}: Die Testperson erhält den Zugang zum standardisierten Usability-Fragebogen. 
  Diesen soll sie alleine bearbeiten. 
  Für eventuelle Rückfragen während der ersten zu bewertenden Aussagen steht der Beobachter jedoch bereit. 
  \item \textbf{Auswertung}: Die Usability-Tests werden vom Beobachter im Rahmen dieser Arbeit ausgewertet. 
\end{enumerate}

Der formale Usability-Test umfasst dabei Aufgaben, die quasi jede verfügbare Funktion testen. 
Sie reichen von der Erstellung und Modifizierung von Lieferketten und Templates über das Anordnen und Verknüpfen von Template-Instanzen bis hin zur Ausgabe des Ergebnis-Graphen. 

Der Umfang des IsoMetrics-Fragebogen wurde hingegen bereits in \ref{usability tests} beschrieben und sei deshalb an dieser Stelle nicht weiter ausgeführt. 

\section{Testpersonen} \label{testpersonen}

Die folgenden und in Tabelle \ref{person-table} aufgelisteten Testpersonen konnten für die Durchführung der Usability-Tests engagiert werden. 
Dabei werden ebenfalls die in der Vorbefragung erhaltenen Informationen angegeben. 

\begin{table}[H]
\begin{tabular}{|llll|lll|}
\hline
\multicolumn{4}{|l|}{\textbf{Angaben zur Person}}                                                                                                                                                                                           & \multicolumn{3}{c|}{\textbf{Erfahrungen mit}}                                                                                                                                                                                                              \\ \hline
\multicolumn{1}{|l|}{\textbf{ID}} & \multicolumn{1}{l|}{\textbf{Alter}} & \multicolumn{1}{l|}{\textbf{\begin{tabular}[c]{@{}l@{}}Mutter-\\ sprache\end{tabular}}} & \textbf{\begin{tabular}[c]{@{}l@{}}Englisch-\\ kenntnisse\end{tabular}} & \multicolumn{1}{l|}{\textbf{\begin{tabular}[c]{@{}l@{}}Liefer-\\ ketten\end{tabular}}} & \multicolumn{1}{l|}{\textbf{\begin{tabular}[c]{@{}l@{}}Web-\\ anwendungen\end{tabular}}} & \textbf{\begin{tabular}[c]{@{}l@{}}Daten-\\ modellierung\end{tabular}} \\ \hline
\multicolumn{1}{|l|}{\textbf{1}}  & \multicolumn{1}{l|}{25-30}          & \multicolumn{1}{l|}{Deutsch}                                                            & gut bis fließend                                                        & \multicolumn{1}{l|}{keine}                                                             & \multicolumn{1}{l|}{täglich}                                                             & gelegentlich                                                           \\ \hline
\multicolumn{1}{|l|}{\textbf{2}}  & \multicolumn{1}{l|}{25-30}          & \multicolumn{1}{l|}{Chinesisch}                                                         & gut bis fließend                                                        & \multicolumn{1}{l|}{keine}                                                             & \multicolumn{1}{l|}{täglich}                                                             & gelegentlich                                                           \\ \hline
\multicolumn{1}{|l|}{\textbf{3}}  & \multicolumn{1}{l|}{26}             & \multicolumn{1}{l|}{\begin{tabular}[c]{@{}l@{}}Deutsch, \\ Polnisch\end{tabular}}       & fließend                                                                & \multicolumn{1}{l|}{keine}                                                             & \multicolumn{1}{l|}{täglich}                                                             & gelegentlich                                                           \\ \hline
\multicolumn{1}{|l|}{\textbf{4}}  & \multicolumn{1}{l|}{53}             & \multicolumn{1}{l|}{Deutsch}                                                            & anfänglich                                                              & \multicolumn{1}{l|}{keine}                                                             & \multicolumn{1}{l|}{gelegentlich}                                                        & keine                                                                  \\ \hline
\end{tabular}
\caption{Übersicht der Testpersonen für die Usability-Tests} \label{person-table}
\end{table}

\section{Formaler Usability-Test}

In diesem Abschnitt sollen die während der Durchführung des formalen Usability-Tests angefertigten Notizen zu Beobachtungen und Rückmeldungen zusammengefasst werden. 
Dabei wird auf die Testpersonen anhand der in Tabelle \ref{person-table} gegebenen Nummer referenziert. 
Es sei dabei angemerkt, dass hierbei ein Fokus auf die gefundenen Usability-Probleme und -Verbesserungsvorschläge gelegt wird. 

\subsubsection{Testperson 1}

Testperson 1 hatte keine großen Probleme, die Komponenten der Benutzeroberfläche zu verstehen. 
Beim Öffnen einer Lieferkette wurde jedoch angemerkt, dass sich die Benutzeroberfläche nicht ausreichend verändert. 
Als Vorschlag wurde eine andere Farbe für die App Bar diskutiert, welche sich von jener des Lieferketten-Menüs unterscheidet. 

Da bereits eine Lieferkette für den Test vorgegeben war und ebenfalls bereits Lieferketten-Instanzen auf der graphischen Oberfläche zu sehen waren, hielt die Testperson die Liste der Templates auf der linken Seite im Drawer für die Liste der angezeigten Template-Instanzen. 
So war es für die Testperson verwunderlich, als beim Löschen einer Template-Instanz nicht auch der gleichnamige Eintrag in der Liste verschwand. 

Bezüglich der graphischen Oberfläche erwartete die Testperson eine feste Arbeitsfläche und versuchte deshalb, die gezeigten Template-Instanzen durch das Aufziehen einer Box zu markieren und so alle Elemente gleichzeitig zu verschieben. 
Erst nach einigen Versuchen wurde der Testperson klar, dass die Arbeitsfläche selbst verschiebbar ist und damit die gleichzeitige Verschiebung aller gezeigten Elemente möglich ist. 

Das Löschen von Template-Instanzen und Verknüpfungen, indem das zu löschende Element erst einmal angeklickt werden muss, hielt die Testperson zwar für übersichtlicher, aber dadurch auch schwerer zu erlernen und zu merken. 
Hier wurde als Vorschlag angegeben, die Entfernen-Icons bereits beim \textit{hovern} anzuzeigen. 
Allgemein empfiehlt die Testperson dabei, die Icons farbig in Rot zu gestalten. 

Die Verknüpfungspunkte der Template-Instanzen empfindet die Testperson als zu klein.
Es gelang ihr erst bei einer höheren Zoomstufe, die Verknüpfungsfunktion fehlerfrei auszuführen. 

Die automatische Layout-Funktion hat die Testperson nicht gefunden. 
Das liegt zum einen an einem zu kleinen und unbekannten Icon in der Funktionsleiste in der unteren linken Ecke der graphischen Oberfläche. 
Zum anderen bemerkte die Testperson jedoch auch das fehlende Label, welches bei den anderen Einträgen der Leiste erscheint, wenn der Cursor darüber gelegt wird. 

Insgesamt monierte die Testperson fehlende Bedienhinweise, die ihr die Verwendung der Benutzeroberfläche ohne Hilfe beibringen könnten. 
Dass nur Englisch als Sprache für die Benutzeroberfläche angeboten wird, hielt die Testperson für angemessen. 

\subsubsection{Testperson 2}

Testperson 2 lobte zunächst das klare und aufgeräumte Design der Benutzeroberfläche.
Die Funktionen zum Erstellen von neuen Lieferketten und Templates konnte sie ohne Probleme selbstständig lösen. 
Was jedoch die Einträge in der im Drawer gezeigten Liste darstellen, hat die Testperson nicht erkannt. 
Die darüber stehende Überschrift ist für sie nicht ersichtlich genug. 
Als Vorschlag wurde angemerkt, dass diese Überschrift zentriert direkt über der Liste sein sollte und nicht in einem abgetrennten Bereich darüber. 

Ebenso wie die erste Testperson konnte auch Testperson 2 die automatische Layout-Funktion nicht finden. 
Allgemein versuchte sie für die meisten gestellten Aufgaben zur grafischen Oberfläche ein Aktionsmenü per Rechtsklick zu öffnen. 
Dies war auch der Fall beim Löschen von Template-Instanzen und Verknüpfungen. 
Die tatsächliche Funktion hat die Testperson erst nach einigen Versuchen gefunden. 
Als Lösungsansatz merkte die Testperson an, dass es eine Möglichkeit geben sollte, die Entfernen-Icons dauerhaft anzeigen zu lassen. 

Bezüglich der Instanziierung von Templates schlug die Testperson darüber hinaus eine Möglichkeit vor, diese per \textit{Drag and Drop} von der Liste der Templates aus auf die grafische Oberfläche ziehen zu können. 

Den Experten-Modus hat die Testperson nicht verstanden. 
Sie hat lange versucht herauszufinden, warum bestimmte Aktionen nicht möglich sind und konnte erst mit einem Hinweis die entsprechende Checkbox finden. 
Als Vorschlag gab die Testperson an, diese prominenter zu platzieren. 

Ähnlich zu Testperson 1 gab Testperson 2 an, dass sie beim Löschen eines Templates erwarten würde, dass die entsprechenden Template-Instanzen auf der graphischen Oberfläche verschwinden würden. 
Es scheint also ein Problem beim Verständnis von Templates und Template-Instanzen vorzuliegen. 

Bezüglich der Verwendung von Englisch als Sprache der Benutzeroberfläche war die Testperson einverstanden. 
Sie vermutet, dass sie aufgrund ihrer Muttersprache Chinesisch mit einer deutschsprachigen Benutzeroberfläche Probleme gehabt hätte. 

\subsubsection{Testperson 3}

Testperson 3 hatte direkt bei einer der ersten Aufgaben einen Verbesserungsvorschlag: 
Beim Erstellen einer Lieferkette soll diese direkt im Anschluss geöffnet werden. 

Allgemein hat sie analog zu Testperson 2 die Überschrift über der Liste des Drawers nicht wahrgenommen und deshalb den Typ der gezeigten Elemente nicht verstanden. 

Auch Testperson 3 hatte große Probleme beim Finden der Funktion zum Löschen von Template-Instanzen und Verknüpfungen und benötigte einen Hinweis. 
Auch sie gab an, dass die Entfernen-Icons entweder immer sichtbar sein sollten, oder beim Darüberlegen des Cursors erscheinen sollten. 

Insgesamt vermisste die Testperson Möglichkeiten, unterstützende Erklärungen bei der Verwendung der Benutzeroberfläche zu erhalten. 
Sie regt dazu etwa die Verwendung von \textit{Helfe}-Icons an, die beim Auswählen eine Textbox öffnen, sodass die Benutzeroberfläche weiterhin übersichtlich bleibt. 

Dass die Benutzeroberfläche ausschließlich auf Englisch verfügbar ist, findet die Testperson in Ordnung. 
Sie vermutet jedoch auch, dass weitere Sprachen die Benutzerfreundlichkeit verbessern würden. 

\subsubsection{Testperson 4}

Testperson 4 hatte insgesamt große Probleme bei der Verwendung der Benutzeroberfläche. 
Sie selbst begründete dies mit einer seltenen Nutzung von vergleichbaren Webanwendungen in Beruf und Alltag. 

Hinzu kam, dass ausschließlich Englisch als Sprache angeboten wurde. 
Eine solche Anwendung würde die Testperson nach eigenen Angaben generell nicht selbstständig versuchen zu bedienen und zu verstehen, sondern immer Hilfe holen. 

Die Erstellung von Lieferketten und Templates gelang noch mit wenig Hilfe, jedoch benötigte die Testperson bei der Bedienung der grafischen Oberfläche so umfangreiche Hilfestellungen, dass aus dem Usability-Test eher eine Art Vorstellung mit Rückfragen an die Testperson wurde. 

Die Testperson würde sich wünschen, dass ihr entweder durch ein Video oder eine Person die Verwendung einer solchen Webanwendung beigebracht wird. 
Dass es einen Experten-Modus gibt, fiel der Testperson dabei positiv auf, da somit weniger Fehler bei der Bedienung begangen werden können. 

Abschließend gefiel der Testperson jedoch die insgesamte Darstellung der Template-Instanzen und Verknüpfungen. 
 
\section{Usability-Befragung}

Das Ergebnis der Usability-Befragung sei mit der Tabelle \ref{anhang-fragebogen} im Anhang dieser Arbeit dargestellt. 
Da es sich um eine kleine Teilnehmerzahl handelt, seien im Folgenden bestimmte Aussagen aus dem Fragebogen herausgegriffen und anhand der von den Testpersonen gegebenen Rückmeldungen ausgewertet. 
Den Testpersonen wurde der in \ref{usability befragung} beschrieben IsoMetrics-Fragebogen mit 72 Aussagen in 5 Themenbereichen zur Bewertung vorgelegt.  

\subsection{Aufgabenangemessenheit}

\begin{itemize}
  \item \textbf{Die für die Aufgabenbearbeitung notwendigen Informationen befinden sich immer am richtigen Platz auf dem Bildschirm}: Bei dieser Aussage vergeben die Testpersonen einen Mittelwert von 3,25. 
  Das mittelmäßige Ergebnis lässt sich vermutlich auf die während des formalen Usability-Tests erhaltenen Rückmeldungen zur Überschrift der Lieferketten bzw. Template-Liste und die Platzierung der Experten-Modus Checkbox zurückführen. 
  \item \textbf{Auf dem Bildschirm finde ich alle Informationen, die ich gerade benötige}: Die Teilnehmer vergeben hier im Schnitt eine Bewertung von 3,5. 
  Dies ist durch die fehlenden Hilfestellungen zu begründen. 
  \item \textbf{Die Software ermöglicht es mir, Daten so einzugeben, wie es von der Aufgabenstellung gefordert wird}: Alle Testpersonen vergeben hier die beste Bewertung (5), was auf die beschriebenen Modale für die Erstellung und Änderung von Lieferketten und Templates zurückzuführen ist.  
  \item \textbf{Die Software ist auf die von mir zu bearbeitenden Aufgaben zugeschnitten}: Die Testpersonen vergeben hier im Mittel eine sehr gute Bewertung von 4,75. 
\end{itemize}

\subsection{Selbstbeschreibungsfähigkeit}

\begin{itemize}
  \item \textbf{Bei Bedarf können für die Benutzung des Systems Erläuterungen abgerufen werden}: Hier vergeben alle Testpersonen eine Bewertung von 1, was angesichts des gänzlichen Fehlens einer solchen Funktion zu erwarten war. 
  \item \textbf{Die Meldungen der Software sind für mich sofort verständlich}: Nur drei der viel Teilnehmer vergeben hier eine Bewertung. 
  Diese liegt dabei im Mittel bei 2,67. 
  Es wird vermutet, dass diese schlechte Bewertung darin begründet ist, dass es keine Meldungen in der Oberfläche gibt, etwa wenn etwas erfolgreich erstellt oder gelöscht wurde. 
  \item \textbf{Wenn ich Informationen zu einem bestimmten Eingabefeld benötige, lassen sich diese einfach abrufen}: Für diese Aussage vergeben die Testpersonen zwar eine durchschnittliche Bewertung von 2,25, jedoch hat eine Testperson, vermutlich fälschlicherweise, eine 5 statt einer 1 angegeben, was auf einen Fehler beim Verständnis hindeutet. 
  Denn, wie bereits angemerkt, gibt es kaum Hilfestellungen in Form von Labels oder Texten. 
  \item \textbf{Wenn Befehle in bestimmten Situationen nicht zur Verfügung stehen (gesperrt sind), ist dies leicht erkennbar}: Hier vergeben die Teilnehmer eine Durchschnittsbewertung von 4,75. 
  Dies ist mit der deutlichen Kennzeichnung von inaktiven Funktionen bei ausgeschaltetem Experten-Modus zu begründen. 
  \item \textbf{Die von der Software verwendeten Begriffe sind für mich sofort verständlich}: Hier vergeben die Testpersonen einen maximalen Wert von 5, einen minimalen Wert von 2 und im Mittel einen Wert von 4. 
  Es wird vermutet, dass hier die mit dem formalen Usability-Test erkannten sprachlichen Barrieren eingeflossen sind. 
\end{itemize}

\subsection{Steuerbarkeit}

\begin{itemize}
  \item \textbf{Die Software bietet mir gute Bedienungsmöglichkeiten, um mich in Dokumenten (Texten, Datenbanken, Kalkulationsblättern etc.) zu bewegen}: Die vier Testpersonen vergeben für diese Aussage im Mittel eine Bewertung von 4,75. 
  Das sehr gute Ergebnis ist vor allem auf die intuitive Bedienung der graphischen Oberfläche zurückzuführen. 
  \item \textbf{Mit der Software ist für mich ein einfaches Bewegen zwischen den unterschiedlichen Menüebenen möglich}: Alle Testpersonen vergeben hier die beste Bewertung (5). 
  Zu begründen ist dies einerseits mit dem Vorhandensein einer entsprechenden Schaltfläche in der unteren linken Ecke der Lieferketten-Ansicht. 
  Andererseits gestaltet sich darüber hinaus durch die Verwendung von den in Abschnitt \ref{frontend technologien} beschriebenen Bibliotheken React Router DOM und Redux Persist das Vor- und Zurückspringen mit der Browser-Historie intuitiv. 
\end{itemize}

\subsection{Erwartungskonformität}

\begin{itemize}
  \item \textbf{Die Software erschwert meine Aufgabenbearbeitung durch eine uneinheitliche Gestaltung}: Die Testpersonen vergeben für diese Aussage eine durchschnittliche Bewertung von 1,25, was nahezu dem für diese Aussage bestem Wert entspricht. 
  Da sich die grundlegende Struktur zwischen Lieferketten-Menü und Lieferketten-Ansicht nicht verändert, war eine gute Bewertung zu erwarten. 
  \item \textbf{Gleiche Funktionen lassen sich in allen Teilen der Software einheitlich ausführen}: Drei von vier Testpersonen vergeben hier eine Bewertung, die dabei jedoch jeweils eine 5 ist. 
  Es wird vermutet, dass dieses positive Ergebnis durch die einheitliche Benutzeroberfläche und die Ähnlichkeit der Elemente der Lieferketten- und Template-Liste entstanden ist. 
\end{itemize}

\subsection{Fehlerrobustheit}

\begin{itemize}
  \item \textbf{Bei der Arbeit mit der Software kann es passieren, dass auch kleine Fehler schwerwiegende Folgen nach sich ziehen}: Die vier Testpersonen vergeben hier als minimalen Wert eine 1, als maximalen Wert eine 3 und im Mittel eine 1,75. 
  Es wird vermutet, dass hier eine Testperson an versehentliches Auslösen von Entfernungs-Funktionen auf Lieferketten oder Templates gedacht hat. 
  Tatsächlich gibt es keine Rückfrage, ob der Löschvorgang wirklich durchgeführt werden soll und im schlechtesten Fall ist damit bei einem vermeintlich kleinem Fehler eine gesamte Lieferkette gelöscht. 
  \item \textbf{Die Software ist so gestaltet, dass das versehentliche Auslösen von Aktionen verhindert wird (z.B. durch Sicherheitsabstände zwischen kritischen Tasten, durch geeignete Benennung, durch Hervorhebungen etc.)}: Hier ist das Ergebnis mit 2, 2, 4 und 5 sehr unterschiedlich. 
  Bereits beim formalen Usability-Test gab eine Testperson an, dass etwa die Entfernen-Icons farblich mit der Signalfarbe Rot hervorgehoben werden sollten. 
  Zudem muss auch hier angemerkt werden, dass ein versehentliches Auslösen von Aktionen direkt ohne Rückfrage ausgeführt wird. 
  \item \textbf{Es hat lange gedauert bis ich die Bedienung der Software erlernt habe}: Die Testpersonen vergeben hier drei Mal eine 1 und ein Mal eine 3. 
  Es wird vermutet, dass die 3, aus offensichtlichen Gründen, von der Testperson mit den größten Schwierigkeiten bei der Bedienung der Benutzeroberfläche stammt. 
  \item \textbf{Ich konnte die Software von Anfang an alleine bedienen, ohne dass ich Kollegen fragen musste}: Hier vergeben die Testpersonen als minimalen Wert eine 2 und als maximalen Wert eine 5. Im Mittel vergeben sie eine 3,25. 
  Das ist kein gutes Ergebnis, ist jedoch angesichts der während des formalen Usability-Tests erhaltenen Rückmeldungen zu erwarten gewesen. 
\end{itemize}

\chapter{Zusammenfassung} \label{zusammenfassung}

\section{Fazit}

Im Rahmen dieser Arbeit sollte eine Webanwendung entwickelt werden, die es auch IT-fremden Personen ermöglicht, Lieferketten auf der Basis von \ac{shex} Templates zu modellieren. 
Es wurden dafür zunächst die benötigten konzeptionellen und technischen Grundlagen in Kapitel \ref{grundlagen} erarbeitet. 
Dazu gehörte die Klärung von Grundbegriffen wie \textit{Lieferkette} und \textit{Template}. 

Ebenfalls wurde mit Abschnitt \ref{rdf-grundlagen} eine Einführung in das \acl{rdf} gegeben und dabei Probleme mit Blank Nodes erläutert, die es nicht möglich machen, diese für die Zwecke dieser Arbeit zu verwenden. 
Dabei wurde auch auf das Thema der Skolemisation eingegangen, welches zuvor bereits in einem eigenen Abschnitt ausgeführt wurde. 
Es wurde dabei das Konzept der Skolem \ac{iri}s erarbeitet, welche eine wichtige Rolle bei der Generierung von \ac{rdf} Graphen spielen. 

Was die Bestandteile der \acl{shex} Sprache sind wurde umfangreich in Abschnitt \ref{shex-grundlagen} beschrieben, da diese für die Beschreibung der Template-Definition relevant sind. 
Es wurde in Kapitel \ref{templates mit shex} dargelegt, wie genau \ac{shex} Shapes bzw. Schemas für die Generierung von \ac{rdf} Graphen verwendet werden können. 
Dabei wurde auf Gemeinsamkeiten zu \ac{rdf} und neue Konzepte eingegangen, die dies ermöglichen. 
Zu diesen Konzepten gehört etwa die Verwendung von Skolem \ac{iri}s für anonyme Shapes und für die existenzquantifizierenden Variablen. 
Wie Templates dabei verknüpft werden können, wurde durch die Einführung von Eingangs- und Ausgangsvariablen beschrieben. 

Die Webanwendung wurde auf Basis der soeben zusammengefassten Konzepte implementiert. 
Dabei wurde auf bekannte Technologien zurückgegriffen, welche entweder im Rahmen der Grundlagen-Abschnitte \ref{backend-technologien} und \ref{frontend technologien} oder dem Kapitel \ref{implementierung} zur Implementierung beschrieben wurden. 

Um die Benutzerfreundlichkeit der entwickelten Webanwendung beurteilen zu können, wurden bereits im Grundlagen-Abschnitt \ref{usability-grundlagen} standardisierte Interaktionsprinzipien und zwei Usability-Tests erarbeitet. 
Diese wurden in Kapitel \ref{durchführung nutzerstudie} mithilfe von vier Testpersonen angewendet und dabei nützliche Erkenntnisse erhalten. 
Dabei wurde ersichtlich, dass sich die Verwendung von bekannten Komponenten für die Benutzeroberfläche positiv auf die Usability auswirkt. 
Jedoch fanden die Testpersonen auch eine Reihe von Usability-Probleme und gaben Anregungen für Verbesserungen. 

Insgesamt kann die gestellte Forschungsfrage deshalb nicht nur positiv beantwortet werden. 
Der entstandene Prototyp erfüllt zwar seine Aufgabe, jedoch nicht unbedingt gegenüber IT-fremden Personen, was anhand der Testperson 4 zu sehen war. 

\section{Limitationen}

Anknüpfend an den letzten Abschnitt sei dabei jedoch erwähnt, dass vier Personen eine nicht allzu große Zahl von Test-Teilnehmern ist. 
Sie ist zwar groß genug, um die meisten Usability-Probleme zu finden und macht deshalb auch in Bezug auf den Entwicklungsstand der Prototyp-Webanwendung Sinn, jedoch kann damit keine abschließende Beantwortung der Forschungsfrage stattfinden. 
Denn hinzu kommt die schwache Heterogenität der Gruppe der Testpersonen. 
Es hat sicherlich geholfen eine in Bezug auf IT-Themen größtenteils unerfahrene Person im Teilnehmerkreis zu haben, jedoch war diese Person nicht mit den \textit{IT-fremden Personen} aus der Forschungsfrage gemeint. 
Vielmehr wäre eine Personen, die sich etwa im Rahmen des Supply Chain Managements in einem Unternehmen mit Lieferketten beschäftigt, für die Nutzerstudie geeigneter gewesen. 
Bei einer solchen Person ist ohnehin von einem regelmäßigem Gebrauch von Webapplikationen auszugehen. 
Zudem sind ihr Grundbegriffe aus der Logistik bestens bekannt, womit die von dieser Person erhaltenen Rückmeldungen, nicht nur in Bezug auf Usability, aber in Bezug auf die Gesamtidee der Webanwendung, umso gewichtiger ausfallen würden. 
Womöglich könnte damit ein größerer positiver Einfluss auf die Gestaltung dieser Webanwendung genommen werden, als mit nicht mit dem Thema in Verbindung stehenden Personen. 

\section{Ausblick} \label{ausblick}

Nichtsdestotrotz kann diese Arbeit und die entwickelte Webanwendung als eine gute Basis für weiterführende Forschungs- und Entwicklungstätigkeiten gesehen werden. 
Es liegen durch die Nutzerstudie genügend Idee für Verbesserungen der Benutzeroberfläche vor, von denen viele ohne allzu großen Aufwand umgesetzt werden können. 
Dabei kann auf eine ebenfalls solide Code-Basis zurückgegriffen werden, da nahezu alle Komponenten nach gängigen Konventionen aufgesetzt und entwickelt wurden. 
Eine Erweiterung ohne umfangreiche Umbauarbeiten sollte deshalb ohne weiteres möglich sein. 

Hinzu kommt eine Erweiterung der in Kapitel \ref{templates mit shex} beschriebenen Generierung von \ac{rdf} Graphen anhand der \ac{shex} Templates. 
Denn aktuell werden, wie dargelegt, nur die Value Sets von Value Constraints für die Generierung berücksichtigt, da alle übrigen zu abstrakt scheinen. 
Doch zumindest für Validierungszwecke könnten beispielsweise Node Kind Constraints oder Datatype Constraints verwendet werden. 
Wenn etwa bekannt ist, dass eine Eingangsvariable vom Typ ein \texttt{ex:Truck} sein muss, könnten falsche Verbindungsversuche in der graphischen Oberfläche direkt unterbunden werden. 
Auch könnte somit eine Art automatische Vervollständigung umgesetzt werden, die automatisch die Verknüpfungen zwischen Templates vornimmt oder fehlende Template-Instanzen für unbesetzte Eingangs- und Ausgangsvariablen erzeugt.

\clearpage
\bibliographystyle{unsrt}
\bibliography{literatur}

\chapter*{Abkürzungsverzeichnis}
\begin{acronym}
\acro{iri}[IRI]{Internationalized Resource Identifier}
\acro{rdf}[RDF]{Resource Description Framework}
\acro{w3c}[W3C]{World Wide Web Consortium}
\acro{rdfs}[RDFS]{RDF Schema}
\acro{owl}[OWL]{Web Ontology Language}
\acro{sparql}[SPARQL]{SPARQL Protocol and RDF Query Language}
\acro{turtle}[Turtle]{Terse RDF Triple Language}
\acro{shex}[ShEx]{Shape Expressions}
\acro{xsd}[XSD]{XML Schema Definition}
\acro{shexj}[ShExJ]{ShEx JSON Syntax}
\acro{shexc}[ShExC]{ShEx Compact Syntax}
\acro{exvar}[exVar]{existenzquantifizierende Variable}
\acro{crud}[CRUD]{Create Read Update Delete}
\acro{api}[API]{Application Programming Interface}
\acro{orm}[ORM]{Object Relational Mapping}
\acro{sql}[SQL]{Structured Query Language}
\acro{url}[URL]{Uniform Resource Locator}
\acro{http}[HTTP]{Hypertext Transfer Protocol}
\acro{shacl}[SHACL]{Shapes Constraint Language}
\acro{rest}[REST]{Representational State Transfer}
\acro{cra}[CRA]{Create React App}
\acro{html}[HTML]{Hypertext Markup Language}
\acro{css}[CSS]{Cascading Style Sheets}
\acro{jsx}[JSX]{JavaScript Syntax Extension}
\acro{ide}[IDE]{Integrated Development Environment}
\end{acronym}

\clearpage
\appendix

\chapter{Ergebnisse der Usability-Befragung}

\begin{longtable}{|p{11cm}|llll|}
\hline
\textbf{Aussage}                                                                                                                                                                                         & \multicolumn{4}{l|}{\textbf{Antworten}}                                      \\ \hline
\endfirsthead
\multicolumn{5}{c}%
{{\bfseries Tabelle \thetable\ weitergeführt von vorheriger Seite}} \\
\hline
\textbf{Aussage}                                                                                                                                                                                         & \multicolumn{4}{l|}{\textbf{Antworten}}                                      \\ \hline
\endhead
Die Software zwingt mich, überflüssige Arbeitsschritte durchzuführen.                                                                                                                                    & \multicolumn{1}{l|}{1} & \multicolumn{1}{l|}{2} & \multicolumn{1}{l|}{1} & 1 \\ \hline
Mit der Software kann ich zusammenhängende Arbeitsabläufe vollständig bearbeiten.                                                                                                                         & \multicolumn{1}{l|}{5} & \multicolumn{1}{l|}{4} & \multicolumn{1}{l|}{4} & 5 \\ \hline
Die Software bietet mir alle Möglichkeiten, die ich für die Bearbeitung meiner Aufgaben benötige.                                                                                                         & \multicolumn{1}{l|}{5} & \multicolumn{1}{l|}{4} & \multicolumn{1}{l|}{3} & 5 \\ \hline
Die Software ermöglicht es mir, Daten so einzugeben, wie es von der Aufgabenstellung gefordert wird.                                                                                                       & \multicolumn{1}{l|}{5} & \multicolumn{1}{l|}{5} & \multicolumn{1}{l|}{5} & 5 \\ \hline
Die für die Aufgabenbearbeitung notwendigen Informationen befinden sich immer am richtigen Platz auf dem Bildschirm.                                                                                       & \multicolumn{1}{l|}{3} & \multicolumn{1}{l|}{3} & \multicolumn{1}{l|}{3} & 4 \\ \hline
Es müssen zu viele Eingabeschritte für die Bearbeitung mancher Aufgaben durchgeführt werden.                                                                                                                & \multicolumn{1}{l|}{1} & \multicolumn{1}{l|}{1} & \multicolumn{1}{l|}{2} & 1 \\ \hline
Die vom Programm erzeugten Ausgaben passen zum einen Aufgabenstellungen, d.h. sie erhalten keine überflüssigen, zu knappen oder unverständlich formulierten Informationen.                                  & \multicolumn{1}{l|}{K} & \multicolumn{1}{l|}{5} & \multicolumn{1}{l|}{K} & 5 \\ \hline
Die Software ist auf die von mir zu bearbeitenden Aufgaben zugeschnitten.                                                                                                                                 & \multicolumn{1}{l|}{5} & \multicolumn{1}{l|}{4} & \multicolumn{1}{l|}{5} & 5 \\ \hline
Auf dem Bildschirm finde ich alle Informationen, die ich gerade benötige.                                                                                                                                 & \multicolumn{1}{l|}{4} & \multicolumn{1}{l|}{3} & \multicolumn{1}{l|}{3} & 4 \\ \hline
Die in der Software verwendeten Begriffe und Bezeichnungen entsprechen denen meiner Arbeitstätigkeit.                                                                                                      & \multicolumn{1}{l|}{5} & \multicolumn{1}{l|}{3} & \multicolumn{1}{l|}{K} & 5 \\ \hline
Die Software bietet mir eine Wiederhol-Funktion für wiederkehrende Arbeitsschritte.                                                                                                                       & \multicolumn{1}{l|}{5} & \multicolumn{1}{l|}{5} & \multicolumn{1}{l|}{3} & 5 \\ \hline
Auch nicht routinemäßig auftretende Arbeitsaufgaben lassen sich mit der Software einfach bearbeiten.                                                                                                       & \multicolumn{1}{l|}{K} & \multicolumn{1}{l|}{K} & \multicolumn{1}{l|}{K} & 5 \\ \hline
Für meine Arbeit wichtige Befehle werden von der Software so dargeboten, dass sie sich leicht auffinden lassen.                                                                                             & \multicolumn{1}{l|}{4} & \multicolumn{1}{l|}{5} & \multicolumn{1}{l|}{3} & 5 \\ \hline
Die mit der Software erzeugten Ergebnisse lassen sich meinen Anforderungen entsprechend darstellen bzw. ausgeben.                                                                                          & \multicolumn{1}{l|}{5} & \multicolumn{1}{l|}{4} & \multicolumn{1}{l|}{5} & 5 \\ \hline
Die Darstellung der Informationen auf dem Bildschirm unterstützt mich bei der Bearbeitung meiner Aufgaben.                                                                                                 & \multicolumn{1}{l|}{5} & \multicolumn{1}{l|}{4} & \multicolumn{1}{l|}{2} & 5 \\ \hline
Bei Bedarf können für die Benutzung des Systems Erläuterungen abgerufen werden.                                                                                                                           & \multicolumn{1}{l|}{1} & \multicolumn{1}{l|}{1} & \multicolumn{1}{l|}{1} & 1 \\ \hline
Die Meldungen der Software sind für mich sofort verständlich.                                                                                                                                             & \multicolumn{1}{l|}{2} & \multicolumn{1}{l|}{3} & \multicolumn{1}{l|}{3} & K \\ \hline
Wenn ich Informationen zu einem bestimmten Eingabefeld benötige, lassen sich diese einfach abrufen.                                                                                                        & \multicolumn{1}{l|}{1} & \multicolumn{1}{l|}{2} & \multicolumn{1}{l|}{1} & 5 \\ \hline
Wenn Befehle in bestimmten Situationen nicht zur Verfügung stehen (gesperrt sind), ist dies leicht erkennbar.                                                                                              & \multicolumn{1}{l|}{5} & \multicolumn{1}{l|}{4} & \multicolumn{1}{l|}{5} & 5 \\ \hline
Auf Wunsch bietet mir die Software neben allgemeinen Erklärungen auch Beispiele an.                                                                                                                       & \multicolumn{1}{l|}{1} & \multicolumn{1}{l|}{1} & \multicolumn{1}{l|}{1} & 5 \\ \hline
Ich kann die Rückmeldungen, die ich von der Software erhalte, eindeutig dem auslösenden Vorgang zuordnen.                                                                                                  & \multicolumn{1}{l|}{1} & \multicolumn{1}{l|}{4} & \multicolumn{1}{l|}{3} & K \\ \hline
Die Software stellt mir auf Wunsch Informationen über die aktuellen Bedien- und Nutzungsmöglichkeiten zur Verfügung.                                                                                       & \multicolumn{1}{l|}{1} & \multicolumn{1}{l|}{3} & \multicolumn{1}{l|}{1} & 1 \\ \hline
Die Software liefert für mich in ausreichendem Maße Informationen darüber, welche Eingaben gerade zulässig sind.                                                                                           & \multicolumn{1}{l|}{1} & \multicolumn{1}{l|}{4} & \multicolumn{1}{l|}{3} & 5 \\ \hline
Es ist für mich unmittelbar ersichtlich, was die Befehle des Systems bewirken.                                                                                                                            & \multicolumn{1}{l|}{5} & \multicolumn{1}{l|}{4} & \multicolumn{1}{l|}{3} & 5 \\ \hline
Die von der Software verwendeten Begriffe sind für mich sofort verständlich.                                                                                                                              & \multicolumn{1}{l|}{5} & \multicolumn{1}{l|}{5} & \multicolumn{1}{l|}{2} & 4 \\ \hline
Die Software bietet mir stets visuelle Hinweise auf die aktuelle Eingabestelle (z.B. durch Markierung,Farbe, Cursorblinken, Mauscursor etc.).                                                             & \multicolumn{1}{l|}{3} & \multicolumn{1}{l|}{4} & \multicolumn{1}{l|}{3} & 4 \\ \hline
Es ist für mich eindeutig unterscheidbar, ob die Software Rückmeldungen, Sicherheitsabfragen, Warnungen oder Fehlermeldungen ausgibt.                                                                      & \multicolumn{1}{l|}{1} & \multicolumn{1}{l|}{4} & \multicolumn{1}{l|}{1} & K \\ \hline
Die Software bietet mir gute Bedienungsmöglichkeiten, um mich in Dokumenten (Texten, Datenbanken, Kalkulationsblättern, etc.) zu bewegen.                                                                 & \multicolumn{1}{l|}{5} & \multicolumn{1}{l|}{5} & \multicolumn{1}{l|}{4} & 5 \\ \hline
Mit der Software ist für mich ein einfaches Bewegen zwischen den unterschiedlichen Menüebenen möglich.                                                                                                     & \multicolumn{1}{l|}{5} & \multicolumn{1}{l|}{5} & \multicolumn{1}{l|}{5} & 5 \\ \hline
Die Software bietet mir die Möglichkeit, von jeder beliebigen Menüebene direkt zum Hauptmenü zurückzuspringen.                                                                                             & \multicolumn{1}{l|}{5} & \multicolumn{1}{l|}{5} & \multicolumn{1}{l|}{4} & 5 \\ \hline
Es besteht jederzeit die Möglichkeit, bei einer Befehlseingabe abzubrechen.                                                                                                                               & \multicolumn{1}{l|}{4} & \multicolumn{1}{l|}{5} & \multicolumn{1}{l|}{5} & 5 \\ \hline
Es ist immer einfach, ein gerade benötigtes Bearbeitungsprogramm auszuführen.                                                                                                                             & \multicolumn{1}{l|}{K} & \multicolumn{1}{l|}{5} & \multicolumn{1}{l|}{K} & 5 \\ \hline
Es ist für mich einfach, zwischen unterschiedlichen Bearbeitungsbildschirmen zu wechseln.                                                                                                                 & \multicolumn{1}{l|}{K} & \multicolumn{1}{l|}{4} & \multicolumn{1}{l|}{K} & 5 \\ \hline
Die Software erlaubt mir eine Unterbrechung des Bearbeitungsschrittes, obwohl sie eine Eingabe erwartet.                                                                                                   & \multicolumn{1}{l|}{1} & \multicolumn{1}{l|}{5} & \multicolumn{1}{l|}{K} & 5 \\ \hline
Die Bedienmöglichkeiten der Software unterstützen eine optimale Nutzung des Systems.                                                                                                                      & \multicolumn{1}{l|}{K} & \multicolumn{1}{l|}{4} & \multicolumn{1}{l|}{3} & 5 \\ \hline
Das System lässt sich nur in einer starr vorgegebenen Weise bedienen                                                                                                                                       & \multicolumn{1}{l|}{5} & \multicolumn{1}{l|}{K} & \multicolumn{1}{l|}{3} & 3 \\ \hline
Die Auswahl von Menübefehlen kann wahlweise durch die Eingabe von Abkürzungen (Buchstaben oder Transaktionscodes) vorgenommen werden.                                                                      & \multicolumn{1}{l|}{K} & \multicolumn{1}{l|}{3} & \multicolumn{1}{l|}{1} & K \\ \hline
Die Software erlaubt es, einen laufenden Vorgang abzubrechen.                                                                                                                                             & \multicolumn{1}{l|}{5} & \multicolumn{1}{l|}{5} & \multicolumn{1}{l|}{4} & 5 \\ \hline
Die Software erschwert meine Aufgabenbearbeitung durch eine uneinheitliche Gestaltung.                                                                                                                    & \multicolumn{1}{l|}{1} & \multicolumn{1}{l|}{1} & \multicolumn{1}{l|}{2} & 1 \\ \hline
Die Bildschirmdarbietungen (Bedienelemente, Eingabemasken, Fenster, etc.) in einer Bearbeitungssequenz sind für mich vorhersagbar.                                                                          & \multicolumn{1}{l|}{4} & \multicolumn{1}{l|}{5} & \multicolumn{1}{l|}{4} & 5 \\ \hline
Die Bearbeitungszeiten der Software sind für mich gut abschätzbar                                                                                                                                         & \multicolumn{1}{l|}{5} & \multicolumn{1}{l|}{K} & \multicolumn{1}{l|}{4} & 5 \\ \hline
Begriffe und graphische Darstellungen werden in allen mir bekannten Softwareteilen einheitlich benutzt.                                                                                                    & \multicolumn{1}{l|}{K} & \multicolumn{1}{l|}{5} & \multicolumn{1}{l|}{3} & 5 \\ \hline
Gleiche Funktionen lassen sich in allen Teilen der Software einheitlich ausführen.                                                                                                                        & \multicolumn{1}{l|}{K} & \multicolumn{1}{l|}{5} & \multicolumn{1}{l|}{5} & 5 \\ \hline
Die Ausführung einer Funktionen führt immer zudem erwarteten Ergebnis                                                                                                                                    & \multicolumn{1}{l|}{5} & \multicolumn{1}{l|}{5} & \multicolumn{1}{l|}{4} & 5 \\ \hline
Die Möglichkeiten zur Bewegung innerhalb und zwischen allen Teilen der Software empfinde ich als einheitlich.                                                                                              & \multicolumn{1}{l|}{5} & \multicolumn{1}{l|}{5} & \multicolumn{1}{l|}{5} & 5 \\ \hline
Die Meldungen der Software erscheinen immer an der gleichen Stelle.                                                                                                                                       & \multicolumn{1}{l|}{1} & \multicolumn{1}{l|}{K} & \multicolumn{1}{l|}{K} & K \\ \hline
Bei der Arbeit mit der Software kann es passieren, dass auch kleine Fehler schwerwiegende Folgen nach sich ziehen.                                                                                            & \multicolumn{1}{l|}{1} & \multicolumn{1}{l|}{2} & \multicolumn{1}{l|}{3} & 1 \\ \hline
Eingegebene Informationen (Daten, Texte, Grafiken) gehen selbst bei einer Fehlbedienung nicht verloren.                                                                                                   & \multicolumn{1}{l|}{1} & \multicolumn{1}{l|}{K} & \multicolumn{1}{l|}{2} & K \\ \hline
Fehler bei der Eingabe von Daten (z.B. in Bildschirmmasken oder Formulare) können leicht rückgängig gemacht werden.                                                                                        & \multicolumn{1}{l|}{K} & \multicolumn{1}{l|}{5} & \multicolumn{1}{l|}{4} & 5 \\ \hline
Befehle, die Daten unwiderruflich löschen, sind mit einer Sicherheitsabfrage gekoppelt                                                                                                                    & \multicolumn{1}{l|}{1} & \multicolumn{1}{l|}{K} & \multicolumn{1}{l|}{1} & 1 \\ \hline
Ich empfinde den Korrekturaufwand bei Fehlern als gering                                                                                                                                                  & \multicolumn{1}{l|}{5} & \multicolumn{1}{l|}{K} & \multicolumn{1}{l|}{3} & 5 \\ \hline
Eingaben, die ich mache, werden auf ihre Richtigkeit hin überprüft, bevor die Daten weiter verarbeitet werden.                                                                                             & \multicolumn{1}{l|}{1} & \multicolumn{1}{l|}{K} & \multicolumn{1}{l|}{1} & K \\ \hline
Bei meiner Arbeit mit der Software treten Systemfehler (z.B. Absturz) auf.                                                                                                                            & \multicolumn{1}{l|}{5} & \multicolumn{1}{l|}{2} & \multicolumn{1}{l|}{1} & 1 \\ \hline
Mache ich bei der Bearbeitung einer Aufgabe einmal einen Fehler, kann ich die fehlerhafte Operation leicht zurücknehmen.                                                                                   & \multicolumn{1}{l|}{1} & \multicolumn{1}{l|}{K} & \multicolumn{1}{l|}{2} & 5 \\ \hline
Eine Eingabe von mir hat noch nie zu einem Systemfehler (z.B. Absturz) geführt.                                                                                                                       & \multicolumn{1}{l|}{1} & \multicolumn{1}{l|}{2} & \multicolumn{1}{l|}{5} & 5 \\ \hline
Die Software ist so gestaltet, dass das versehentliche Auslösen von Aktionen verhindert wird (z.B. durch Sicherheitsabstände zwischen kritischen Tasten, durch geeignete Benennung, durch Hervorhebungen etc.). & \multicolumn{1}{l|}{2} & \multicolumn{1}{l|}{5} & \multicolumn{1}{l|}{4} & 2 \\ \hline
Die Fehlermeldungen sind gut verständlich und hilfreich.                                                                                                                                                  & \multicolumn{1}{l|}{K} & \multicolumn{1}{l|}{K} & \multicolumn{1}{l|}{K} & K \\ \hline
Bei fehlerhaften Eingaben gibt die Software in einigen Fällen zu spät Rückmeldungen.                                                                                                                      & \multicolumn{1}{l|}{K} & \multicolumn{1}{l|}{K} & \multicolumn{1}{l|}{K} & K \\ \hline
Vor der Ausführung möglicherweise problematischer Aktionen gibt die Software eine Warnung aus.                                                                                                            & \multicolumn{1}{l|}{1} & \multicolumn{1}{l|}{K} & \multicolumn{1}{l|}{1} & K \\ \hline
Die Software bietet mir die Möglichkeit, trotz der Veränderung von Daten, die Orginaldaten weiterhin verfügbar zu halten.                                                                                  & \multicolumn{1}{l|}{K} & \multicolumn{1}{l|}{K} & \multicolumn{1}{l|}{K} & 3 \\ \hline
Die Software bietet mir die Möglichkeit der Anpassung (z.B. bei Menüs, Bildschirmdarstellungen) an meine individuellen Bedürfnisse und Anforderungen.                                                    & \multicolumn{1}{l|}{5} & \multicolumn{1}{l|}{K} & \multicolumn{1}{l|}{1} & 1 \\ \hline
Die Software bietet einfache Möglichkeiten, sie an meinen individuellen Kenntnisstand anzupassen.                                                                                                         & \multicolumn{1}{l|}{5} & \multicolumn{1}{l|}{K} & \multicolumn{1}{l|}{1} & 1 \\ \hline
Ich habe die Möglichkeit, die Menge der auf dem Bildschirm dargestellten Informationen (Daten, Graphiken, Texte, etc.) meinen Erfordernissen anzupassen.                                                  & \multicolumn{1}{l|}{5} & \multicolumn{1}{l|}{5} & \multicolumn{1}{l|}{3} & 1 \\ \hline
Die Software bietet die Möglichkeit, Kommandos,Funktionen, etc. individuell zu benennen.                                                                                                                 & \multicolumn{1}{l|}{5} & \multicolumn{1}{l|}{K} & \multicolumn{1}{l|}{1} & 5 \\ \hline
Spezielle Eigenschaften (z.B. Geschwindigkeit) der Eingabegeräte (Maus, Tastatur, etc.) sind individuell einstellbar.                                                                                    & \multicolumn{1}{l|}{1} & \multicolumn{1}{l|}{K} & \multicolumn{1}{l|}{1} & 1 \\ \hline
Ich kann die Reaktionszeiten der Software an meine individuelle Arbeitsgeschwindigkeit anpassen.                                                                                                          & \multicolumn{1}{l|}{1} & \multicolumn{1}{l|}{5} & \multicolumn{1}{l|}{1} & 1 \\ \hline
Es hat lange gedauert bis ich die Bedienung der Software erlernt habe.                                                                                                                                    & \multicolumn{1}{l|}{1} & \multicolumn{1}{l|}{1} & \multicolumn{1}{l|}{3} & 1 \\ \hline
Auch bei seltenem Gebrauch ist es kein Problem sich wieder in die Software hineinzufinden.                                                                                                                & \multicolumn{1}{l|}{5} & \multicolumn{1}{l|}{5} & \multicolumn{1}{l|}{3} & 5 \\ \hline
Bei Bedarf bekomme ich Hilfestellungen, die das Erlernen der Software erleichtern.                                                                                                                        & \multicolumn{1}{l|}{5} & \multicolumn{1}{l|}{2} & \multicolumn{1}{l|}{1} & 1 \\ \hline
Bisher war es für mich nicht schwer die Bedienung des Software zu erlernen.                                                                                                                               & \multicolumn{1}{l|}{5} & \multicolumn{1}{l|}{5} & \multicolumn{1}{l|}{3} & 5 \\ \hline
Ich konnte die Software von Anfang an alleine bedienen, ohne dass ich Kollegen fragen musste                                                                                                                & \multicolumn{1}{l|}{2} & \multicolumn{1}{l|}{3} & \multicolumn{1}{l|}{3} & 5 \\ \hline
Die Software ist so gestaltet, dass bisher unbekannte Funktionen durch ausprobieren erlernt werden können.                                                                                                  & \multicolumn{1}{l|}{5} & \multicolumn{1}{l|}{4} & \multicolumn{1}{l|}{3} & 5 \\ \hline
Um die Software bedienen zu können, muss ich mir viele Details merken                                                                                                                                      & \multicolumn{1}{l|}{1} & \multicolumn{1}{l|}{1} & \multicolumn{1}{l|}{3} & 1 \\ \hline
Die Bedienmöglichkeiten (z.B. Programmbefehle, Kommandos, etc.) kann ich mir gut merken.                                                                                                                   & \multicolumn{1}{l|}{5} & \multicolumn{1}{l|}{K} & \multicolumn{1}{l|}{3} & 5 \\ \hline

\caption{Aussagen und Ergebnisse der Usability-Befragung (K steht für \textit{Keine Angabe})} \label{anhang-fragebogen}
\end{longtable}

\clearpage
\pagenumbering{gobble}

\end{document}